\documentclass[twocolumn,showpacs,aps,prd,amssymb,nofootinbib,nobibnotes,floatfix,longbibliography]{aastex631}

\usepackage{newtxtext,newtxmath}
\usepackage{ae,aecompl}
\usepackage[mathscr]{euscript}

\usepackage[T1]{fontenc}
\usepackage{hyperref}
\usepackage{amsfonts}
\usepackage{amsmath}
\usepackage{mathrsfs}
\usepackage{epsfig}
\usepackage{graphicx}
\usepackage{url}
\usepackage{hyperref}
\usepackage{float}
\usepackage{color}
\usepackage[utf8]{inputenc} 
\usepackage{epsf}
\usepackage{comment}
\usepackage{slashed}
\usepackage{footmisc}
\usepackage{lipsum}
\definecolor{linkcolor}{rgb}{0.0,0.3,0.5}
\usepackage[caption=false]{subfig}

\usepackage{color}
\definecolor{cerulean}{rgb}{0.0, 0.48, 0.65}

\definecolor{navy}{rgb}{0.2, 0.0, 1.0}

\definecolor{jungle}{rgb}{0.0, 0.5, 0.0}

\usepackage{color}

\definecolor{orange}{rgb}{1,0.5,0}
\definecolor{orangeB}{rgb}{1,0.7,0}

\usepackage[mathlines]{lineno}

\begin{document}


\title{Environmental effects in stellar mass gravitational wave sources I:\\
Expected fraction of signals with significant dephasing in the dynamical and AGN channels.}

\author{Lorenz Zwick}
\email{lorenz.zwick@nbi.ku.dk}
\affiliation{Niels Bohr International Academy, The Niels Bohr Institute, Blegdamsvej 17, DK-2100, Copenhagen, Denmark}
\author{János Takátsy}
\affiliation{Niels Bohr International Academy, The Niels Bohr Institute, Blegdamsvej 17, DK-2100, Copenhagen, Denmark}
\author{Pankaj Saini}
\affiliation{Niels Bohr International Academy, The Niels Bohr Institute, Blegdamsvej 17, DK-2100, Copenhagen, Denmark}
\author{Kai Hendriks}
\affiliation{Niels Bohr International Academy, The Niels Bohr Institute, Blegdamsvej 17, DK-2100, Copenhagen, Denmark}
\author{Johan Samsing}
\affiliation{Niels Bohr International Academy, The Niels Bohr Institute, Blegdamsvej 17, DK-2100, Copenhagen, Denmark}
\author{Christopher Tiede}
\affiliation{Niels Bohr International Academy, The Niels Bohr Institute, Blegdamsvej 17, DK-2100, Copenhagen, Denmark}
\author{Connar Rowan}
\affiliation{Niels Bohr International Academy, The Niels Bohr Institute, Blegdamsvej 17, DK-2100, Copenhagen, Denmark}
\author{Alessandro A. Trani}
\affiliation{Niels Bohr International Academy, The Niels Bohr Institute, Blegdamsvej 17, DK-2100, Copenhagen, Denmark}
\affiliation{National Institute for Nuclear Physics – INFN, Sezione di Trieste, I-34127, Trieste, Italy
}%


%


\date{\today}

\begin{abstract}
\noindent  We present the first overview of the expected quantity of signals which will showcase significant gravitational wave phase shifts caused by astrophysical environments, considering the upcoming A+ and A\# LIGO/Virgo/KAGRA, Cosmic Explorer and Einstein Telescope detectors. We construct and analyse two general families of dephasing prescriptions with extensions to eccentric sources, as well as collect five specific prescriptions for the fundamental smoking gun physical mechanisms at play in the dynamical and AGN formation channel for stellar mass binary black holes: Roemer delays, tidal forces and hydrodynamical interactions. We compute the expected fraction of signals containing astrophysical dephasing, as a function of environmental properties and based on observed distributions of binary parameters. We find that next generation detectors can expect to find environmental effects in hundreds of detected signals.
\end{abstract}


\section{Introduction}
\label{sec:introduction}
\noindent Almost a decade after the first detection of gravitational waves (GWs), many fundamental aspects of binary black hole (BBH) astrophysics remain unanswered: Where are binaries preferentially formed? Which astrophysical mechanisms shape observable binary properties? How do such mechanisms couple to in-spiralling GW sources? In the current framework of population based hierarchical inference, properties of many observed GW sources are aggregated and analysed statistically in order to produce constraints on their astrophysical formation pathways \citep{2002belczynski,2007oleary,2008Sadowski,2016antonini,2017vitale,kavanagh2020,zevin2020,2021zevin,2021kimball,2023santini}. However, despite the increasing number of signals in the LIGO-Virgo-KAGRA (LVK) catalogue, progress is proving to be challenging, as smoking gun features from different environments are hard to distinguish from intrinsic variability of black hole (BH) binary properties \citep{zevin2020}. In recent years, two pathways have emerged to overcome the limitations of population based statistics, which focus instead on extract astrophysical information from a small number of GW signals. The first is the study of eccentricity as a smoking gun signature of dynamical formation channels \citep[e.g.][]{2006ApJ...640..156G, 2014ApJ...784...71S, 2017ApJ...840L..14S, Samsing2018,Takatsy:2018euo, 2018ApJ...855..124S, 2018MNRAS.tmp.2223S, 2018PhRvD..98l3005R, 2019ApJ...881...41L,2019ApJ...871...91Z, 2019PhRvD.100d3010S, 2020PhRvD.101l3010S,2024A&A...689A..24T}, which is becoming more and more relevant as the low-frequency sensitivity of ground based detector improves \citep{samsing2014,samsing2018a,zevin2017,zevin2019}. The second is more elusive, though perhaps hides even more potential: it is the study of so called “Environmental Effects” (EE).

EEs refer to small deviations from an expected vacuum waveform, caused by astrophysical perturbations to the GW source’s dynamics \citep{1993chakrabarti,1995ryan,2008barausse,2007levin,kocsis,2014barausse,inayoshi2017,2017meiron,2017Bonetti,2019alejandro,2019randall,2020cardoso,DOrazioGWLens:2020,2022liu,2022xuan,garg2022,2022cole,2022chandramouli,2022sberna,2023zwick,2023Tiede,2024dyson,2022destounis,2022cardoso,2020caputo,2024zwicknovel,Derdzinksi:2021,2024basu,2024santoro}. The most studied type of EE in GWs consists of a phase-shift, or dephasing, between the vacuum waveform and the perturbed waveform. Dephasing has typically only been considered to be realistically detectable in extreme-mass-ratio inspirals in the milli-Hz band, due to the large number of available GW cycles \citep[see e.g.][]{2019lisa,2024lisa}. However, with the advent of third generation GW detectors it is worth to reconsider whether dephasing may arise in stellar mass binary black holes, as a natural consequence of both the dynamical and the AGN formation channel. In the former, Roemer delays \citep{2017meiron,2024samsing,kai2024,kai22024,2018PhRvD..98f4012R} and tidal fields \citep{liu2015,2021toubiana,2023zwick,2024barandiaran,2024zwick} due to third bodies affect the observed GW frequency and the inspiral rate of binaries. In the latter, gas accretion and drag \citep{2007levin,Derdzinksi:2021,garg2022,2022speri,2024duque} will similarly induce modifications to the secular evolution of binary orbital parameters.

\begin{figure}
    \centering
\includegraphics[width=0.7\columnwidth]{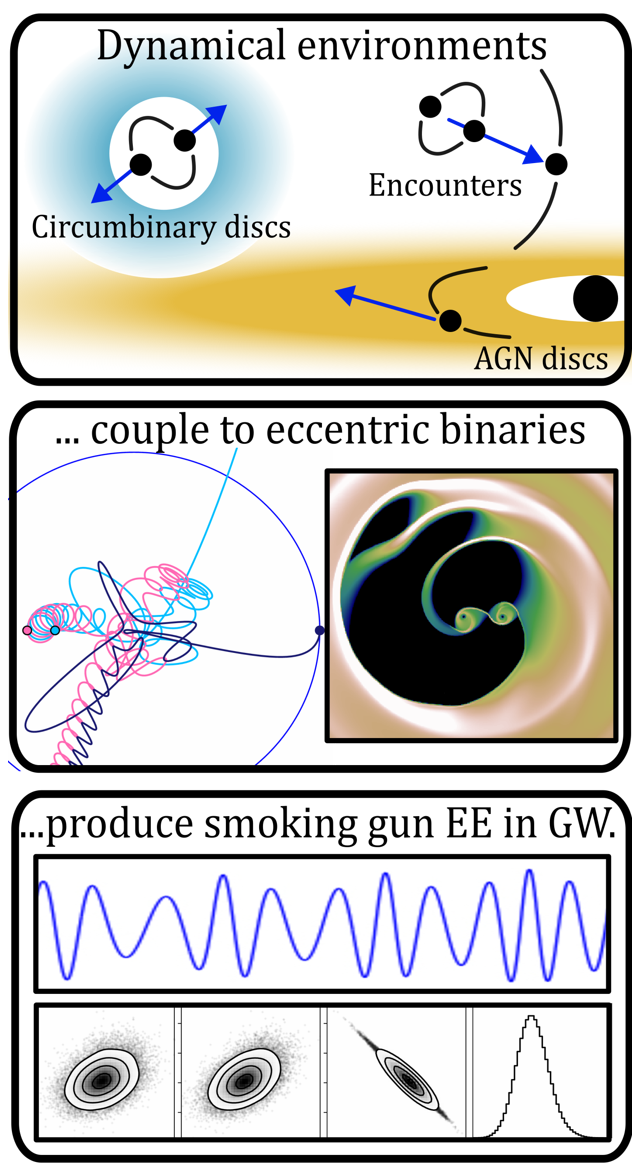}
    \caption{Illustration of the motivating principles for this work. Identifying and inferring eccentricity and/or environmental perturbations in individual binary sources of GW can reveal smoking gun signatures of binary environments.}
    \label{fig:img}
\end{figure}

Despite the existence of many physical mechanisms that induce GW phase shifts, it is still unclear which have realistic prospects for detection, and which regions of binary parameter space are most likely to showcase potentially detectable EEs. While a number of individual studies on the detectability of de-phasing exist, no work has collected and directly compared their relevance for stellar mass binary sources in light of near future ground based detectors. Furthermore, the interaction between environmentally caused de-phasing and eccentricity is largely unexplored (with the exception e.g. promising work on Roemer delays in \cite{2024samsing}, and in \cite{2024duque} on extreme mass ratio systems). This aspect is crucial, as eccentricity is itself considered a smoking gun signature of more complex BH binary formation channels \citep[see e.g.][]{2020shaw,trani2022,samsing2022,trani2024,tra2024b,pin2025,2024stegmann}, and it strongly affects both the GW emission of binaries \citep{peters1964} and their coupling to external forces. As the sensitivity of ground based detectors improves, it is therefore of great importance to revisit the topic of EE and its interaction with eccentricity in order to prepare appropriate waveform models that can extract the maximal amount of astrophysical information from GW signals (see Fig. \ref{fig:img} for an illustration of this principle).

In this work we present an overview of the significance and detectability of dephasing in stellar mass BBHs, which are most likely to arise in both the dynamical \citep{2000ApJ...528L..17P, Lee:2010in,
2010MNRAS.402..371B, 2013MNRAS.435.1358T, 2014MNRAS.440.2714B,
2015PhRvL.115e1101R, 2015ApJ...802L..22R, 2016PhRvD..93h4029R, 2016ApJ...824L...8R,
2016ApJ...824L...8R, 2017MNRAS.464L..36A, 2017MNRAS.469.4665P, 2018MNRAS.tmp.2223S, 2020PhRvD.101l3010S, 2021MNRAS.504..910T, 2013ApJ...773..187N, 2014ApJ...785..116L, 2016ApJ...816...65A, 2016MNRAS.456.4219A, 2017ApJ...836...39S, 2018ApJ...864..134R, 2019ApJ...883...23H, 2021MNRAS.502.2049L, 2022MNRAS.511.1362T} and AGN channel \citep{2017ApJ...835..165B,  2017MNRAS.464..946S, 2017arXiv170207818M, 2020ApJ...898...25T, 2022Natur.603..237S,
2023arXiv231213281T, trani2024, Fabj24}. We construct two families of dephasing prescriptions that have characteristic extensions to high eccentricities, and thoroughly discuss their properties in terms of detectability. Then, we collect or develop five dephasing prescriptions which specifically model smoking gun physical mechanisms at play in the aforementioned formation channels, focusing on Roemer delays, tidal forces and various type of hydrodynamical interactions with a gaseous medium.  We analyse their expected magnitude and detectability considering four different near future ground-based detector configurations, i.e LVK A+ \citep{2022ligopl,2025ligopl} and A\# \citep{2024ligosharp} sensitivities, Cosmic Explorer \citep{2023arXiv230613745E} and the Einstein Telescope \citep{2020maggioreet}, for sources with a small but non-vanishing eccentricity. Furthermore, we briefly comment on the detectability of such EE in LISA. We then estimate the expected fractions of sources with significant environmental dephasing based on currently observed BH binary parameter distributions, providing an expectation as a function of the parameters of the environment. Our results provide a first estimate of the expected fraction of signals containing EEs in ground based GW detectors, based on observed population distribution. They can be used to guide further development of EE models and waveforms containing dephasing prescriptions. An extension of this work to highly eccentric sources is left for paper II. 

The paper is structured as follows: In section \ref{sec:methods} we explain the methodology used to model gravitational waveforms and gravitational wave sources, including a criterion to define significant GW perturbations. In section \ref{sec:dephasing}, we derive and discuss the properties of a set of general dephasing prescriptions and their extensions to higher eccentricity. In section \ref{sec:dephasing_astro} we collect or derive explicit dephasing prescriptions for expected EE in the dynamical and AGN formation channels. In \ref{sec:GB_detectors} we present the results for the expected detectability fractions of the EE as a function of the environmental parameters. We summarise the results and present some concluding remarks in section \ref{sec:conclusion}.

\section{Methodology}
\label{sec:methods}
\subsection{Dephasing SNR criteria}
\label{sec:methods:GW}
\begin{figure}
    \centering
\includegraphics[width=\columnwidth]{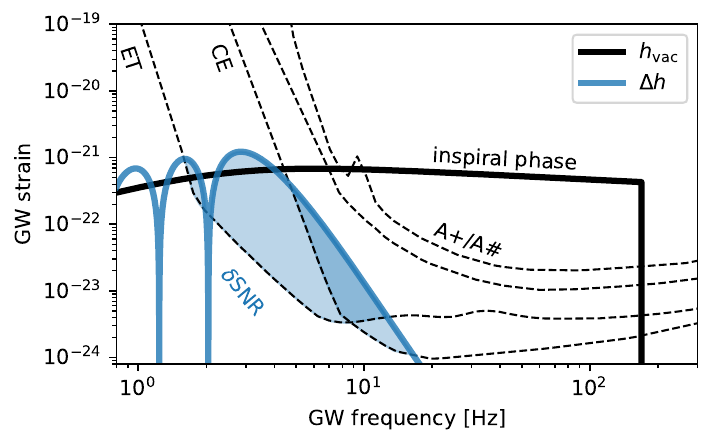}
    \caption{The inspiral phase characteristic strain $h_{\rm vac}$ of a 20 M$_{\odot}$, equal mass binary source at $z=0.3$ (black line), according to our waveform model (see Eqs. \ref{eq:GWnew} and \ref{eq:supphc}). The binary has a reference eccentricity of $e_{10 \,\rm{Hz}}=0.1$. Note how the radiated power at lower frequencies is suppressed due to eccentricity. The blue line denotes the the residual $\Delta h$ between $h_{\rm vac}$ and a phase shifted GW where strong Roemer delays are present (see Eq. \ref{eq:Romerdeph}). The strains are compared to the sensitivities of near future ground based detectors, where the shaded blue areas denote the available $\delta$SNR of the signal (Eq. \ref{eq:dSNRcrit}).}
    \label{fig:strain}
\end{figure}

\noindent The purpose of this work is to extensively survey the parameter space of stellar mass BH binaries in order to identify regions in which EE are likely to be present. In particular we want to finely sample source redshift $z$, rest-frame chirp mass $\mathcal{M}$, reduced mass $\mu$ and eccentricity $e$, as well as any parameters required to describe the binary's environment. We require an efficient method to estimate the relevance of environmentally caused GW perturbations given the properties of the source, environment, and a GW detector. Therefore we employ the $\delta$SNR criterion \citep[see e.g.][]{kocsis,2023zwick}, which states that a waveform perturbation is significant if the following inequality is satisfied:
\begin{align}
\label{eq:dSNRcrit}
   \delta \text{SNR}^2 &\equiv \left<\Delta h,\Delta h \right> > \mathcal{C}^2, \\
   \Delta h&= h_{\rm vac} - h_{\rm EE}
\end{align}
where $h_{\rm vac}$ is the unperturbed vacuum waveform and $h_{\rm EE}$ is the perturbed waveform including EE, and $\mathcal{C}$ is threshold commonly chosen to be $\mathcal{O}(10)$. An example of a reference waveform and the waveform difference $\Delta h$ can be seen in Fig. \ref{fig:strain}, compared to the sensitivity of ground based detectors. The inner product $\left< \cdot , \cdot \right>$ represents the noise weighted SNR of a waveform $h$:
\begin{align}
    \label{eq:innerprod}
    \left< h_1 , h_2\right> = 2\int_0^{\infty} \frac{\tilde{h}_1\tilde{h}_2^{*} + \tilde{h}_1^{*}\tilde{h}_2}{S_{n}(f')}{\rm{d}}f'
\end{align}
where $\tilde{h}_i$ is the Fourier frequency space representation of the waveform $h_i$ and $S_{n}$ is the noise power spectral density of a given detector.

The interpretation of Eq.~\eqref{eq:dSNRcrit} is that the SNR of the difference between unperturbed and perturbed waveforms should reach a certain threshold in order to be significant. In many works, the value of $\mathcal{C}$ is chosen to be $8$, in order to match the typical minimum SNR required for a confident detection in LVK, accounting for various statistical considerations such as the false-positive rate and the “look-elsewhere effect" \citep{2018maggiore}. {This choice is arbitrary, and in reality the actual value depends on the employed waveforms, the detector properties and the specific form of the EE in question. As a general consideration however, we know that EE can only ever be detectable on high SNR signals which have already been identified, and that adding dephasing to templates only introduces one or two additional parameters with respect to the vacuum waveform. Therefore, the actual required threshold is most likely much lower than the typical value of $\delta\mathrm{SNR}=8$ (without considering the problem of degeneracies). Therefore, in this work we will employ a threshold of $\mathcal{C}=3$ to signify the presence of significant dephasing, as it is still an indication that the perturbation may be distinguishable in some favourable cases, or at least introduce some biases in the inference of vacuum parameters.}

Satisfying Eq. \ref{eq:dSNRcrit} does not assure that a perturbation may be distinguished in the full parameter estimation procedure, due to the presence of degeneracies. However, leaving the problem of degeneracies aside, the $\delta$SNR criterion suffices for the purposes of this work, as it indicates whether a waveform perturbation is in principle large enough to result in the presence of significant residual power when the best fit vacuum model is removed. {As we will later show, the $\delta$SNR of dephasing scales approximately linearly with the dephasing amplitude. Therefore, our results can be easily rescaled if a more stringent criterion is required. Given this fact and within the context of the large astrophysical uncertainties, we opt to not analyse the importance of the exact value of the threshold $\mathcal{C}$ in detail.} However, we do note that further work is required to go from the criterion suggested by Eq. \ref{eq:dSNRcrit} to the full parameter estimation problem. Preliminary work on distinguishing EE in a Bayesian context exists and is promising \citep[e.g.][]{2022cole,2022speri,2024zwicknovel,2024garg,2024soumen} 

\subsection{Vacuum waveform model}
\noindent The expected GW signal from a Newtonian inspiralling binary is determined by the following parameters:
\begin{align}
\label{eq:paramssss}
    \text{Vac. par.:} = &\begin{cases}
        z; & \text{Redshift}\\
        \mathcal{M}; & \text{Chirp mass}\\
        \mu; & \text{Reduced mass}\\
        e(f_0); & \text{eccentricity at $f_0$}
    \end{cases}
\end{align}
where $f_0$ is a reference frequency and we neglected binary spins, and averaged over orientation and sky localisation. Additionally, EE will be typically parameterised by a single additional free parameter $\xi$. After extensive testing, we find that satisfactory resolution for this work, which considers four detector sensitivities and five types of environmental effect, requires the evaluation of the $\delta$SNR criterion for approximately $10^8$ waveforms. Therefore, we are limited to using extremely efficient waveforms that can be evaluated precisely in Eq. \ref{eq:dSNRcrit} in a time of order few ms. This completely excludes the possibility of estimating detectability with more sophisticated methods such as parameter inference with e.g. Monte-Carlo Markhov-Chain pipelines. 

Here we employ a modification to the well know analytical Newtonian waveform result in the stationary phase approximation, which for circular binaries reads \citep{1994cutler,blanchet2014}: 
\begin{align}
\label{eq:GWnew}
    \tilde{h}(f) = \frac{Q}{D(z)}\left(\frac{G\mathcal{M}_z}{c^3}\right)^{5/6}f^{-7/6} \exp \left[ i (- \psi_{\rm vac}) \right],
\end{align}
where $f$ is the observer frame GW frequency of the fundamental mode, $\mathcal{M}_z$ is the red-shifted chirp mass, $D(z)$ is the luminosity distance, $Q = 2/5$\footnote{Third generation detectors may have different configurations \citep{2012PhRvD..86l2001R}, slightly changing the value of Q.} is a geometric pre-factor that accounts for projections of the wave onto an L-shaped interferometric detector in the long wavelength limit \citep{2019robson}. The Fourier domain phase $\psi_{\rm vac}$ is given by:
\begin{align}
    \psi_{\rm vac} &= - \frac{\pi}{4}-  \frac{3}{4}\left(8\pi \mathcal{M}_z f \right)^{-5/3},
\end{align}
where we neglected a constant phase offset $\phi_{\rm c}$ and an integration constant $t_{\rm c}$ representing the phase and time at coalescence, as they do not influence the $\delta$SNR results.

To adapt Eq. \ref{eq:GWnew} to low eccentricities, we note that $\tilde{h}(f)^2 = 2 T_{\rm in}f\times S_{\rm GW}(f)$, where $S_{\rm GW}(f)$ is the GW's power spectral density and $T_{\rm in}$ is the binary's in-spiral timescale \citep[][]{peters1964,2013gair,2020MNRAS.495.2321Z}. For $e\lesssim 0.2$, we can assume that the majority of the orbital binding energy is radiated in the fundamental GW mode at twice the orbital frequency, i.e. $f = 2 \Omega_{\rm K}/(1+z)$, where $\Omega_{\rm K}$ is the binary's orbital frequency in the rest frame \citep{2024vijakumar}. Then, the only effect of eccentricity is to reduce the inspiral timescale, which causes the radiated power to be suppressed for higher eccentricities:
\begin{align}
    \label{eq:supphc}
    \tilde{h}(f) = \tilde{h}(f)^{e=0}\times \sqrt{\frac{\dot{f}_{e=0}}{\dot{f}(e)}}.
\end{align}
Note that the phase of the GW is also modified by eccentricity. In the small eccentricity limit, the phase can be decomposed into circular part and eccentric corrections \citep{Moore:2016qxz}.
\begin{align}
    \psi_{\rm vac} = \psi_{\rm vac}^{e=0} + \Delta\psi^{\rm ecc}(f, e^2) \,.
\end{align}
However, this aspect is irrelevant in our $\delta$SNR calculations as we are looking at differences in phases, and the vacuum phase evolution is subtracted away. We note that here, and for the remainder of our paper, the eccentricity $e$ specifically denotes the value of the binary's eccentricity at a given fundamental GW frequency $f$. For the eccentricity evolution, we use the following fit, adapted from \cite{2009yunes}:
\begin{align}
    e(f) &\approx \frac{16.83 - 3.814\, b(f)^{0.3858}}{16.04 + 8.1b(f)^{1.637}} \\
    b(f)&=\left(\frac{f}{f_0}\right)^{2/3}\frac{(1+\frac{121}{304}e_0^2)^{870/2299}}{1-e_0^2}e_0^{12/19},
\end{align}
where $e_0$ is the eccentricity at a reference frequency $f_0$. We note again that the adopted modeling of eccentricity in this paper is only appropriate for small values, and that a more thorough analysis of eccentric sources will be performed in paper II. In this regime, the majority of energy is radiated in the fundamental GW mode at twice the orbital frequency. In fact, Newtonian waveforms of the form shown in Eq. \ref{eq:GWnew} typically give a slightly conservative estimate of the optimal SNR for binaries as they do not account for the additional energy radiated via higher order harmonics and post-Newtonian fluxes \citep{2009yunes}. Additionally, higher modes caused by the presence of eccentricity may enter the band of a given detector while the binary is orbiting at lower frequencies, which can lead to higher overall SNRs \citep{2009yunes}. Furthermore, in terms of $\delta$SNR results for phase shifted waveforms, the precise form of the vacuum waveform does not affect our results beyond providing a slightly different overall SNR normalisation, i.e. $= \left<h_{\rm vac},h_{\rm{vac}}\right>$.

It is crucial to state that a complete analysis of the extensions of de-phasing prescriptions to $e \sim 0.2$ and above do in fact require eccentric waveforms, for the following two reasons. Firstly, dephasing will accumulate faster in the phase of higher order GW modes since they scale with multiples of the binary orbital frequency. Secondly, highly eccentric binaries may enter the band of a given detector at comparatively large separations (lower orbital frequencies), where EE are typically stronger. The effect of including higher order eccentric harmonics on the $\delta$SNR of dephasing is highly interesting and will be the subject of paper II.

\subsection{Binary population}
\label{sec:methods:distributions}
\noindent In this work, we wish to estimate the fraction of sources which will present a significant dephasing depending on a choice of detector, environmental properties and binary parameters. For this reason, we require estimates for the distributions of intrinsic binary parameters over which we can marginalise $\delta$SNR results. As mentioned in the introduction, the determination of the intrinsic distributions of BH binary parameters is a great achievement of the current LVK instruments. It is the current understanding that the observed distributions emerge as a combination of a small number of distinct  binary formation pathways, most commonly separated into the isolated, dynamical and AGN binary formation channels \citep[see e.g.][]{2021zevin}\footnote{Other pathways such as the primordial BH formation channel may also contribute to the detected rates \citep{deluca2021,2016PhRvD..94h3504C}}. A significant amount of population synthesis works have produced estimates of the expected rates of detectable GW signals originating from these channels, often resulting in rates comparable with the overall LVK detections. However, disentangling the individual contribution of each formation channel within the data remains a challenge, and many conflicting claims exist in the literature.

\begin{figure}
    \centering
\includegraphics[width=0.9\columnwidth]{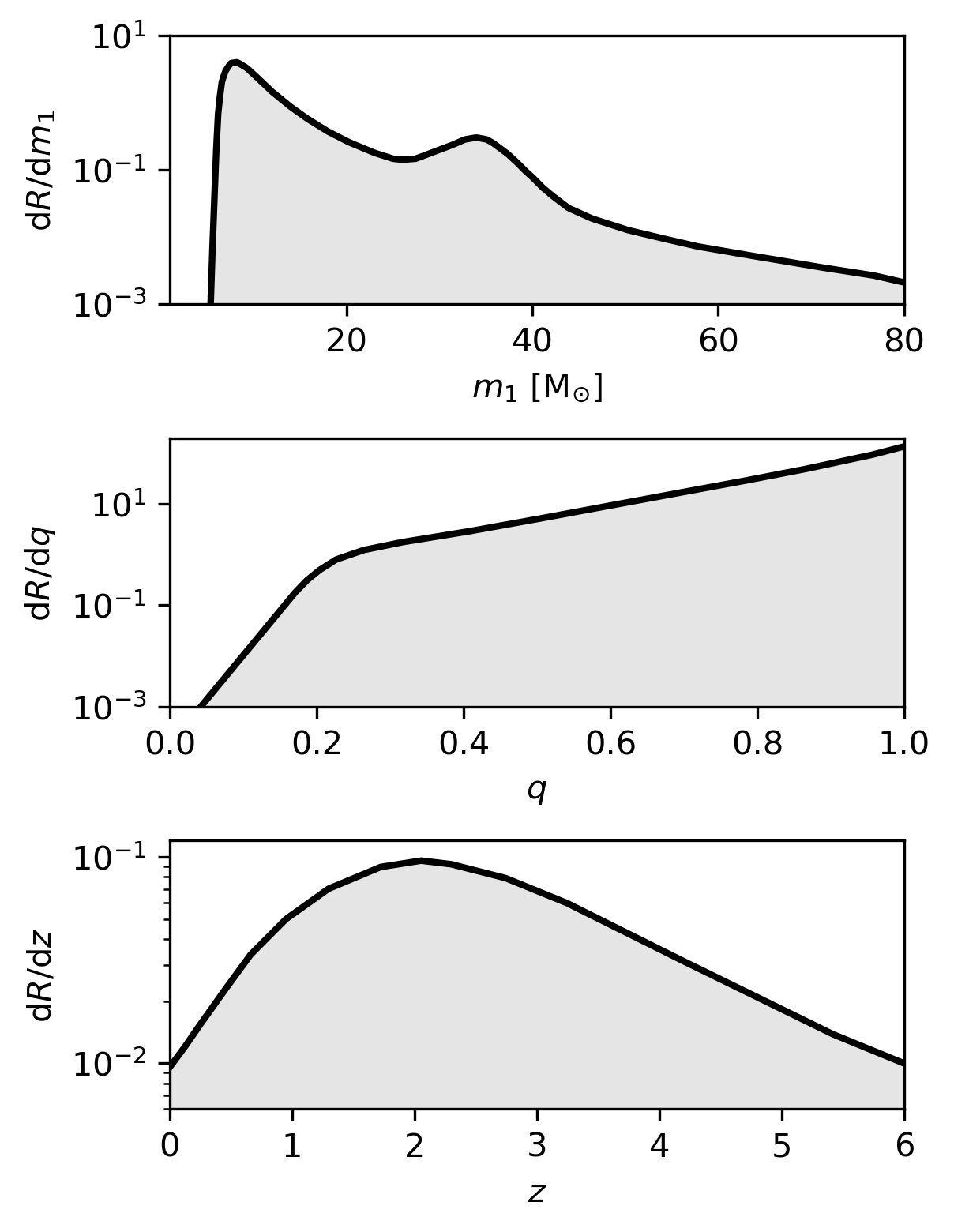}
    \caption{Differential merger rates $R$ of BH binaries with primary mass $m_1$, mass ratio $q$ at redshift $z$. The panels above are reconstructions of the third LVK observation run \citep{Abbot2023}. The distribution in redshift is extended by extrapolating with a simple fit of the observed star formation rate \citep{2014madau,2021jo}. {The units here are irrelevant as we only use these rates as normalised distributions.}}
    \label{fig:Ligodistr}
\end{figure}

For the purposes of this work, we make the following simplifying assumptions regarding the distribution of binary parameters. Firstly, we assume that the overall observed LVK distributions in binary mass, mass ratio and redshift is representative of any given formation channel that showcases a certain EE. Then, our results for the expected detection rates for different EE can be adjusted by simply multiplying the total detection fraction with an efficiency parameter $\epsilon$ that accounts for the fraction of sources that originate from that particular channel {(or more precisely the fraction of sources that present an EE of a given strength, see section \ref{sec:GB_detectors})}. Secondly, we assume that the distributions for the parameters are independent of each other, i.e.:
\begin{align}
    \mathcal{P}(z\lvert\mathcal{M}\lvert\mu\lvert e) \sim \mathcal{P}(z) \mathcal{P}(\mathcal{M})\mathcal{P}(\mu)\mathcal{P}(e),
\end{align}
where $\mathcal{P}(x)$ is the probability of finding a binary with parameters $x$. While both of these assumptions are most likely inaccurate \citep[see e.g.][]{belczynski2002,compas2022}, they serve as a first order approximation to compute expectations for the detectability of de-phasing within the rather large uncertainty of the properties of binary environments. The distributions of intrinsic binary properties are taken from \cite{Abbot2023}, and plotted in Fig. \ref{fig:Ligodistr}. Here we make the standard identification $\mathcal{P}(x) \sim {\rm{d}}R/{\rm{d}}x$, where $R$ is the binary merger rate as reconstructed from LVK observations. Note that here we are referring to the intrinsic binary merger rate rather than the observed one.

In this particular work, we do not perform an extensive analysis of the variation of the detection fractions as a function of the population parameter distributions. Instead, we base the results on the observed LVK binary merger rates. Since the eccentricity distributions are still unconstrained, we simply assume a half-Gaussian PDF between 0 and 0.2 for the binary eccentricity. We can then vary the standard deviation $\sigma _{e}$ to model more or less eccentric binary populations:
\begin{align}
\mathcal{P}\left(e\left(f_0\right)\right) \sim \exp\left(-\frac{e(f_0)^2}{2\sigma_e^2} \right).
\end{align}
Our baseline choice will be  $\sigma_e =0.01$, as it is lower than the currently estimated limit of detectability for LVK \citep{Favata:2021vhw, Saini:2023wdk}, and there are only potential signs of eccentricity in a small number of detected sources \citep{2021ApJ...921L..31R}. Another sensible choice for the eccentricity distribution is a log-normal distribution with $\sim -2$ as the mean and a standard deviation $1$ dex. We note here that neither of these represents the eccentricity distribution of the underlying populations. Here we only consider a choice of $\mathcal{P}(e)$ as a baseline and find that eccentricity distribution plays a minor role in determining the detectability of EE at this level of analysis. We reserve a more thorough treatment of eccentricity distributions for paper II, in which we also consider proper eccentric waveforms. Secondly, we perform tests in which we vary the SNR threshold required to consider a GW source to be detectable. The implications of varying this parameter are discussed in section \ref{sec:GB_detectors}.

\section{General de-phasing prescriptions}
\label{sec:dephasing}
\subsection{De-phasing families}
\noindent GW dephasing can arise in a wide range of astrophysical scenarios. Despite the large variety, we can collect the vast majority of cases into three distinct mechanisms (see \cite{2014barausse} for a review of EE in small mass ratio systems, (and \cite{pedpaper} for an extensive first principles derivation of the properties of phase shifts). The first is through direct perturbations to the sources inspiral rate $\dot{f} \to \dot{f}_{\rm vac}+\delta \dot{f}$, such that:
\begin{align}
\label{eq:secshift}
\delta \phi \sim \int \int \left( \delta \dot{f} \, {\rm{d}}t\right) \, \, {\rm{d}}t \sim  \frac{1}{2} \delta \dot{f} \times T_{\rm obs}^2,
\end{align}
where $\delta \phi$ is the GW dephasing in the time domain, and $T_{\rm obs}$ is the observation time. Note that here we used a almost-monochromatic approximation where $\delta \dot{f}$ is constant. This is often appropriate as de-phasing tends to accumulate primarily in the very early inspiral phases of binary evolution.

The second mechanism is through the accumulation of a time dependent Doppler shift $\delta f \sim v/c$, where $v$ is the line of sight velocity and $c$ is the speed of light. While one would expect de-phasing of the form:
\begin{align}
    \label{eq:dopplshift}
  \delta \phi \sim \int v/c\, {\rm{d}}t \sim \delta f \times  T_{\rm obs},
\end{align}
a constant Doppler shift is fully degenerate with the binary's source mass \footnote{This limitation may be overcome in the case where multiple de-phased images of the same source are present, such as the case with strongly lensed GW sources \citep{2024samsinglensing1,2025samsinglensing2,zwicklensing}.}. Therefore, the significant de-phasing due to Doppler shifts is more appropriately defined as the one accumulating through the source acceleration, rather than velocity \citep[see e.g.][]{2017meiron,2024MNRAS.527.8586T}:
\begin{align}
    \label{eq:dopplshift}
  \delta \phi \sim \int\left(\int \dot{v}/c\, {\rm{d}}t\right) \, {\rm{d}}t \sim \frac{1}{2}\delta \dot{f} \times  T_{\rm obs}^2.
\end{align}
Despite the different origin, this effect reduces to the same scaling with the inspiral timescale as in Eq. \ref{eq:secshift}.

The third type of mechanism to produce de-phasing occurs whenever some external potential causes the binary's binding energy to be modified $E_{\rm B} \to E_{\rm B}+ \delta E$, while the in-spiral rate remains unaffected \citep{2014barausse,2023zwick}. Then, the additional binding energy requires time to be radiated away, producing a change in the in-spiral timescale $\delta T/T_{\rm in} \sim\delta E/E_{\rm b}$. The total phase $\phi_{\rm tot}$ of the binary is then modified to:
\begin{align}
  \phi_{\rm tot} = \int_0^{T_{\rm obs}+ \delta T_{\rm}} f\, {\rm{d}}t,
\end{align}
which implies that the de-phasing scales linearly with the inspiral timescale, which for chirping binaries is the limiting factor in the observation, $T_{\rm obs} \to T_{\rm in}$:
\begin{align}
    \label{eq:binddephscale}
    \delta \phi \sim f\, T_{\rm in}\times \delta E/E_{\rm b}.
\end{align}
Note here that de-phasing due to constant Doppler-shifts also shares this linear scaling in the in-spiral timescale.

Here we propose a general parametrisation for these various mechanisms of de-phasing in a vein similar to \cite{2020cardoso}. The parametrisation is additionally categorised into two families of functions, each with a distinct extensions to moderate eccentricities, though these will be discussed in more detail in paper II. First, recall that the quantities $\delta \dot{f}$ and $\delta E$ encode information regarding the coupling between the binary and its environment. In general, they are functions of a time variable $t$, as well as the binary and environmental parameters. In this work, we only treat orbit averaged quantities (see \cite{2022zwick} and also \cite{2024zwicknovel} for an initial treatment of non-orbit averaged environmental effects), such that $\delta \dot{f}=\delta \dot{f}(\xi\lvert f,e,\mathcal{M},\mu)$ and $\delta E=\delta E(\xi\lvert f,e,\mathcal{M},\mu)$, where once again $\xi$ is the parameter (or set of parameters) describing the environment.

Recall the inspiral rate of a binary emitting GW at an observer frame frequency $f$:
\begin{align}
    \label{eq:chirpf}
    \dot{f}_{\rm vac} = \frac{96}{5} \pi ^{8/3}\left(\frac{G \mathcal{M}_z}{c^3}\right)^{5/3}f^{11/3} \times F(e),
\end{align}
where:
\begin{align}
     F(e) = \left(1+\frac{73}{24}e^2 + \frac{37}{96}e^4 \right)(1-e^2)^{-7/2}.
\end{align}
The resulting GW driven inspiral timescale $T_{\rm in}$ does not have an explicit analytical solution valid for all initial eccentricities, and several approximation schemes have been proposed. Here we adopt the following approximation from \cite{2020MNRAS.495.2321Z, 2021zwick}:
\begin{align}
    \label{eq:insp_time}
    T_{\rm in} \approx T_{\rm in}\lvert_{e=0} \times F(e)^{-1} \times R(e)
\end{align}
where $T_{\rm in}\lvert_{e=0}$ is Peters' timescale for circular orbits and the factor $R(e)=8^{1 - \sqrt{1-e}}$ approximates the effect of consistently evolving eccentricity and semimajor axis. Combining the latter equations with the two scalings suggested by Eqs. \ref{eq:secshift} and \ref{eq:binddephscale}, we propose two de-phasing families of the form:
\begin{align}
\label{eq:dephfamI}
    \delta \phi^{I}_{i,j,k;n}(f) &=  \mathcal{A} \times \mathcal{M}^{i} \mu^{j} f^{n} \mathcal{F}_k^{I}(e) ,\\
    \label{eq:dephfamII}
    \delta \phi^{II}_{i,j,k;n}(f) &=  \mathcal{A} \times \mathcal{M}^{i} \mu^{j} f^{n}\mathcal{F}_k^{II}(e),
\end{align}
where:
\begin{align}
    \mathcal{F}_k^{I}(e) &= e^k\left(\frac{T_{\rm in}}{T_{\rm in}^{e=0}} \right) ^2 \approx e^k\left(\frac{R(e)}{F(e)}\right)^2 \\\mathcal{F}_k^{II}(e) &= e^k\left(\frac{T_{\rm in}}{T_{\rm in}^{e=0}} \right)  \approx e^k\left(\frac{R(e)}{F(e)}\right).
\end{align}
These two families encode all possible scalings of dephasing prescriptions with binary parameters, provided that the quantities $\delta \dot{f}$ and $\delta E$ are power-laws in $f$, $\mathcal{M}$, and are polynomial in $e$. Crucially, this is typically the case for all current models of astrophysical EEs in the orbit averaged limit.

The eccentricity scaling of these de-phasing families is a mostly novel aspect that requires further discussion. Here we simply comment that if $k=0$ the functions $\mathcal{F}^{I}_k$ and $\mathcal{F}^{II}_k$ limit to the value $1$ for small $e$, and the dephasing prescriptions approach the simple power laws that describe the circular case. This is the typical case for EE and for all of the specific prescriptions analysed in this work. The behavior of the prescriptions for other values of $k$ and the existence of related EE will be discussed in detail in paper II. Finally, we note that the conversion from the GW dephasing $\delta \phi$ to the Fourier domain dephasing $\delta \psi$ appropriate for Eq. \ref{eq:dSNRcrit} is not always trivial, in particular in the presence of a time variable eccentricity \citep[see e.g.][]{2009yunes}. Here we make the assumption that $\delta \psi \sim \delta \phi$, which is correct up to a numerical pre-factor of order unity in the stationary phase approximation, for power-law de-phasing prescriptions \citep{1994cutler}. Of course, this shortcut is only appropriate in the context of a population study, in which the intrinsic uncertainties of the environmental effect range over order of magnitudes.

\subsection{SNR scalings for detectors with power-law sensitivity at low frequency}
\noindent Adding the dephasing prescriptions of Eqs. \ref{eq:dephfamI} and \ref{eq:dephfamII} to eccentric waveform templates may allow to provide constraints on the parameters $i$, $j$, $n$ and $k$, which in most cases uniquely determine the type of environmental effect. The obvious degeneracy between $\mathcal{A}$, $i$ and $j$ may perhaps be resolved if information from higher PN and eccentric harmonics is used, though this requires further investigation. Here we use their general form to calculate the expected scaling of de-phasing $\delta$SNR with binary parameters, for different detectors. This will result in a set of useful inequalities that indicate in what region of parameter space a given EE is likely to be detectable.

From Eq. \ref{eq:dSNRcrit}, one can easily derive that the $\delta$SNR of a de-phased waveform roughly scales as \citep[see e.g][among many other]{kocsis,2021katz,2023zwick}:
\begin{align}
   \delta{\text{SNR}}\propto \delta\phi(f_{\rm in}),
\end{align}
in the small de-phasing limit, and where we assumed that the majority of the dephasing $\delta$SNR accumulates close to the initial frequency $f_{\rm in}$ at which the source enters the band. From this fact alone, we can calculate expected $\delta$SNR scalings as follows. Consider a detector, the low frequency slope of which scales as $f^{d}$. Varying the source chirp mass, $\mathcal{M}$, changes the amplitude of the signal $A \to A'$, which in turn causes the binary to enter in the sensitivity band at a slightly different frequency $f_{\rm in}' = f_{\rm in}\times(A'/A)^{1/d}$. The dephasing is then affected as:
\begin{align}
    \delta \phi &\to \delta\phi' \propto (f_{\rm in}')^{n},
\end{align}
where as before the apostrophe denotes the new values after $\mathcal{M}$ is varied. From the scalings of the de-phasing families (Eqs. \ref{eq:dephfamI} and \ref{eq:dephfamII}) and the strain amplitude (Eq. \ref{eq:GWnew}) we find:
\begin{align}
    \delta {\rm{SNR}} \propto \delta \phi(f_{\rm in}) \propto \mathcal{M}^{(5/6) \times (n/d) + i},
\end{align}
where we only considered the variation in chirp mass and used $A \propto \mathcal{M}^{5/6}$. We can repeat this series of steps for any binary parameter to find:
\begin{align}
    \delta \phi(f_{\rm in})&\propto \left(\frac{\mathcal{M}'}{\mathcal{M}}\right)^{(5/6)\times (n/d) +i } \left(\frac{\mu'}{\mu}\right)^{j}\nonumber \\ &\times \left(\frac{D_l}{D_l'}\right)^{n/d} \left(\frac{1+z'}{1+z}\right)^{n + (5/6)\times(n/d)} \nonumber \\ &\times \frac{\mathcal{F}_k(e_{\rm in}')}{\mathcal{F}_k(e_{\rm in})},
\end{align}
where the scalings originate from the variation in the frequency at which the fundamental GW mode enters the detector band. Note the complex scaling in redshift which accounts for the transformations between frequency and chirp mass in the source vs observer frame. In terms of total mass and mass ratio we have  $\mathcal{M}= M (q/(1+q)^2)^{3/5}$ and $\mu = Mq/(1+q)^2$. Then the scalings read:
\begin{align}
   \delta \phi(f_{\rm in})&\propto  M^{(5/6)\times (n/d) +i +j } \nonumber \\ & \times \left(\frac{q}{(1+q)^2}\right)^{(3/5)i + (1/2)(n/d) +j}.
\end{align}
We can now include the effect of the binary signal itself, and integrate over the chirp. We limit our calculations to detectors with an approximately flat sensitivity curve over several frequency decades, which approximates all detectors for sources that chirp significantly. For a detector with flat sensitivity, $S_{\rm n} \propto f^{-1}$, and according to Eq. \ref{eq:innerprod} and Eq. \ref{eq:GWnew}, SNR$_{\rm vac}$$^2 \propto \int f^{-14/6 +1}$. Therefore, the vacuum SNR of vacuum waveforms scales as $  \propto  f^{-1/6}$ \citep[see also][]{}. This is a very shallow power law, which shows that the total SNR depends strongly on both the initial ($f_{\rm in}$) and final ($f_{\rm fin}$) frequency of the observation. In terms of the $\delta$SNR we have instead:
\begin{align}
    \frac{\rm{d}}{{\rm{d}}f} \left(\delta \rm{SNR}(f)^2\right) \propto \frac{\tilde{h}(f)^2 \delta \phi(f)^2}{S_{n}(f)} \propto f^{-14/6+1 + 2n}.
\end{align}
Performing the integral we have:
\begin{align}
    \delta \rm{SNR}(f) \propto f_{\rm fin}^{n-1/6} - f_{\rm in}^{n-1/6}.
\end{align}
Considering that the parameter $n$ is typically a negative value, we see that the integral is dominated by the initial frequency and therefore $\delta$SNR $\propto f_{\rm in}^{n-1/6}$. Given that $f_{\rm in}\propto \mathcal{M}^{5/6\times1/d}D_{\rm L}^{1/d}(1+z)^{5/6\times 1/d}$ and that the vacuum SNR $\propto \mathcal{M}^{5/6}D_{\rm L}^{-1}$ we obtain, for a mostly white noise detector with a low frequency slope of $f^{-d}$, a set of $\delta$SNR scalings:
\begin{align}
    \delta {\rm SNR} &\propto D_{\rm l}^{-N/d -1}(1+z)^{n + (5/6)\times(N/d)}, \\
     \delta {\rm SNR} &\propto M^{5/6\times(1 + N/d) + i +j}, \\
     \delta {\rm SNR} &\propto \left( \frac{q}{(1+q)^2}\right)^{3i/5 + 1/2 \times(1 + N/d )+j},\\
     \delta {\rm SNR} &\propto \mathcal{F}_k(e_{\rm in}),
\end{align}
where we defined $N = n - 1/6$.
These scalings provide us with valuable information about any given EE and dephasing prescription, just by looking at their particular set of parameters $i$, $j$ and $k$ in the dephasing families of Eqs. \ref{eq:dephfamI} and \ref{eq:dephfamII}. For any given detector, we know that an EE will be most significant in a certain region of binary parameter space given by the following inequalities:
\begin{align}
\label{eq:ineq}
   i &\lesssim    -j -\frac{5}{6}\left(1 + \frac{N}{d} \right)  \implies {\rm lower \, \, total\, \, mass}\\
   j &< -\frac{3i}{5} - \frac{1}{2}\left(1 + \frac{N}{d} \right)  \implies {\rm smaller} \, \, {\rm{mass}} \, \, {\rm{ratio}} \\
   k &= [0,1,2,3,...] \implies e_{\rm in}\approx e^{\rm max}_{k}\,\, ,
\end{align}
where we used the fact that the function $q/(1+q^2)$ is monotonic in $q$. These inequalities are only valid in the small de-phasing limit, and eventually lose meaning for sources with low SNR in general. Nevertheless, they shed interesting light on the nature of detecting dephasing. The first inequality states that almost any detector will be more sensitive to de-phasing for lower redshift sources. Inequality \ref{eq:ineq} is only approximate, and more accurate for low mass sources which chirp significantly before merging. It states the un-intuitive result that detectors characterised by different sensitivity slopes at low frequency will be preferentially sensitive to de-phasing in sources of different mass, all else being equal. The behaviour is as follows:
\begin{align}
    d &< -\frac{n-\frac{1}{6}}{1+ \frac{6}{5}(i+j)} &\implies {\text{Det. prefers low mass,}}\\
    d &> -\frac{n-\frac{1}{6}}{1+ \frac{6}{5}(i+j)} &\implies {\text{Det. prefers high mass,}}
\end{align}
where typically $d \in (- \infty, - 1 )$. From these equations, we see that higher masses are always preferred whenever $i+j > -5/6$. For all other cases, a steeper detector sensitivity curve implies that lower masses are preferred. Finally, the third inequality states that whether small or large mass ratios are preferred depends on all parameters $d$, $i$, $j$ and $n$. The behaviour is as follows:
\begin{align}
    d &< -\frac{n-\frac{1}{6}}{1+ \frac{6}{5}i + 2j} &\implies {\text{Det. pref. unequal mass,}}\\
    d &> -\frac{n -\frac{1}{6}}{1+ \frac{6}{5}i + 2j} &\implies {\text{Det. pref. equal mass.}},
\end{align}
Note that in most cases, and in particular for high SNR sources that enter detector at high amplitudes and low frequency, $d\lesssim -3$ and both low mass and unequal mass ratios are preferred due to the corresponding increase in inspiral timescale.

These inequalities open up the intriguing possibility: To define a general a-priori way to inform Bayesian priors on what regions of parameter space are likely to contain dephasing of various forms. The priors would only depend on the detector properties and the specific choice of coefficients $i$, $j$, and $n$ for the dephased waveform model that are used in the parameter inference. While intriguing, this argument is of course only tentative and more work is required to qualify it. Nevertheless, it is indeed the case that the only source that presents evidence for the presence of acceleration in the current LVK catalogue is GW190814 \citep{2024Han}, which showcases a relatively low total mass of $\sim 25$ M$_{\odot}$ and a mass ratio of around 10 \citep{lvcGW190814}.

\section{Astrophysical dephasing prescriptions}
\label{sec:dephasing_astro}
\noindent We now turn our attention to specific models of EE, which are known to affect binaries in realistic astrophysical settings. We select or derive five different dephasing prescriptions, with the justification that their respective EE must necessarily arise as a consequence of the astrophysical formation channel of stellar mass binary BHs, in particular the dynamical channel and the AGN channel. As detailed in section \ref{sec:methods}, we make the simplifying assumption that the population distributions of these channels are broadly consistent with the observed distributions in LVK. Their efficiency, i.e the ratio of the rate at which they produce GW sources, is described simply by an efficiency fraction $\epsilon_{\rm Dyn}$ and $\epsilon_{\rm AGN}$, respectively.
\subsection{Dephasing due to Roemer delays}
\noindent GW phase shifts due to Roemer delays arise in triple systems when the inspiralling binary's centre of mass is accelerated around a tertiary on timescales comparable to the GW driven inspiral. The phase shift is produced by means of a time dependent Doppler shift, as in Eq. \ref{eq:dopplshift}. This means that Roemer dephasing belongs to the dephasing Family I. Identifying Roemer delays in GW signals has first been proposed in \cite{2017meiron} and later extended to eccentric inner binaries and eccentric outer binaries \citep{2019robson,2024samsing,kai22024}. However, their detectability as a function of binary and GW detector properties has not been extensively studied. Here we adopt the following formula for dephasing due to Roemer delays, derived in \cite{2024samsing}:
\begin{align}
    \label{eq:Romerdeph}
    \delta\phi_{\rm R} &= \left(\frac{5}{256 \pi^{13/6}}\right)^2\frac{c^9 m_3 f_{\rm p}^{-13/3}}{R^2G^{7/3}\mathcal{M}^{10/3}}\nonumber \\&\times\frac{\left(1+e(f_{\rm p})\right)^7}  {\left(1-e(f_{\rm p})\right)^{-1/2}}.
\end{align}
Note that here $f_{\rm p}$ is the (redshifted) GW binary's peak frequency, rather than the frequency of the fundamental GW mode. Here $m_3$ is the tertiary's mass and $R$ is the {instantaneous} distance of the tertiary to the centre of mass of the inner binary. We relate the binary's peak frequency to the fundamental GW frequency $f$ (or equivalently 22 frequency) by means of the following fit \citep{hamers2021fit}:
\begin{align}
    f_{\rm p} \approx  f_z\left(1+\sum_{i = 1}^{4} c_i e^i\right)\left(1 - e^2 \right)^{-3/2},
\end{align}
where $c_1 =- 1.01678$, $c_2 =5.57372$, $c_3 =-4.9271$ and $c_4 = 1.68506$. Note that here we introduced $f_z=f(1+z)$ in order to account for cosmological redshift. Evaluated at typical scales, the magnitude of Roemer dephasing for low eccentricity sources reads:
\begin{align}
    \delta \phi_{\rm R}^{e=0} &\approx 0.12 \nonumber \\ &\times \left(\frac{ m_3}{R^2} \frac{\rm{AU}^2}{10\, \rm{M}_{\odot}}\right)\left( \frac{10 \, \rm{M}_{\odot}}{\mathcal{M}}\right)^{10/3}\left( \frac{1\, \rm{Hz}}{f_z}\right)^{13/3},
\end{align}
This particular form of dephasing has a large magnitude, showing how Roemer delays may have ample prospects for detectability \citep{2024samsing,kai22024}, in particular when considering chaotic three-body scattering events \citep{kai2024}.
Reading off the various power laws in the equations above, we identify the dephasing family parameters for Roemer delays: $i_{\rm R}=-10/3$, $j_{\rm R} = 0$, $k_{\rm R}=0$ and $n_{\rm R}=-13/3$. From the inequalities in Eq. \ref{eq:ineq}, we can compute that for detectors with slopes steeper than $d = -11/9$, low total mass is preferred. This is the case for all detectors here considered. Similarly, unequal mass ratios are also preferred for all detectors. Finally, we see that the defining environmental parameter for Roemer delays is $\xi_{\rm R}= m_3/R^2$.

\subsection{Tidal binding energy due to tertiary.}
\noindent Tidal forces arise whenever a third massive body is present in the vicinity of the inspiralling binary. This is a necessary part in both the dynamical formation channel, in which binaries are driven to merger by interactions with other bodies, as well as the AGN formation channel, which necessarily entails the presence of a nearby massive BH. Tidal effects perturb the binary in several distinct ways, which all may induce phase shifts. The most studied is the von Zeipel-Kozai-Lidov \citep[ZKL][]{zeipel1910,koz62,lid62} effect, which consists of an exchange between inner binary eccentricity and inclination and can accelerate the coalescence of compact object binaries  \citep{antonini2016b,antonini2017,silsbee2017,toonen2018,rodriguez2018,vignagomez2021,martinez2020,arcasedda2021,trani2022}. Instead of presenting an exhaustive list of all possible tidally induced perturbations, here we simply analyse the induced change to the binary's binding energy,  $\Delta E_{\rm B}$, as a proxy for any first order tidal EE (Note that ZKL oscillations and their effect on GW in particular have been studied in \cite{2022chandramouli}). The latter is given by \citep{2014will}.
\begin{align}
    \Delta E_{\rm B} = - \frac{1}{4}\frac{G m_3 \mu a^2}{R^3}\left(1-e^2\right) \left( 1 - 3 \sin^2 \iota \sin^2 \omega \right),
\end{align}
where $\iota$ is the inclination between binary and tertiary orbital angular momentum vectors and $\omega$ is the argument of periapsis of the inner binary. As discussed in section \ref{sec:methods} \citep[see also][]{2014barausse,2023zwick}, any additional binding energy requires time to be radiated away via GW. This results in a modification to the GW driven inspiral timescale $T_{\rm in}$ with respect to a vacuum binary \citep{2023zwick}:
\begin{align}
    T_{\rm in} = T_{\rm in}^{\rm vac}\left(1 + 4\frac{\Delta E_{\rm B}}{E_{\rm B}} \right)
\end{align}
Consider now the total phase of a GW signal:
\begin{align}
     \phi \sim \int  f \, {\rm{d}}t  \sim f T_{\rm in},
\end{align}
where once again in the last step we assumed almost monochromatic sources. A change in inspiral timescale automatically produces a change in total phase, and thus a phase shift:
\begin{align}
     \delta \phi \sim  2\pi f\, T_{\rm in} \times 4\frac{\Delta E_{\rm B}}{E_{\rm B}}.
\end{align}
For concreteness, we assume that the binary is co-planar with the tertiary, such that the factor $3 \sin^2 \iota \sin^2 \omega$ vanishes. Then, we can write down a simple prescription for the dephasing that is caused by the dissipation of tidal binding energy:
\begin{align}
    \frac{\delta \phi_{\rm TBE}}{1-e^2} =  \frac{5 f_z^{-11/3}}{64  \pi ^{11/3}}\frac{G m_3}{ R^3 }\left(\frac{G
   \mathcal{M}}{c^3}\right)^{-5/3} \frac{R(e)}{F(e)},
\end{align}
where accounts for the redshift in frequency. For typical parameters, we have:
\begin{align}
    \delta \phi_{\rm TBE}^{e=0} &\approx 0.7 \times 10^{-9} \nonumber \\ &\times\left(\frac{ m_3}{R^3} \frac{(1\,\rm{AU})^3}{10\, \rm{M}_{\odot}}\right)\left( \frac{10 \, \rm{M}_{\odot}}{\mathcal{M}}\right)^{5/3}\left( \frac{1\, \rm{Hz}}{f_z}\right)^{11/3}.
\end{align}
We immediately see that dephasing to the tidal binding energy of the binary is negligible with respect to Roemer delays at relatively large distances. It does however increase rapidly with decreasing tertiary separation $R$. Furthermore, tidal forces may well be produced by a large neighbouring BH with arbitrarily large mass. Finally,  dephasing due to tidal effects presents an interesting angular dependence that can distinguish it from Roemer delay induced dephasing.

Due to the linear scaling with the inspiral timescale, dephasing to to tidal binding energy belongs to Family II. Its dephasing family parameters are $i_{\rm TBE}=-5/3$, $j_{\rm TBE} = 0$, $k_{\rm TBE}=0$ and $n_{\rm R}=-11/3$. The inequalities in Eq. \ref{eq:ineq}, show that, low mass and unequal mass ratio sources are preferred whenever the detector sensitivity curve is steeper than $d< -11/3$. This is the case only for high SNR sources that enter the band of detector high, where sensitivity curves are generally steeper, see Figs \ref{fig:strain} and \ref{fig:deph_scaling}. However, the opposite is true for lower SNR sources, that enter detector bands in a more shallow region of the sensitivity. Finally, we see that the defining environmental parameter for dephasing due to tidal binding energy is $\xi_{\rm R}= m_3/R^3$.

\subsection{Dephasing due to gas accretion and dynamical friction}
\noindent The presence of gas  plays a crucial role in the formation and assembly of compact object binaries in the AGN channel\footnote{And potentially in gas rich clusters as shown in \citep{rozner2022}.} \citep[e.g.][]{Antoni:2019, LiLai:2022, Dempsey3D:2022, rowan2023,2023whitehead, DittmannDempsey:2024,dittmann2024}. With typical densities in the range of $\sim10^{-12}-10^{-9}$g$\,$cm$^{-3}$ within $0.1$pc of the SMBH, gas is known to drive the differential migration of embedded objects through the accretion disc, primarily through Type I or Type II migration torques, analogous to the ones studied in proto-stellar discs \citep{GoldreichTremain:1980, LinPapaloizou:1986}. In the AGN channel, BH binaries may be efficiently formed through the resulting BH-BH scatterings in the plane of the AGN disc through a process known as "gas capture", where the relative two-body energy of the objects is transferred to the local gas \citep{rowan2023,2023whitehead,Rowan2024, LiDempsey:2021}. It has been suggested that gas-capture binaries could account for a significant fraction of merging BBHs in AGN \citep{tagawa2020} and may explain more exotic detections such ass the high mass ratio merger of GW190814 or high mass merger of GW190521 \citep{Rowan2024_rates}. 

If the gas influence is sufficiently strong, it will have a residual effect on the inspiral of binaries even in the GW driven regime, producing potentially detectable EE. In this work, we analyse the two most obvious effects of gas on the orbital elements of an embedded binary, noting however that the hydrodynamics of embedded objects in accretion discs is highly complex, and subject of innumerable theoretical and numerical works \citep{Baruteau:2010bk,LiLai:2022,LiLai:2022,rowan2023,dittmann2024}. The results here presented are therefore only to be understood as representative scalings and order of magnitude approximations of what are inevitably extremely complex physical processes. With this understanding in mind, the first qualitative effect of the presence of a surrounding gas medium is to induce an effective drag force on the binary components as a result of accretion from the medium onto the binary centre of mass \citep{Beckmann2018,fabj2020}. Here we model this with the commonly adopted Bondi-Hoyle-Littleton (BHL) formula \citep{bondi1952}:
\begin{align}
\label{eq:BHLdrag}
    \lvert \vec{F}_{\rm BHL} \rvert = \frac{4\pi G^2 M^2 \rho}{c_{\rm s}^2 + v_{\rm rel}^2},
\end{align}
where $\rho$ is the local gas density, $c_{\rm s}$ is the local speed of sound and $v_{\rm rel}$ is the relative velocity of the binary's centre of mass with respect to the gas. BHL drag represents the case in which accretion is only resisted by gas pressure, and the detailed physics of the interaction of the binary with the gas are neglected. We use it here as a simple limiting case, where more detailed discussions.

In our particular case, the relative velocity term in Eq. \ref{eq:BHLdrag} should be neglectable by only considering binaries that orbit the central SMBH on a close to circular orbit, co-rotating with the accretion disc gas \footnote{In the general case a binary on an eccentric orbit may transition between the sub-sonic and the super-sonic regimes, producing complex accretion patterns and drag forces}. To turn the drag force into a dephasing prescription, we average the power $P$ that is exchanged with the binary over an orbit:
\begin{align}
    P_{\rm BHL} = \frac{1}{2 \pi}\int_{0}^{2 \pi} \vec{F}_{\rm BHL} \cdot \vec{v}(\nu')\, \rm{d}\nu',
\end{align}
where $\nu$ is the true anomaly. {Note that here we are implicitly assuming that gas drag can be applied ``in bulk" to the binary, rather than individually to the binary components. We choose to make this simplification to avoid having to consider how the individual binary components switch between supersonic and subsonic regimes depending on the mass ratio, and to avoid having to distinguish between the induced evolution of the binary and a possible bulk centre of mass acceleration \citep[as in][]{2020cardosoself}. For the purposes of this work, this assumption is justified as the scalings with the physical parameters are preserved.} We can then relate the power to a drift in semi-major axis using the Newtonian relation $E= -GM\mu/(2a)$. Expressed in terms of GW frequency we have:
\begin{align}
    \dot{f}_{\rm BHL} = -\frac{12 f^{2/3}\left(G M \right)^{5/3} \rho}{c_{\rm s}^2 \mu} \times \mathcal{F}_{\rm BHL}^{\rm sub}(e)
\end{align}
where $\dot{f}_{\rm BHL} = P_{\rm BHL}/\left( {\rm{d}} E/{\rm{d}}f \right)$, and $E$ is the binary's energy. Here:
\begin{align}
    \mathcal{F}_{\rm BHL}(e) =  1   + \frac{3 e^2}{4} + \frac{33 e^4}{64} + \frac{107 e^6}{256}+ ...
\end{align}
is a function of eccentricity, that does not have a simple closed form \citep[see also][]{2021zwick}. As expected, BHL drag belongs to the dephasing family I. Similarly to before, we can compute the expected dephasing due to BHL drag in the monochromatic approximation:
\begin{align}
    \delta \phi_{\rm BHL} &= \frac{75}{16384 \pi ^{11/3} }\frac{ c^{10} \mathcal{M}^{5/6} \rho f_z^{-14/3}}{c_{\rm s}^2 
   G^{5/3} \mu ^{7/2}} \nonumber \\& \times\frac{ \mathcal{F}_{\rm BHL}(e) R(e)^2}{F(e)^2}.
\end{align}
For typical values gives:
\begin{align}
    \delta \phi_{\rm BHL}^{ e=0} &\approx 1.3 \nonumber \\ &\times\left(\frac{ \rho}{c_{\rm s}^2} \frac{(10^4\, \rm{m/s})^2}{10^{-10}\, \rm{g/cm}^3}\right)\left( \frac{\mathcal{M}}{10 \, \rm{M}_{\odot}}\right)^{5/6} \nonumber \\
    &\times \left( \frac{5 \, \rm{M}_{\odot}}{\mu}\right)^{7/2}  \left( \frac{1\, \rm{Hz}}{f_z}\right)^{14/3}.
\end{align}
We can read off the BHL drag dephasing family parameters as $i_{\rm BHL} = 5/6$, $j_{\rm BHL} = -7/2$, $k_{\rm BHL}=0$ and $n_{\rm BHL} = -14/3$.  From the inequalites of Eqs. \ref{eq:ineq}, we see BHL drag de-phasing will be preferred for low total masses, provided that the detector sensitivity slope is steeper than $d<-70/33$, and unequal mass ratios for a slope steeper than $d< - 14/15$. The environmental parameter for dephasing due to BHL drag is $\xi_{\rm BHL}=\rho/c_{\rm s}^2$. \newline \newline

\noindent Taking Eq. \ref{eq:BHLdrag} and setting instead the speed of sound to be $c_{\rm s}=0$, we recover the functional form of dynamical friction drag in the supersonic limit, i.e. a $1/v^{2}$ drag \citep{ostriker1999}. This represents the case of drag acting locally on the two individual binary components, which are orbiting around each other at high speed with respect to a hypothetically slow and cold ambient medium. This form of drag represents another limiting case, which is often used as a comparison to numerical investigations, even in the context of fully relativistic treatments of orbiting black holes \citep{2025dyson}. Similarly to before, we derive the resulting dephasing prescription without thoroughly discussing the validity of Eq. \ref{eq:BHLdrag}, as it simply provides well motivated physical scaling and a typical order of magnitude estimate.

Repeating the steps above for supersonic drag, we have:
\begin{align}
    \dot{f}_{\rm sup} = -\frac{12 G M  \rho}{ \mu} \times \mathcal{F}_{\rm BHL}^{\rm sup}(e)
\end{align}
Here:
\begin{align}
    \mathcal{F}_{\rm sup}(e) =  1   - \frac{ e^2}{4} - \frac{7 e^4}{64} - \frac{17 e^6}{256}+ ...
\end{align}
The expected dephasing due to supersonic drag in the monochromatic approximation reads:
\begin{align}
    \delta \phi_{\rm Sup} &= \frac{75}{16384 \pi ^{13/3} }\frac{ c^{10}  \rho f_z^{-16/3}}{ 
   G^{7/3}\mathcal{M}^{5/6} \mu ^{5/2}} \nonumber \\& \times\frac{ \mathcal{F}_{\rm sup}(e) R(e)^2}{F(e)^2}.
\end{align}
For typical values:
\begin{align}
    \delta \phi_{\rm sup}^{ e=0} &\approx 4.5 \times 10^{-12} \nonumber \\ &\times\left(\frac{ \rho}{10^{-10}\, \rm{g/cm}^3}\right)\left( \frac{10 \, \rm{M}_{\odot}}{\mathcal{M}}\right)^{5/6} \nonumber \\
    &\times \left( \frac{5 \, \rm{M}_{\odot}}{\mu}\right)^{5/2}  \left( \frac{1\, \rm{Hz}}{f_z}\right)^{16/3}.
\end{align}
We can read off the supersonic drag dephasing family parameters as $i_{\rm BHL} = -5/6$, $j_{\rm BHL} = -5/2$, $k_{\rm BHL}=0$ and $n_{\rm BHL} = -16/3$.  From the inequalites of Eqs. \ref{eq:ineq}, we see that BHL drag de-phasing will also be preferred in low total masses and unequal mass ratios. The environmental parameter for dephasing due to BHL drag is $\xi_{\rm BHL}=\rho$.

\subsection{Dephasing due to circumbinary disc torques}

\noindent Our analysis of BHL and supersonic drag neglects various aspects of the interaction between the binary and the accretion flow. Binaries surrounded by cold gas are typically described by a viscous circumbinary disc, rather than a radial Bondi flow.  Details of the accretion flow are complex and the subject of numerous theoretical and numerical investigations (see e.g. \cite{LaiMunoz:Review:2022} for a review), but a typical feature is that a near equal-mass binary carves an evacuated central cavity of characteristics size $r_{\rm in } \approx 2a$. A simple analysis (and converse to the pure accretion drag) is that for a system where the tidal barrier of the rotating binary potential identically balances the viscous angular momentum flux at this cavity edge.\footnote{In this picture, no material penetrates the cavity and the binary accretion rate is zero.} If the accretion rate is mediated by viscosity alone, it is given by \citep{Shakura:1973uy,tiede2020}:
\begin{align}
    \dot{M} \sim 3 \pi r^2 \Sigma \tau_{\rm visc}^{-1}
\end{align}
where $\tau_{\rm visc}$ is the viscous timescale, $\Sigma$ is the gas surface density and $r$ is a radial coordinate. For a classical $\alpha$-disk \citep{Shakura:1973uy}, this causes the binary orbit to decay at the associated viscous rate:
\begin{align}
    \frac{a_{\rm CBD}}{a} = -48 \pi \alpha \frac{c_s^2 \Sigma}{\mu \sqrt{1 - e^2}}  f_z^{-1}
\end{align}
where $\alpha \lesssim 1$ characterises the magnitude of disk stresses. Similar to above we can compute the dephasing due to viscous torques in the monochromatic approximation:
\begin{align}
\label{eq:viscdeph}
    \delta \phi_{\alpha} &=- \frac{225 }{8192 \pi ^{10/3}}\frac{\alpha  c^{10} c_{\rm s}^2\Sigma f_z^{-16/3}}{
   G^{10/3}\mathcal{M}^{10/3} \mu  } \nonumber\\ &\times\frac{R(e)^2}{\sqrt{1-e^2}F(e)^2}.
\end{align}
Note that here we have assumed that the cavity edge and the binary remain coupled until the end of the inspiral. Whether or not such binary-disk decoupling occurs in reality remains an active line of study \citep{ArmitageNatarajan:2005, Dittmann:Decoupling:2023, ONeillTiedeDOrazio:2025, EnnoggiCampanelliNoble:2025}, and will not be discussed in detail.

Eq. \ref{eq:viscdeph} represents a model for dephasing for a binary embedded in an isolated, viscous circumbinary disc. However, for binaries embedded in AGN disc, the circumbinary disc will necessarily be embedded in a larger system that may supply gas and angular momentum at higher rates. As an example, mechanisms such as shear or large scale magnetic fields may act to replenish the cavity faster than viscosity alone. In the limiting case, the driving mechanism to drive the angular momentum flux in the circumbinary disc can be characterised by a fast, dynamical timescale. In our setup, the dynamical timescale is the binary's orbital timescale $\tau_{\rm orb}$ around the central massive black hole. Note that this is also equal to the orbital timescale at the binary's Hill sphere, at which the circumbinary disc is coupled to the larger scale AGN disc. With respect to viscosity alone, the angular momentum flux would then be increased by a factor:
\begin{align}
    &\frac{\tau_{\rm visc}}{\tau_{\rm orb}} = \frac{\sqrt{G M_{\rm MBH}}}{2 \pi \alpha c_{\rm s} H R^{1/2}} \nonumber \\ &\approx 6.8 \times10^{4} \frac{0.1}{\alpha}\left(\frac{10^4\, \rm{m/s}}{c_{\rm s}}\right)^2 \left(\frac{M_{\rm MBH}}{10^8\, {\rm{M}}_{\odot}} \right) \left( \frac{0.1\, {\rm{pc}}}{r}\right)^{2},
\end{align}
where $H\sim c_{\rm{S}}/\Omega_{\rm K}$ is the scale height of the disc. Therefore, at typical separations for binaries in the AGN channel, it is exceedingly likely that viscous torques alone are a gross underestimation for the evolution of the binary elements. To model this effect, we introduce a simple fudge factor $f_{\rm CBD}$ to Eq. \ref{eq:viscdeph}, while preserving the scaling appropriate for the evolution of binaries in circumbinary discs:
\begin{align}
    \delta \phi_{\rm visc} &=- \frac{225 f_{\rm CBD} }{8192 \pi ^{10/3}}\frac{\alpha  c^{10} c_{\rm s}^2\Sigma f_z^{-16/3}}{
   G^{10/3}\mathcal{M}^{10/3} \mu  } \nonumber\\ &\times\frac{\left(1+e^2\right)R(e)^2}{F(e)^2}.
\end{align}
Once again, $f_{\rm CBD}$ can potentially be of the order $10^4$ or more depending on the AGN disc model and the detailed hydrodynamics of the embedded binary. For typical values, the dephasing then reads:
\begin{align}
    \delta \phi_{\rm visc}^{ e=0} &\approx 1.3 \times 10^{-4} \nonumber \\ &\times \frac{\alpha f_{\rm CBD}}{0.1}\left( \frac{\Sigma}{10^6\, \rm{g/cm}^2} \right) \left(\frac{c_{\rm s}^2}{10^6 \, \rm{m/s}} \right)\left( \frac{10 \, \rm{M}_{\odot}}{\mathcal{M}}\right)^{10/3} \nonumber \\
    &\times \left( \frac{5 \, \rm{M}_{\odot}}{\mu}\right)  \left( \frac{1\, \rm{Hz}}{f_z}\right)^{16/3}.
\end{align}
Summarising, we see that dephasing due to circumbinary disc torques belongs to family I, and has the parameters $i_{\alpha}= -16/3$, $j_{\alpha}=-1$, $k_{\alpha}=0$ and $n_{\alpha}=-16/3$. It will be preferred for lower mass binaries for detectors with a slope steeper than $d < -80/63$, and unequal mass ratios for detectors with $d<-16/15$. The environmental parameter for dephasing due to circumbinary torques is $\xi_{\alpha}= \alpha \beta \Sigma c_{\rm s}^2$.

\subsection{Summary}
\begin{figure}
    \centering
\includegraphics[width=0.9\columnwidth]{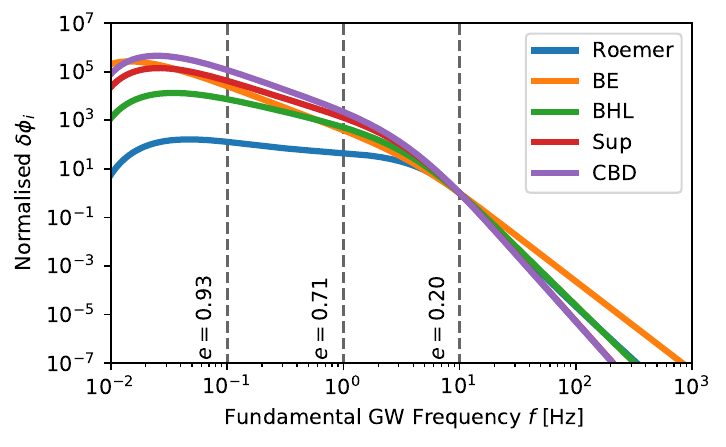}
\vspace{0cm}
\includegraphics[width=0.9\columnwidth]{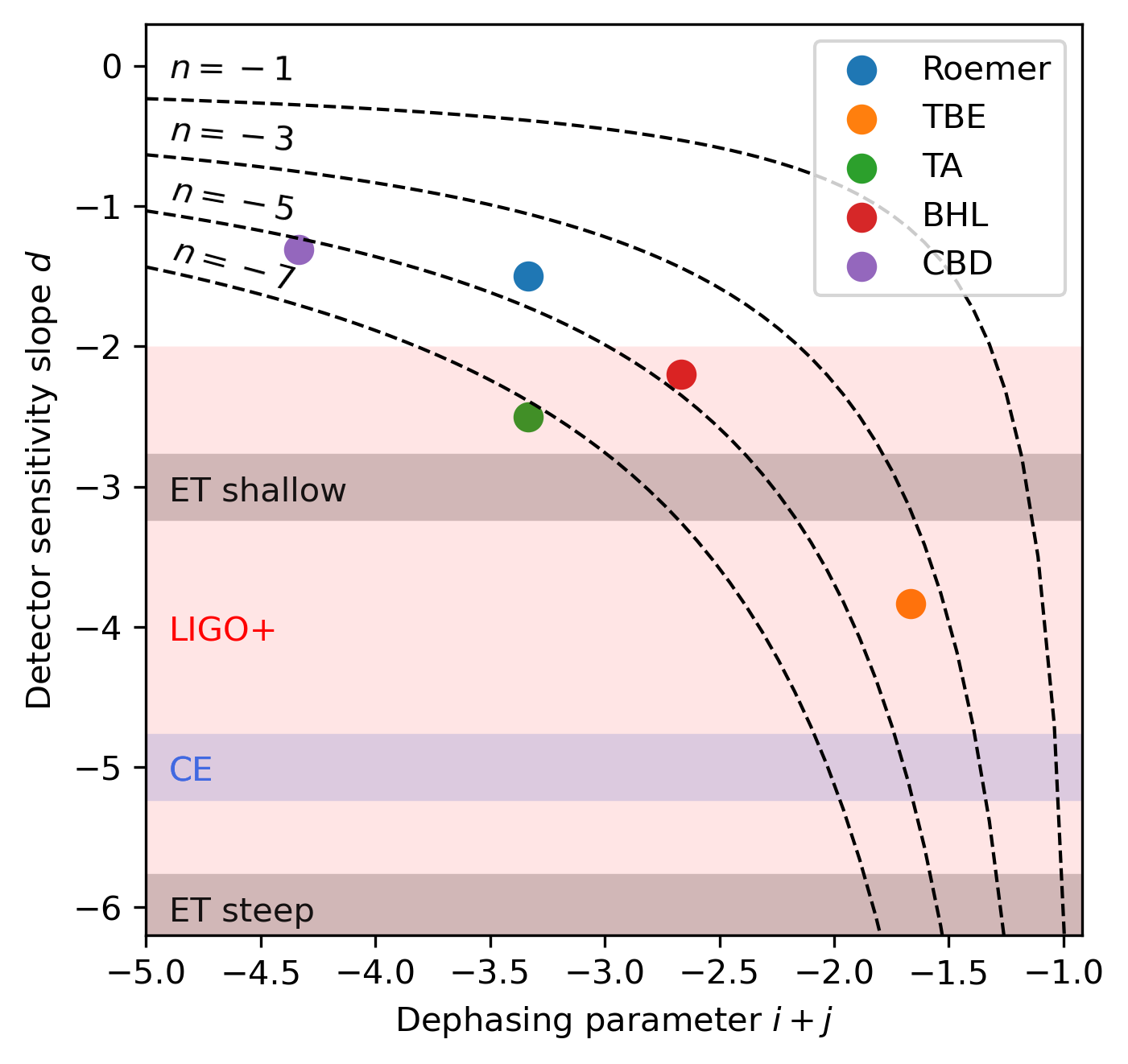}
    \caption{Top panel: evolutionary paths for the five dephasing prescriptions analysed in this work, for a binary with an eccentricity of 0.2 at 10 Hz. The magnitude is normalised. Bottom panel: parameter space plot comparing detector slopes and the dephasing family parameters (see Eqs. \ref{eq:dephfamI} and \ref{eq:dephfamII}). Plotted are the lines (dashed black) under which a given detector will be more sensitive to dephasing for lower mass sources, given a dephasing prescription with fixed $n$ and $i+j$. The five specific dephasing prescriptions analysed in this work are shown as coloured dots. The shaded areas denote the approximate ranges for the slopes of three ground based detectors ET, CE and LVK (A+ and A\#) (see also Fig. \ref{fig:strain}). LVK sensitivity curves span through a large range of slopes due to the characteristic curved shape (see Fig. \ref{fig:strain}). The plot shows how, depending on the specific detector-dephasing prescription combination, it will be more promising to look for EE in higher vs lower mass sources. In the typical case, lower mass sources are more promising in terms of dephasing for steep detector sensitivity curves.}
    \label{fig:deph_scaling}
\end{figure}

\noindent Table \ref{tab:coeff} summarises the family parameters of the five astrophysical dephasing prescriptions we collected or derived. Fig. \ref{fig:deph_scaling} shows their evolutionary paths for an exemplary eccentric binary, as a function of the fundamental GW frequency. Note how the different combination of frequency scaling and eccentricity extensions give rise to unique and potentially distinguishable behaviour. Furthermore, we show a parameter space plot comparing detector slopes and the dephasing family parameters for our choice of prescriptions. Typical detectors will have sensitivity curves with relatively steep slopes at lower frequencies, meaning that dephasing is typically more easily detectable for lower mass sources. In some cases however, e.g. the shallow part of ET's sensitivity curve between $\sim 1$ Hz and $\sim 8$ Hz, an increase in mass can disproportionately increase the total amount of accumulated dephasing by lowering the initial frequency of the observation. These results indicate how the dephasing prescriptions of several crucial EE have unique high eccentricity extensions as well as interactions with different detectors.

\begin{table}
    \centering
    \begin{tabular}{c|c|c|c|c|c}
      EE &Family & $\xi$ & i  & j &  n  \\
      \hline
        Roemer & I & $m_3/R^2$ & - 10/3  & 0  &-13/3 \\
        Tidal BE & II & $m_3/R^3$ & -5/3  & 0 &  -11/3 \\
       
        BHL drag & I & $\rho/c_{\rm s}^2$ & 5/6 & -7/2 &  -14/3\\

        Sup drag & I & $\rho$ & -5/6 & -5/2 &  -16/3\\

         CBD torques& I &$\alpha f_\mathrm{CBD} \Sigma c_{\rm s}^2$ & -10/3 & -1 &  -16/3 \\
        
    \end{tabular}
    
    \caption{Dephasing family parameters of the five dephasing prescriptions treated in this paper. The parameters uniquely identify the EE, and also indicate the preferred parameter space in which to search for the corresponding de-phasing.}
    \label{tab:coeff}
\end{table}

\section{Fraction of sources with significant dephasing}
\label{sec:GB_detectors}
\noindent We now detail our strategy to estimate the fraction of sources that will showcase EE for different detectors. We can evaluate Eq. \ref{eq:dSNRcrit} for an appropriate grid of parameters, after selecting a given dephasing prescription and GW detector. To expedite calculations, we base the results on a reference value of the appropriate environmental parameter $\xi_{\rm ref}$. The parameter $\xi _{\rm ref}$ is chosen such that the evaluated $\delta$SNR is small, and respectively the dephasing caused by the EE is $<< \pi$. In this way, we can rescale the $\delta$SNR results for any choice of $\xi$ by simply multiplying by the factor $\xi/\xi_{\rm ref}$. The result of this procedure is a series of 20 re-scalable, four-dimensional cubes containing the $\delta$SNR contours for the dephasing prescriptions detailed in section \ref{sec:dephasing_astro}.

In Appendix \ref{appendix_GB} (and \ref{appendix_SB} for LISA), we show several 2-d slices of these "$\delta$SNR cubes", focusing on the detector ET. We see how the $\delta$SNR of different dephasing prescriptions scale as a function of the binary parameters, producing complex contours. The shapes of these contours are already quite rich and informative. They represent the numerical counterpart to the discussion of the interaction between dephasing family parameters and detector sensitivity slopes in section \ref{sec:dephasing}. Here we refrain from a thorough discussion, and focus instead on integrating the $\delta$SNR cubes over the population parameter distributions.

Our goal is to estimate the detection fraction of sources with significant dephasing as a function of the environmental parameters $\xi$. To this purpose we employ a Monte Carlo integration scheme. First, we wish to identify the phase space volume $V_{\rm det}$ of all sources that are detectable by a given detector. We sample binary mass, mass ratio, redshift and eccentricity randomly from the following ranges:
\begin{align}
    M_i &\in \left(5\, {\rm M}_{\odot}, 200\, {\rm M}_{\odot} \right) \\
    q_i &\in (0, 1) \\
    z_i &\in (0,5) \\
    e^{10\, {\rm{Hz}}}_i &\in (0,0.2).
\end{align}
Then, we evaluate the SNR of a vacuum waveform with the sampled parameters. Note here that ET and CE are sensitive to sources with masses larger than the current upper limit of $200$ M$_{\odot}$. We introduce this cutoff because the distribution of such sources is entirely unconstrained, and to expedite the convergence of the integration. The phase space volume of detectable sources, for which we use the standard criterion of SNR $>$ 8, is given by the following weighted sum:
\begin{align}
    V_{\rm det} \approx \sum_i^N \Theta({\rm{SNR}_i-8})\times w(M_i,q_i,z_i,e^{10\,\rm{Hz}}_i),
\end{align}
where $\Theta$ is a Heaviside function the $w$ are weights that encode the probability of finding a source with a given set of parameters. In our case, we have $w\equiv \mathcal{P}$, as detailed in section \ref{sec:methods:distributions}. We proceed similarly to compute the phase-space volume $V_{\rm EE}$ for which an EE dephasing prescription with a given parameter $\xi$ is significant (and perhaps detectable assuming that appropriate waveforms exist), i.e. it reaches a $\delta$SNR=3:
\begin{align}
    V_{\rm EE}  \approx \sum_i^N \Theta({\rm{\delta SNR}_i-3})\times w(M_i,q_i,z_i,e^{10\,\rm{Hz}}_i).
\end{align}
The fraction of sources that will showcase a significant dephasing is calculated simply by dividing the two volumes, while multiplying by the appropriate formation channel's efficiency parameter $\epsilon$.
\begin{align}
    {\rm Detection \,\, fraction} = \epsilon\frac{V_{\rm EE}}{V_{\rm det}}.
\end{align}
We test the convergence of the integration by evaluating the volumes for a few random realisations at a given resolution and checking if the results remain unaffected, which is achieved for $N \sim 10^8$. {Note that in this formulation, the value of the efficiency parameter $\epsilon$ is dependent upon the typical magnitude of the EE that is influencing that particular population (denoted by the subscript ``pop"), as described by the environmental parameter $\xi$:
\begin{align}
    \epsilon_{\rm pop}(\xi) \equiv {\text{Fraction of sources in a pop. with } \xi_{\rm EE}= \xi}.
\end{align}
Even within a single formation channel we will in fact have a distribution of values for $\xi$. Therefore, a more accurate assessment of the detection fractions must be performed once such distributions are known, by performing an additional integrals over environmental parameter distributions:
\begin{align}
    \xi \to  \frac{{\rm d \xi}}{{\rm d} \mathcal{J}}(\mathcal{J}),
\end{align}
where the distribution of $\xi$ itself depends on various properties of the environment (here denoted by the variable $\mathcal{J}$). We do not speculate here on the properties of these distributions, and simply comment on plausible ranges of the environmental parameters $\xi$ while assuming a constant value for $\epsilon$. For simplicity, we now omit to specify the efficiencies of the dynamical and AGN channel, i.e. our results are quoted for $\epsilon_{\rm pop}\sim 1$}
\subsection{Detection fractions and required environmental properties}
\begin{figure*}

    \centering
\includegraphics[width=0.9\columnwidth]{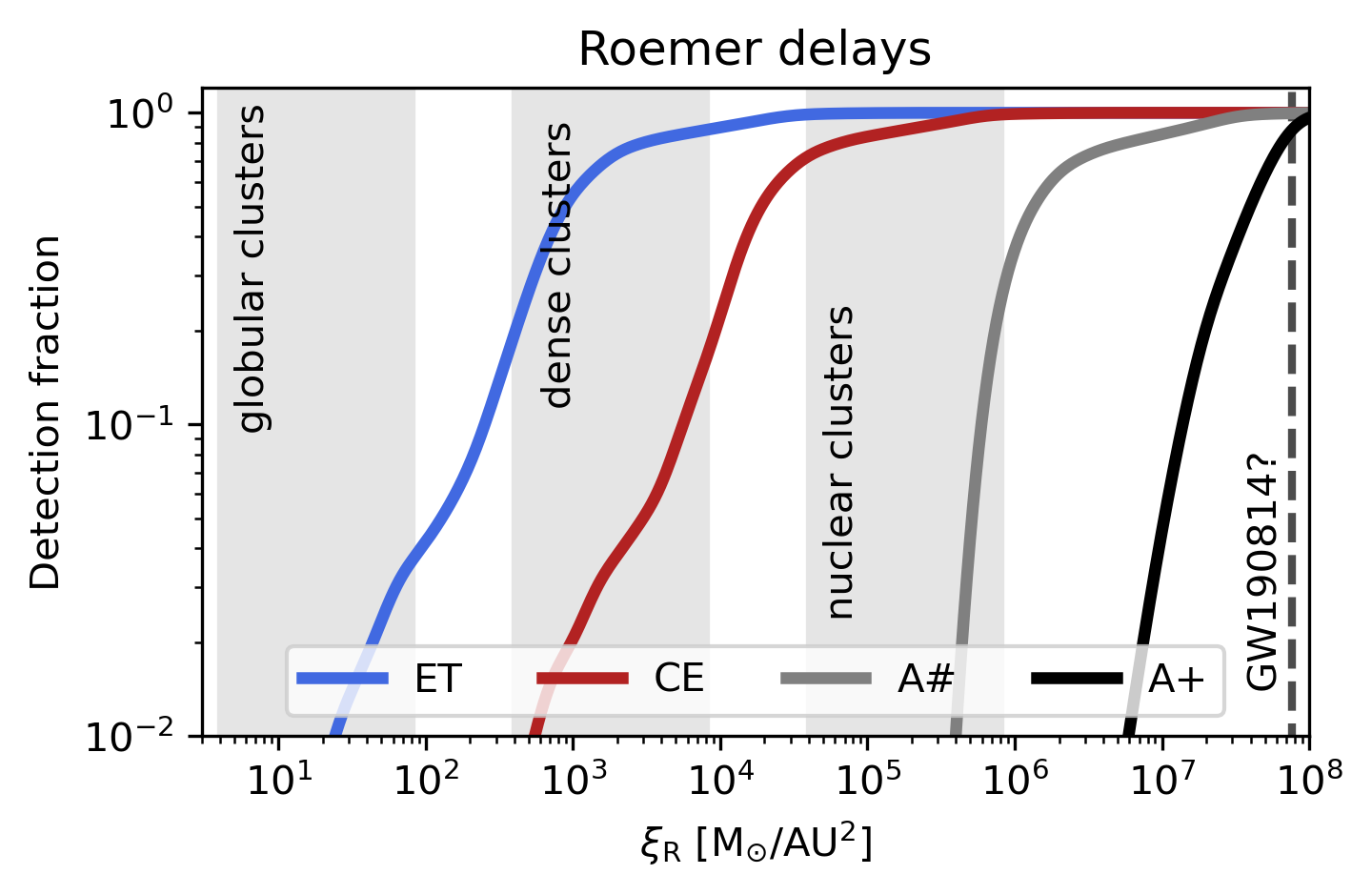}

\includegraphics[width=0.9\columnwidth]{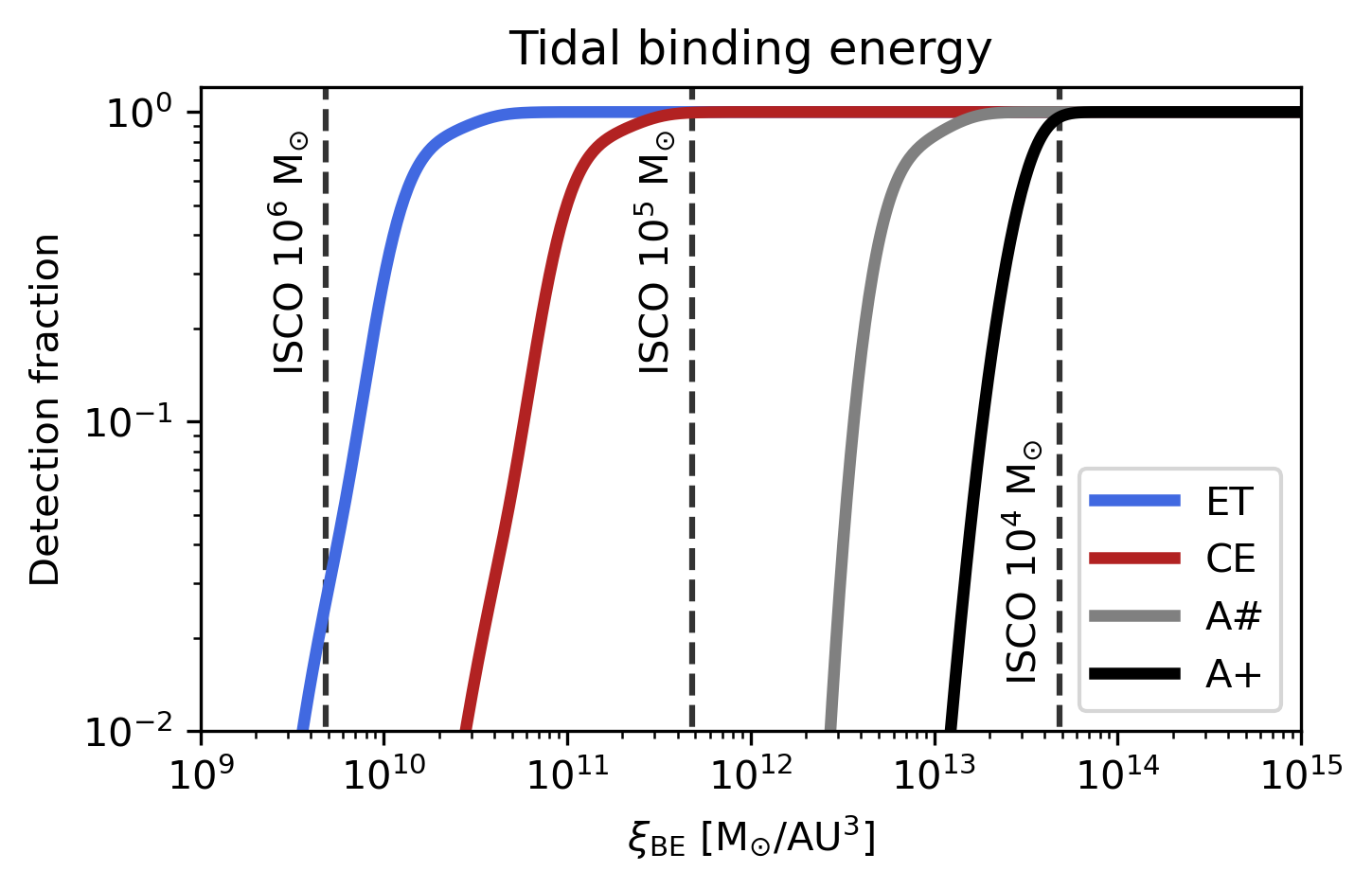}\includegraphics[width=0.9\columnwidth]{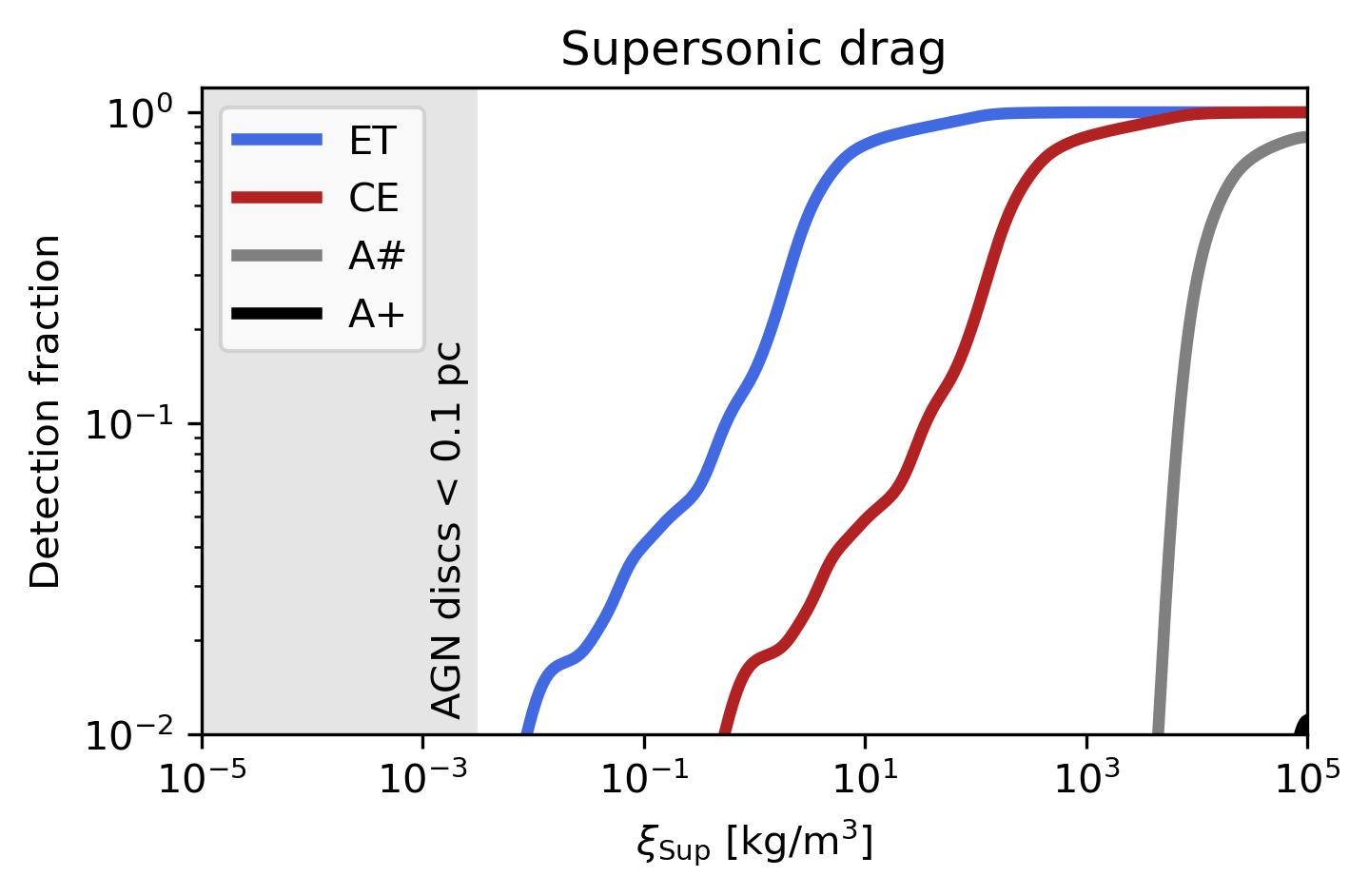}

\includegraphics[width=0.9\columnwidth]{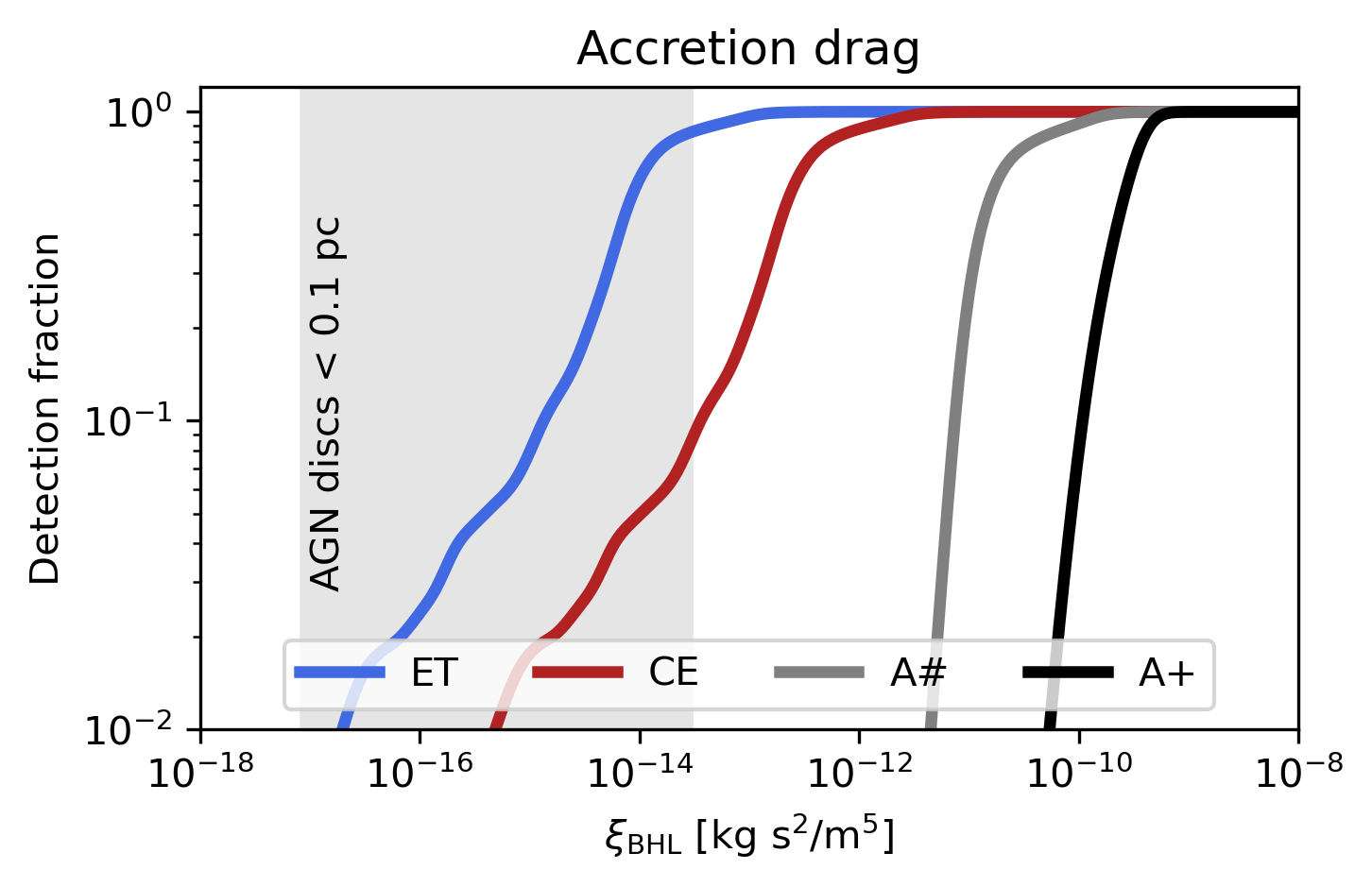}\includegraphics[width=0.9\columnwidth]{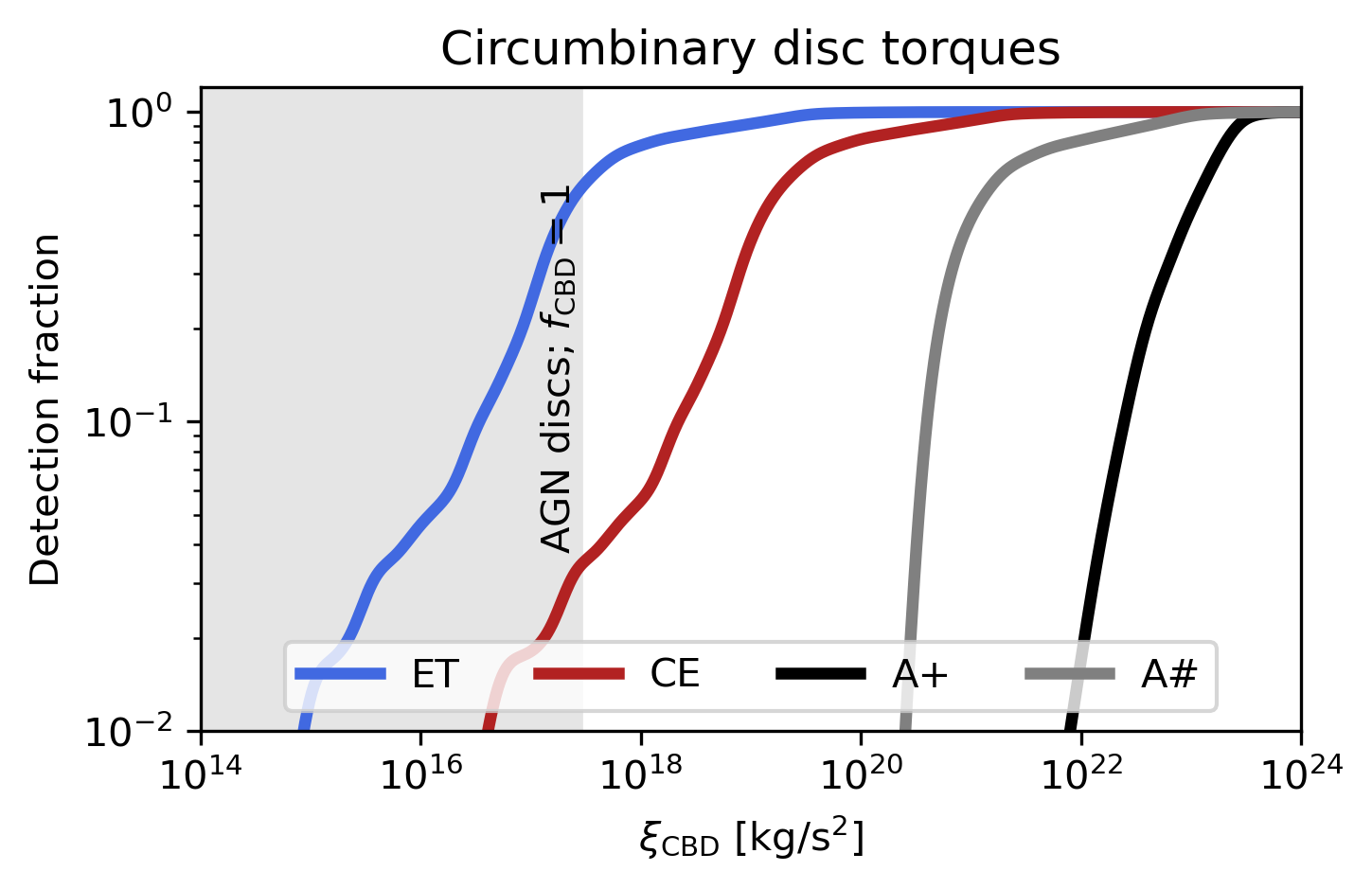}

\caption{The expected detection fractions of sources for LIGO A+ (black), CE (red) and ET (blue), as a function of the appropriate environmental parameter $\xi$. The grey shaded areas and vertical grey lines denote reference ranges and values that could be expected in typical astrophysical systems (see text). The graphs assume the observed LVK population distributions and a formation channel efficiency of $\epsilon=1$. We see that third generation ground based such as ET and CE are will be expected to showcase dephasing in a substantial fraction of sources.}
\label{fig:detfrac}
\end{figure*}

\begin{figure*}

    \centering
\includegraphics[width=0.9\columnwidth]{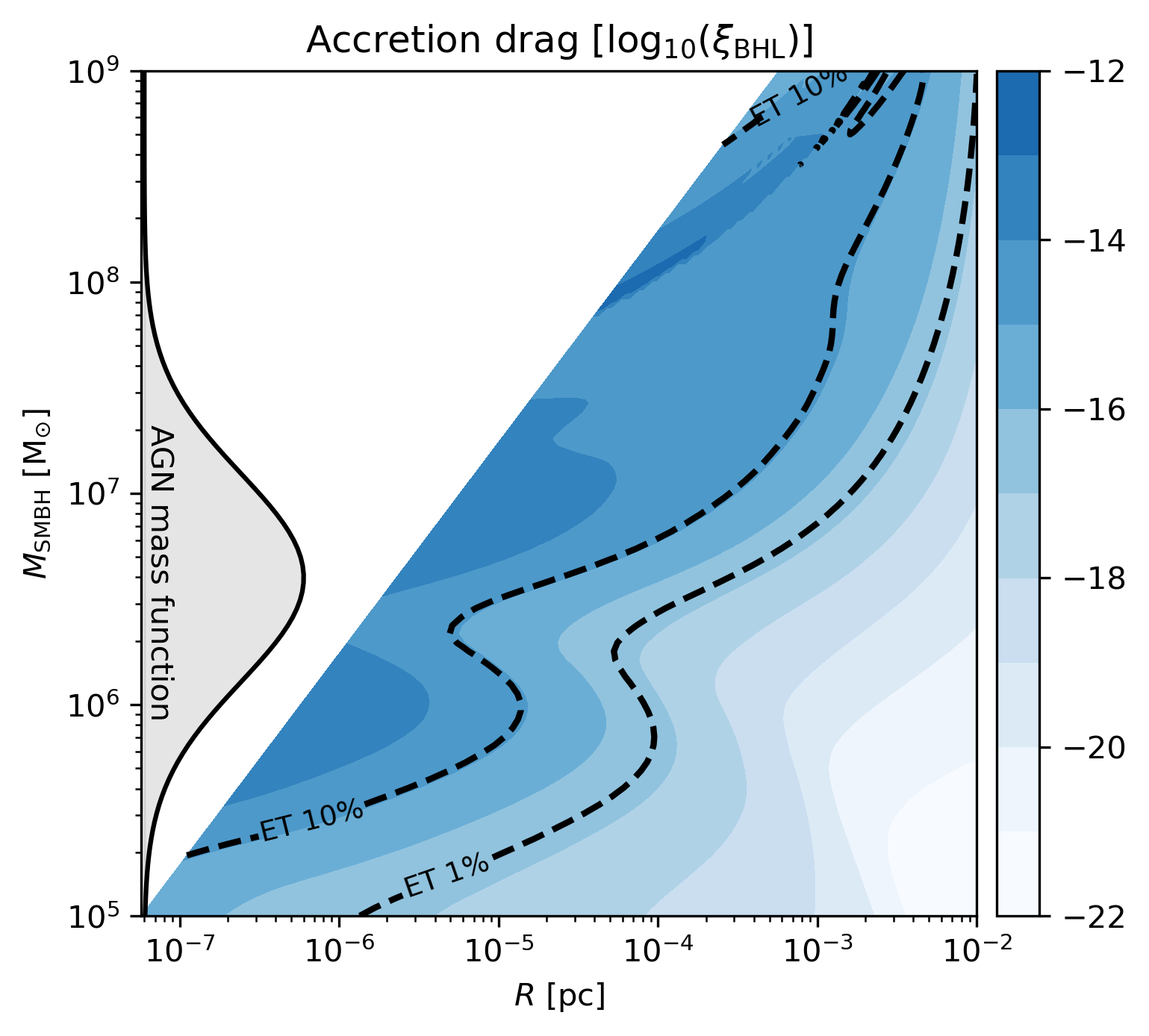}\includegraphics[width=0.9\columnwidth]{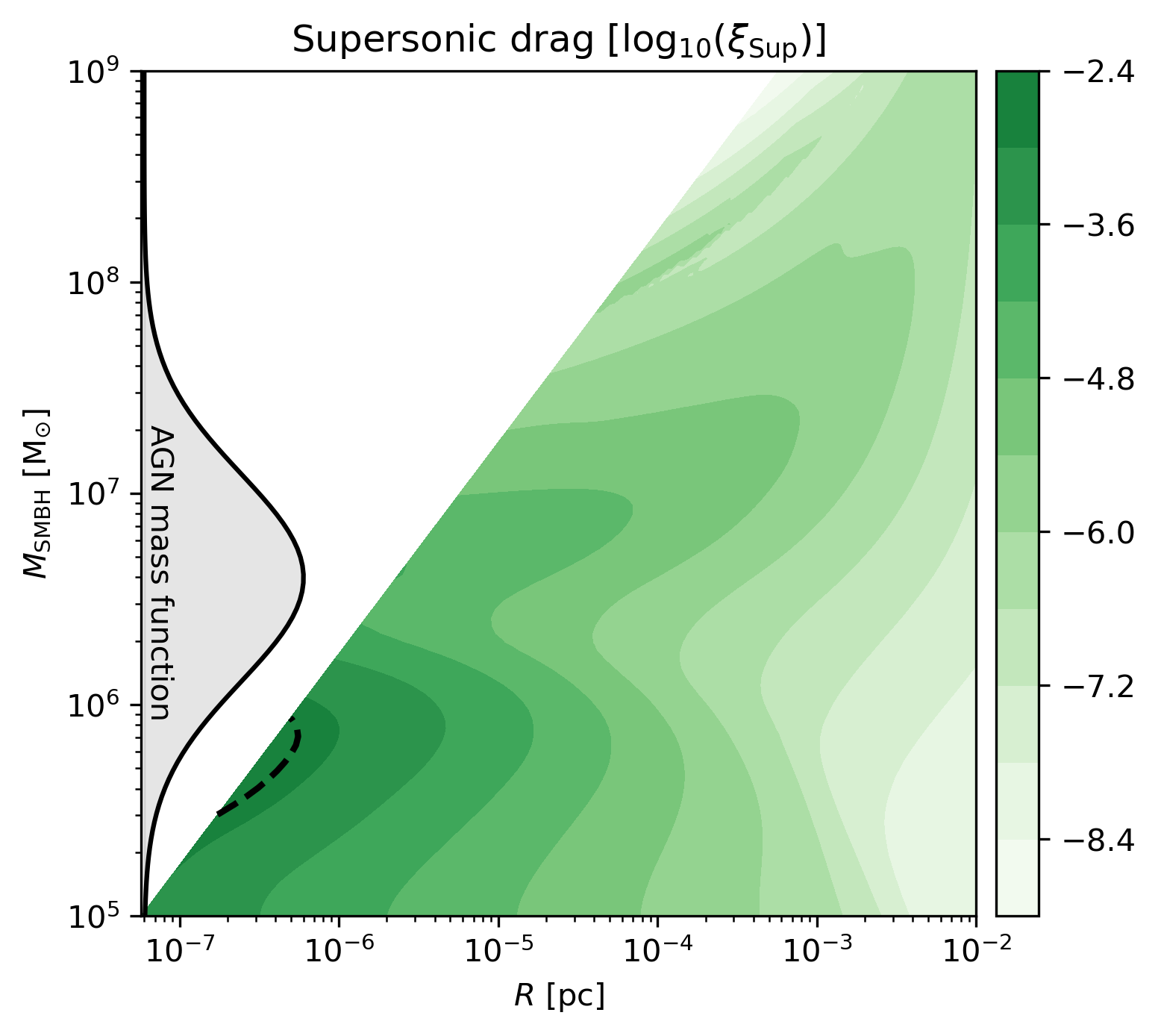}

\includegraphics[width=0.9\columnwidth]{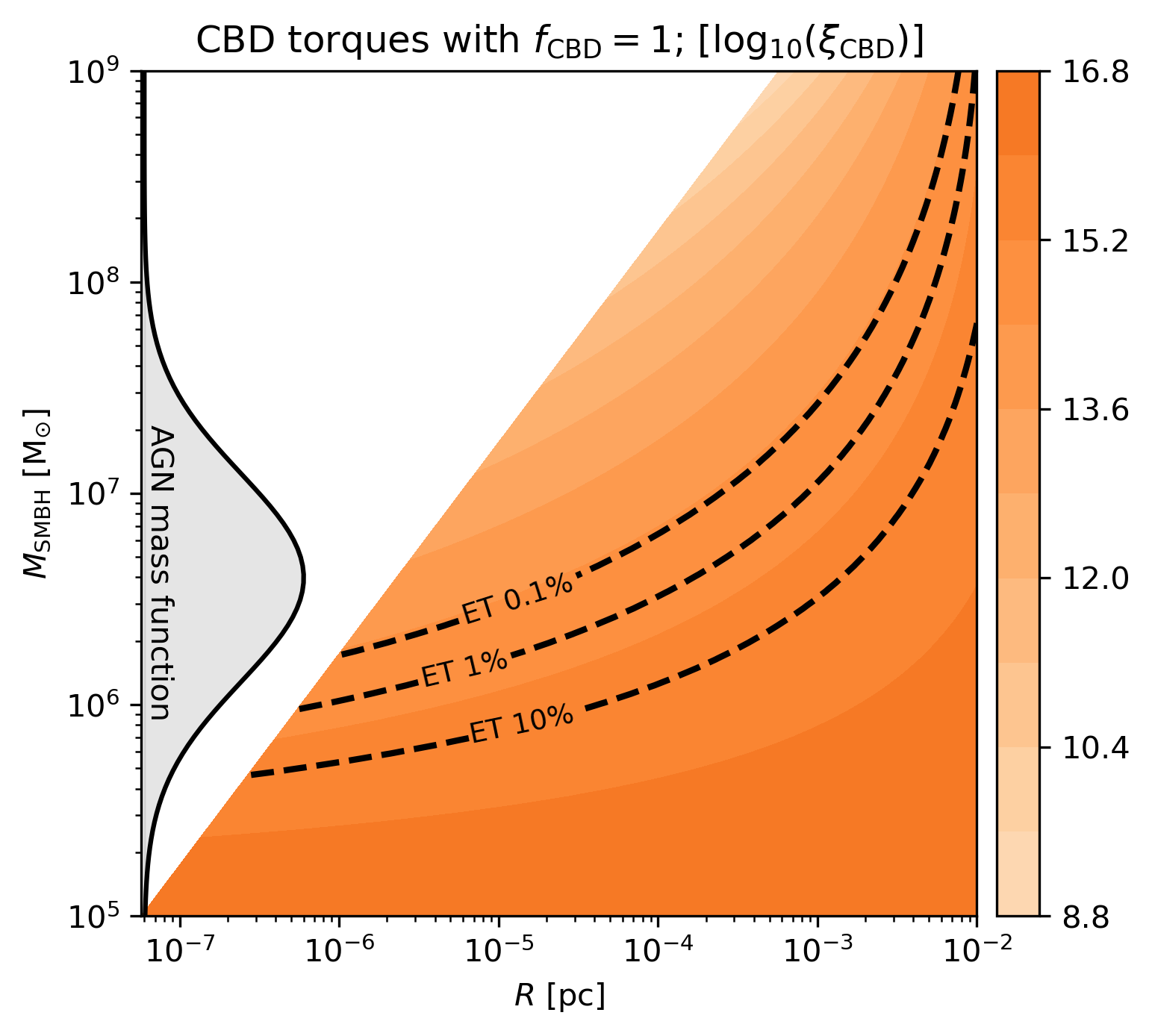}\includegraphics[width=0.9\columnwidth]{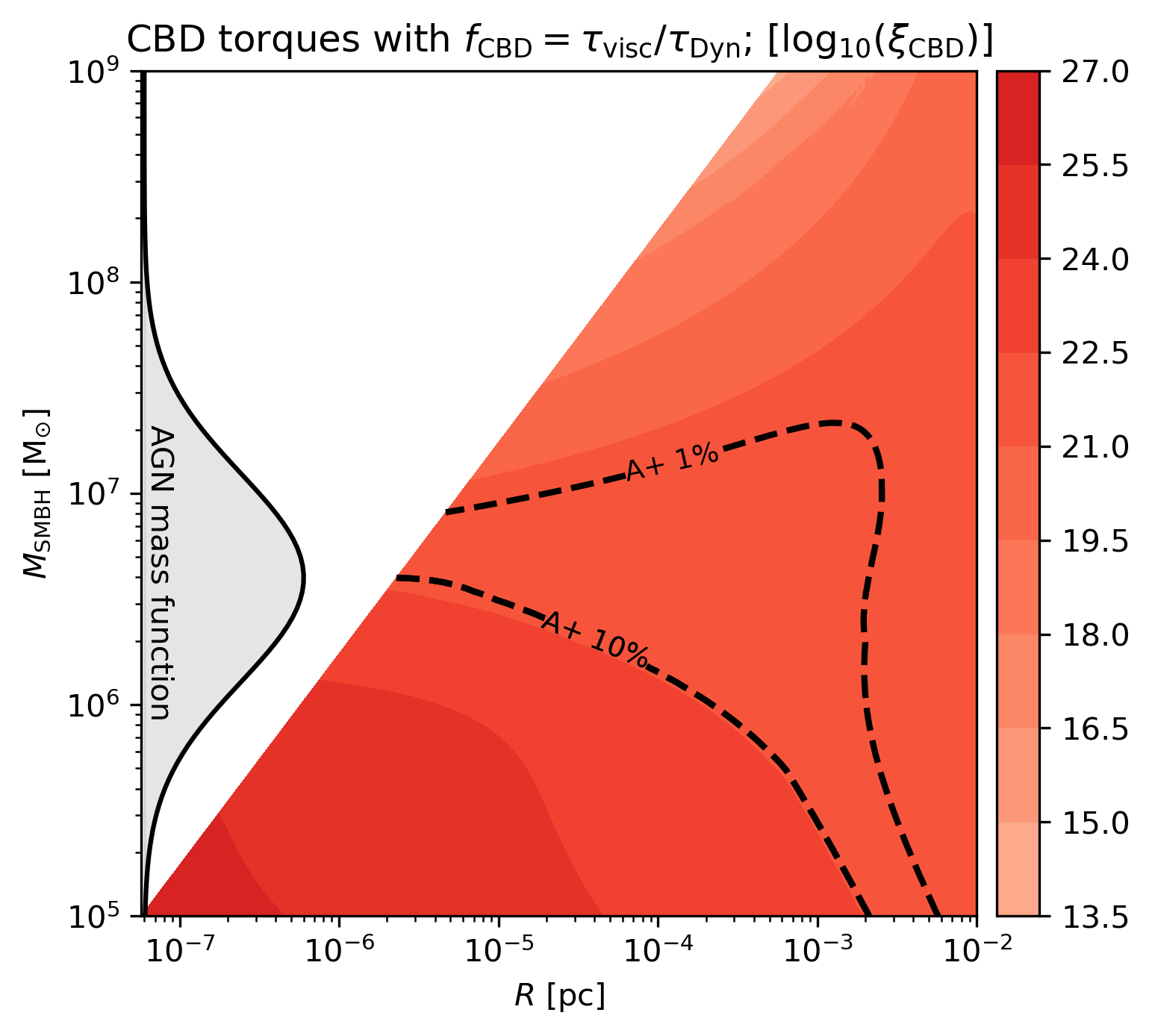}

\caption{Contour plots of the values of the environmental parameters $\xi_{\rm Sup}$,$\xi_{\rm BHL}$ and $\xi_{\rm CBD}$ (in log$_{10}$, where the units are as in Table \ref{tab:coeff}) for a Sirko $\&$ Goodman AGN alpha disc model with $l_\text{E}=0.5$, $X=0.7$, $\alpha=0.1$ and $\varepsilon_\text{rad}=0.1$. We highlight contours depicting the detectability fractions of certain detectors. The plot can be understood as follows: if all BH-BH mergers occurred within the regions of the AGNs discs that corresponds to a percentage contour, then we would expect at least that percentage of sources to showcase dephasing (due to the given hydro effect for the given detector). In the case of ET and CE, 1\% of sources still corresponds to hundreds of signals with significant dephasing. To suggest the most likely case we also plot the AGN mass function as fitted by \citep{Greene:2007dz}, in gray. The contour plots are truncated at $3$ Schwarzschild radii.}
\label{fig:AGN}
\end{figure*}

\noindent We repeat the procedure detailed above for each combination of detector and dephasing prescription. The results of this are shown in Fig. \ref{fig:detfrac}, Fig. \ref{fig:AGN} as well as Table \ref{tab:results} and most conclusions can be drawn directly from the plots. We will now briefly discuss the results and highlight some additional points for each individual dephasing prescription, while overall conclusion is presented in section \ref{sec:conclusion}. Here we note once again that the results presented in this section are informed by the following aspects:
\begin{itemize}
    \item The scalings and functional form of the various dephasing prescriptions (see Eqs. \ref{eq:dephfamI} and \ref{eq:dephfamII}.)

    \item The interaction of said dephasing prescriptions with the detector sensitivity curves (see Fig. \ref{fig:deph_scaling}).

    \item The observed distribution of stellar mass BH binaries and their detectability as a function of redshift, mass and mass ratio (see section \ref{sec:methods:distributions}).
\end{itemize}

\begin{table*}
    \centering
    \begin{tabular}{c|c|c|c|c}
       Det. fraction (Roem.) & A+; $\xi_{\rm R}$ [M$_{\odot}$/AU$^2$] & A\#; $\xi_{\rm R}$ [M$_{\odot}$/AU$^2$] & CE; $\xi_{\rm R}$ [M$_{\odot}$/AU$^2$]& ET; $\xi_{\rm R}$ [M$_{\odot}$/AU$^2$]  \\ 
      \hline
        0.1\% & $3.2\times 10^{6}$ & $2.2\times 10^{6}$ & $1.1\times 10^{2}$  & $7.9\times 10^{0}$  \\
        1\% & $6.4\times 10^{6}$ &$4.9\times 10^{6}$ & $5.8\times 10^{2}$  & $2.2\times 10^{1}$  \\
        10\% & $1.4\times 10^{7}$ &$6.2\times 10^{6}$ & $5.4\times 10^{3}$  & $2.7\times 10^{2}$  

    \end{tabular}
    
\vspace{8px}   

\begin{tabular}{c|c|c|c|c}
     Det. fraction (TBE) & A+; $\xi_{\rm BE}$ [M$_{\odot}$/AU$^3$] & A\#; $\xi_{\rm BE}$ [M$_{\odot}$/AU$^3$] & CE; $\xi_{\rm BE}$ [M$_{\odot}$/AU$^3$]& ET; $\xi_{\rm BE}$ [M$_{\odot}$/AU$^3$]  \\
      \hline
        0.1\% & $1.1\times 10^{13}$ & $3.3\times 10^{12}$ & $1.6\times 10^{10}$  & $2.4\times 10^{9}$  \\
        1\% & $1.4\times 10^{13}$ & $3.8.0\times 10^{12}$ & $2.0\times 10^{10}$  & $3.9\times 10^{9}$  \\
        10\% & $2.1\times 10^{13}$ & $4.2\times 10^{12}$ & $6.5\times 10^{10}$  & $8.1\times 10^{9}$    

    \end{tabular} 
    
\vspace{8px}

    \begin{tabular}{c|c|c|c|c}
     Det. fraction (BHL) & A+; $\xi_{\rm BHL}$ [kg s$^2$/m$^5$] & A\#; $\xi_{\rm BHL}$ [kg s$^2$/m$^5$] & CE; $\xi_{\rm BHL}$ [kg s$^2$/m$^5$]& ET; $\xi_{\rm BHL}$ [kg s$^2$/m$^5$]  \\
      \hline
        0.1\% & $3.4\times 10^{-11}$ & $2.9\times 10^{-12}$ & $2.2\times 10^{-17}$  & $4.6\times 10^{-18}$  \\
        1\% & $6.0\times 10^{-11}$ & $4.9\times 10^{-12}$ & $5.6\times 10^{-16}$  & $2.1\times 10^{-17}$  \\
        10\% & $1.1\times 10^{-10}$ & $8.2\times 10^{-12}$ & $1.2\times 10^{-13}$  & $1.2\times 10^{-15}$   

    \end{tabular}
    
\vspace{8px}

    \begin{tabular}{c|c|c|c|c}
     Det. fraction (Sup) & A+; $\xi_{\rm Sup}$ [kg/m$^3$] & A\#; $\xi_{\rm Sup}$ [kg/m$^3$] & CE; $\xi_{\rm Sup}$ [kg/m$^3$]& ET; $\xi_{\rm Sup}$ [kg/m$^3$]  \\
      \hline
        0.1\% & $4.1\times 10^4$ & $3.4\times 10^{3}$ & $4.0\times 10^{-2}$  & $1.6\times 10^{-3}$  \\
        1\% &$8.5\times 10^4$ & $5.4\times 10^{3}$ & $6.1\times 10^{-1}$  & $9.3\times 10^{-3}$  \\
        10\% &$1.0\times 10^5$ & $7.4\times 10^{3}$ & $3.6\times 10^{1}$  & $5.6\times 10^{-1}$   

    \end{tabular}

\vspace{8px}
    
\begin{tabular}{c|c|c|c|c}
     Det. fraction (CBD) & A+; $\xi_{\rm CBD}$ [kg/s$^2$] & A\#; $\xi_{\rm CBD}$ [kg/s$^2$] & CE; $\xi_{\rm CBD}$ [kg/s$^2$]& ET; $\xi_{\rm CBD}$ [kg/s$^2$]  \\
      \hline
        0.1\% & $3.4\times 10^{21}$ & $2.1\times 10^{20}$ & $2.4\times 10^{15}$  & $2.9\times 10^{14}$  \\
        1\% &$8.4\times 10^{21}$ & $3.1\times 10^{20}$ & $4.2\times 10^{16}$  & $8.7\times 10^{14}$  \\
        10\% &$2.6\times 10^{22}$ & $4.0\times 10^{20}$ & $2.0\times 10^{18}$  & $3.2\times 10^{15}$    
    \end{tabular} 
    
    \caption{Required magnitude of the various environmental parameters $\xi_i$ required for various detectors to expect significant dephasing in 0.1\%, 1\% and 10\% of detected signals. The results assume the observed LVK population distributions and a formation channel efficiency of $\epsilon=1$.}
    
    \label{tab:results}
\end{table*}
\subsubsection{Roemer Delays}
\noindent The typical value of the Roemer delay environmental parameter $\xi_{\rm R} = m_3/R^2$ depends on the third body in the vicinity of the binary. Here we focus on mergers resulting from 3-body scatterings in clusters, which are known to be able to produce a significant rate of mergers as well as mergers with residual eccentricity. As a rough estimate, we expect $m_3$ to range from a few to a few hundreds of M$_{\odot}$, while $R$ is related to the velocity dispersion of the cluster in which the 3-body scattering take place. As discussed in \cite{kai2024}, we expect $R\sim$1 AU, $R\sim$0.1 AU and $R\sim$0.01 AU for typical globular clusters, dense stellar clusters and nuclear star clusters, respectively. {In reality, $R$ and $m_3$ will of course be described by distributions, however modeling this ulterior effect is beyond the scope of the present work.} As seen in Fig. \ref{fig:detfrac} and Table \ref{tab:results} a significant fraction of GW signals (1\% to 10\% ) will showcase significant dephasing due to Roemer delays for a parameter $\xi_{\rm R} \sim 100$ M$_{\odot}$/AU$^2$ to $\sim 1000$ M$_{\odot}$/AU$^2$, for next generation space based detectors. This is not only possible, but \textit{expected} for binaries that merge as a result of scatterings in stellar clusters. We note here, that dephasing due to Roemer delays should be significant in \textit{all} binary mergers that occur as a result of scattering in nuclear clusters, which may account for merger rates of up to $\mathcal{O}(10)$ Gpc$^{-1}$ yr$^{-1}$ \citep{mapelli2021review,2021Wang}. For mergers due to binary-single scatterings in an AGN disc, the dephasing could be significantly increased as the surrounding gas can harden the triple system, substantially reducing $R$ in the lead-up to merger \citep{Rowan2025_trips}.

The magnitude of the Roemer dephasing may be even larger than expected from our model, in which we consider the binary orbit around the third object to be circular. Curvature effects from \textit{eccentric} orbits around the third object, which are a common occurrence in clusters, can give rise to increased Roemer dephasing several orders of magnitude higher than the circular case \citep{kai22024}. Additionally statistical outliers that result from strong chaotic scatterings, as discussed in \cite{kai2024}. We also note that, without considering the effect of eccentric waveforms, differences in the underlying binary eccentricity distribution do not influence our conclusions.

Roemer delays also play a significant role in the AGN channel, in which binaries may be orbiting close to the central BH. rescaled to our typical units, we find the value for the parameter $\xi_{\rm R}$ for a binary orbiting a SMBH at a radius $R$:
\begin{align}
    \xi_{\rm R} &\approx 2.6\times 10^3 \times \left(\frac{10^3\ r_{\rm S}}{R}\right)^2 \\ \nonumber &\times
\left(\frac{M_{\rm SMBH}}{10^6\, \rm{M}_{\odot}}\right) [\rm{M}_{\odot}/\rm{A}U^2]
\end{align}
We see that Roemer delays should lead to a significant fraction of signals with detectable phase-shifts even for typical distances of thousands of Schwarzschild radii.

The emerging picture is that Roemer delays represent the most likely candidate for the detection of an EE in stellar mass BH binaries, and they are expected to be present in a significant fraction of the detected signals for third generation GW detectors for a wide range of the parameter $\xi_{\rm R}$. In fact, for the large but physically plausible values of $\xi_{\rm R} > 10^6$ M$_{\odot}$/AU$^2$ expected for scatterings in nuclear clusters or in the AGN channel at hundreds of Schwarzschild radii, we would even expect a few percent of future LVK sources to showcase detectable dephasing signatures, in particular for light binaries with unequal mass ratios. It is intriguing to note that currently there exists a single candidate signal that showcases evidence of dephasing with a power law of $f^{-13/3}$, originating from a binary with exactly the expected properties \citep{2024Han}. The signal in question is GW190814, with a potential acceleration of 0.0015  $\pm 0.007$ $c/s$, corresponding to a parameter $\xi_{\rm R}=7.6\times 10^7$ M$_{\odot}$/AU$^2$. We note here that an acceleration of this magnitude would be detectable for almost 100 $\%$ of LVK sources with A+ sensitivity, and the most likely case is that this is in fact an outlier, or simply a wrong interpretation of a feature in the data. In any case, our results provide a simple way to assess whether the majority of mergers occur in environments characterised by accelerations similar to the ones experienced (allegedly) by GW190814.
\subsubsection{Tidal effects}
\noindent As a reference value for the environmental parameters $\xi_{\rm TBE}$, we consider a binary orbiting at the innermost stable circular orbit of a supermassive black hole, which yields the following tidal parameter:
\begin{align}
   \xi_{\rm TBE}&\to \frac{m_3 c^6}{27\times 8 G^3m_3^3} \nonumber \\&\approx 4.8\times 10^9\left(\frac{10^6{\rm{M}}_{\odot}}{m_3}\right)^{2} \frac{\rm{M}_{\odot}}{\rm{AU}^3}.
\end{align}
Placing the binary further away would then reduce the environmental parameter with a cubic scaling. Comparing with the results displayed in Fig. \ref{fig:detfrac}, we see that the only scenario in which we may expect a significant fraction of sources to showcase dephasing is for binaries orbiting in the close vicinity of low mass SMBHs or around intermediate mass black holes. Such systems can be realised in the AGN formation channel, due to the presence of migration traps \citep{2016bellovary}, and would represent interesting laboratories to study relativistic dynamics \citep{2021MNRAS.502.2049L,pica2023}. Nevertheless, our analysis suggests that dephasing due to Roemer delays is almost certainly dominant with respect to tidal effects, because of the more forgiving scaling of $m_3/R^2$ rather than $m_3/R^3$.

\subsubsection{Hydrodynamical effects}
\noindent The density and speed of sound expected in accretion discs vary significantly depending on the central object mass, the specific disc model and the radius under consideration. In our analysis, we considered three prescriptions that scale differently with gas properties, as detailed in Table \ref{tab:coeff}. Here we base our analysis of hydrodynamical effects in AGN discs on the commonly used Sirko \& Goodman $\alpha$-disc model \citep{sirko2003,goo03}, as implemented in the publicly available \texttt{pAGN} package. We assume for our disc models an Eddington ratio of $l_\text{E}=0.5$, hydrogen fraction $X=0.7$, alpha viscosity of $\alpha=0.1$ and a radiative efficiency of $\varepsilon_\text{rad}=0.1$. 

Fig. \ref{fig:AGN} displays the values of the environmental parameters $\xi_{\rm Sup}$,$\xi_{\rm BHL}$ and $\xi_{\rm CBD}$ evaluated for such a disc model. Additionally, we highlight the contours that correspond to a certain detection fraction for detectors. We summarise the results qualitatively as follows:
\begin{itemize}
    \item For BHL drag, we find that for next generation detectors such as ET and CE, values of the environmental parameters that correspond to significant detection fraction are reached in the inner regions of the disc.

    \item For supersonic drag, dephasing is too small to be present in a significant fraction of sources for any of the considered detectors.

    \item For CBD torques, {values of the environmental parameters that correspond to significant detection fraction are reached in the outer regions of AGN discs around lighter massive BH.}. For a fudge factor proportional to the ratio of viscous to dynamical time, CBD torques are so strong that significant detection fractions are also expected with A+ sensitivity, in the inner regions of AGN around low mass SMBH.
\end{itemize}
Overall, the conclusions regarding the significance of dephasing due to hydrodynamical effects depend on the prescription that is used. Taking BHL drag {or CBD drag with no fudge factor} as a baseline, we conclude that significant detection fractions are expected in the AGN channel for third generation detectors, though not for LVK upgrades.

\section{Summary and Conclusion}
\label{sec:conclusion}
\noindent Our work can be summarised as follows:
\begin{itemize}
    \item We have constructed two general families of dephasing prescriptions and derived a set of inequalities that detail what regions of parameter space are more likely to showcase dephasing with different scalings. The latter may be used to inform priors in an a-priori way in paramter inference studies.

    \item We have collected and analysed five dephasing prescriptions that represent smoking gun physical mechanisms at play in the dynamical and AGN formation channel for stellar mass BH binary systems.

    \item We have computed the expected fraction of sources that will showcase significant dephasing for four upcoming ground based detector sensitivities, as a function of environmental properties. The results are visualised in Fig. \ref{fig:detfrac} and \ref{fig:AGN}, as well as Table \ref{tab:results}. These results can be used as a reference to prioritise the efforts in GW modeling work, as well as inform future studies that aim to model binary environments and their coupling to binaries.
    
\end{itemize}
We have found that Roemer delays and hydrodynamical effects in particular can induce strong dephasing for third body properties and gas properties that are considered typical for the dynamical and the AGN formation channels, respectively. Summarising:
\begin{itemize}
    \item For Roemer delays, accelerations that are entirely consistent with the typical environments of binaries will cause significant dephasing in $\sim \epsilon\times1\%$ to $\sim \epsilon\times 100\%$ of binaries in the dynamical or AGN channel for ET and CE. It is likely that hundreds of signals will showcase such signatures (see Fig \ref{fig:detfrac} and Table \ref{tab:results}). Roemer delay may also be relevant for a significant amount of sources with A\# sensitivity, depending on the dominance of the nuclear cluster sub-channel or the AGN channel.

    \item For third generation detectors such as ET and CE one can expect to find significant fractions of signals with gas induced dephasing, {where BHL drag dephasing dominates in the inner regions of the disc while CBD torques in the outer part of the disc} (see Fig. \ref{fig:AGN} in combination with Table \ref{tab:results}). {This suggests that gas induced EE will play a significant role in the data analysis of sources in the AGN channel, \textit{regardless} of the typical scale at which such binaries merge}. While this is conclusion is robust, more work is required to properly model the details of accretion flows for binaries embedded in AGN, as commonly used prescriptions yield wildly varying results.
\end{itemize}

The third generation of ground based GW detectors is expected to become operational within the next decade. We believe that our work strongly indicates that EEs are not only a possibility, but are almost certainly going to be present in a significant fraction of the thousands of expected signals. To take advantage of this fact, it will be necessary to fully develop GW templates that contain EE prescriptions, and are able to extract and distinguish between EE in real GW signals. In particular, we recommend focusing on Roemer delays and its eccentricity extensions, as well as dephasing prescriptions for gas accretion and CBD torques.

Another aspect that is often neglected in discussions of binary environments is the coupling between EE and eccentricity, which is considered a smoking gun for the dynamical formation channel and perhaps for the AGN channel \citep{dittmann2024}. Our preliminary analysis of the two dephasing families in section \ref{sec:dephasing} further solidifies the claim that eccentric dephasing prescriptions carry additional information that can be used to identify and distinguish unique properties of EE \citep[see e.g.][ for the similar considerations in the case of extreme mass ratio systems]{2024duque}. We will explore this aspect in detail in paper II.
Overall, we find that the emerging picture is that the joint study of eccentricity and environmental dephasing in individual systems is likely to become a reality in the next decade. The prospects to complement current population based inferences with the detection of smoking gun signatures in individual signals are extremely rich, and will add a new tool in the rapidly growing field of gravitational wave astronomy.

\begin{acknowledgments}
L.Z., J.T., P.S., J.S and K.H. acknowledge support from  ERC Starting Grant No. 121817–BlackHoleMergs led by J.S. L.Z. and J.S. are also supported by the Villum Fonden grant No. 29466. A.A.T. acknowledges support from the Horizon Europe research and innovation programs under the Marie Sk\l{}odowska-Curie grant agreement no. 101103134. The research leading to this work was supported by the Independent Research Fund Denmark via grant ID 10.46540/3103-00205B.

\end{acknowledgments}

\appendix{}
\section{Phase space slices for the Einstein telescope}
\begin{figure*}
\label{fig:app_BG}
    \centering
\includegraphics[scale=0.51]{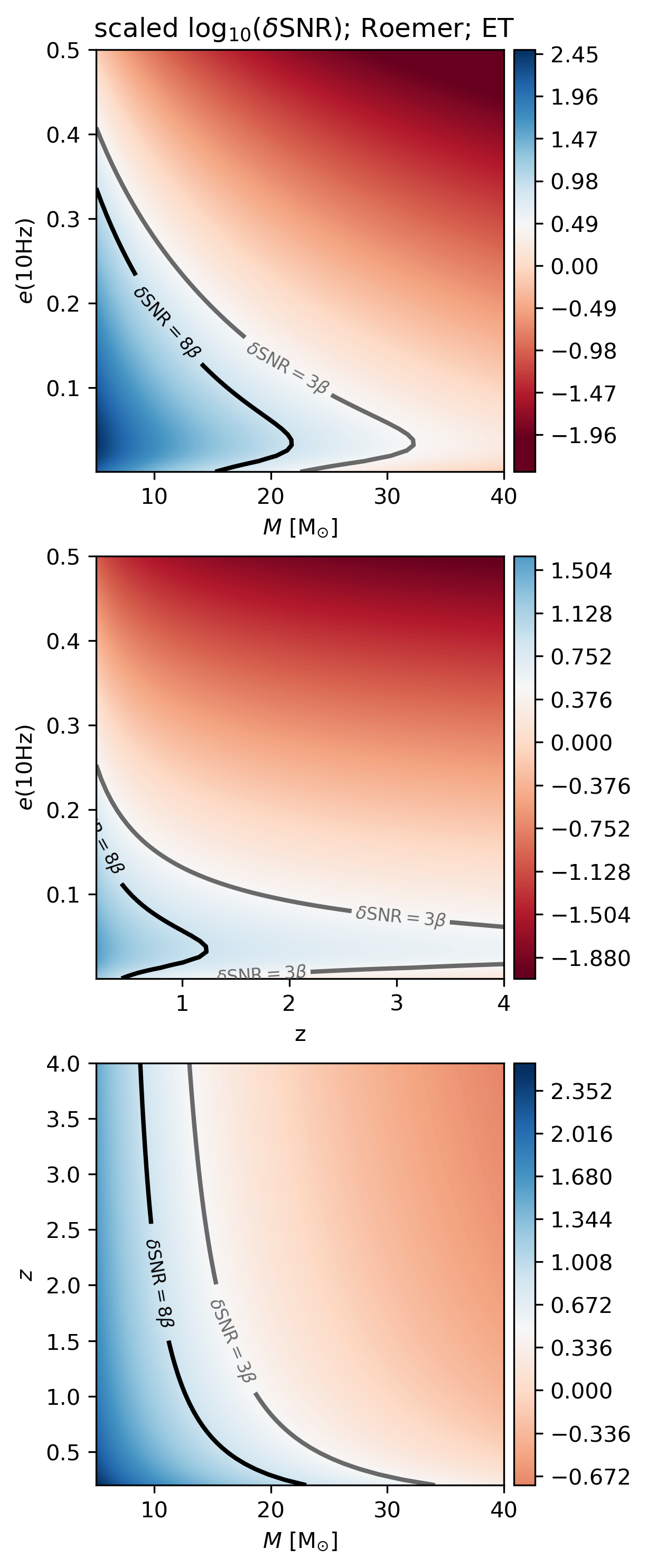}
\includegraphics[scale=0.51]{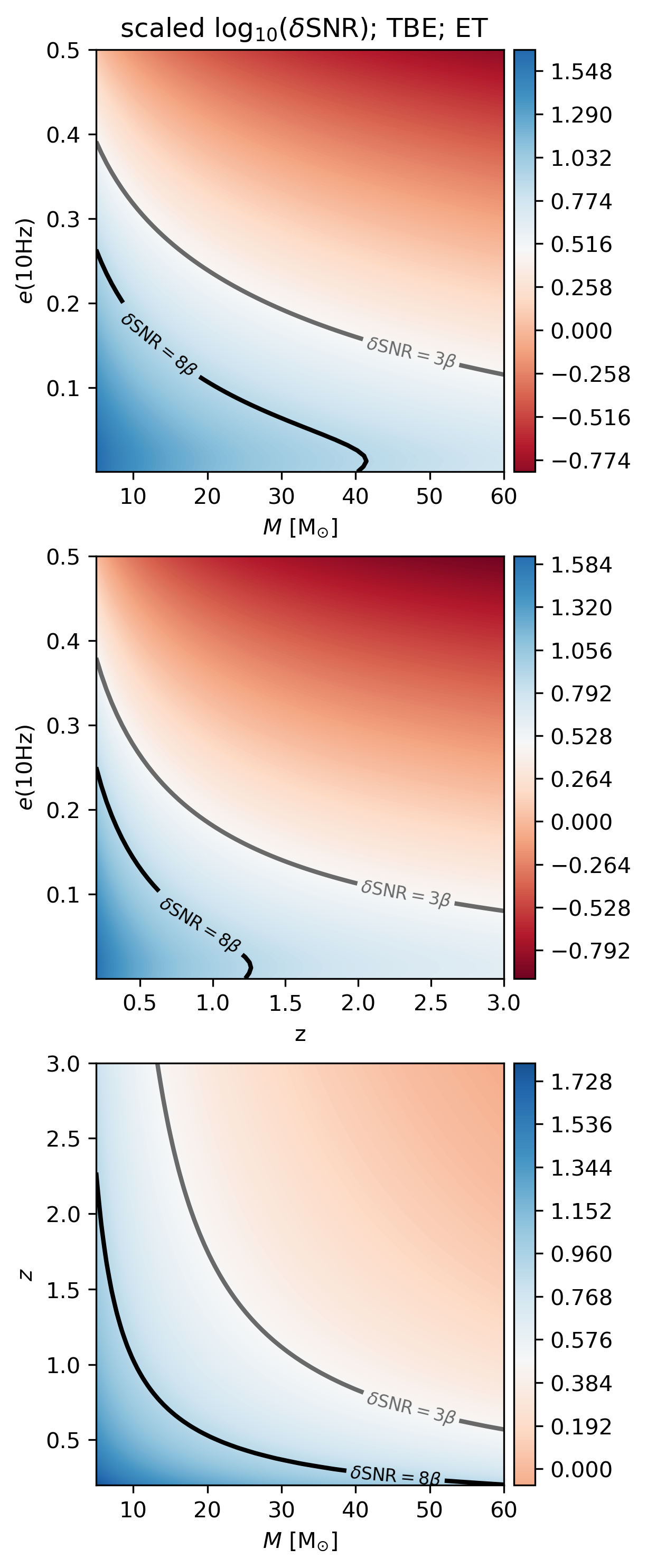}
\includegraphics[scale=0.51]{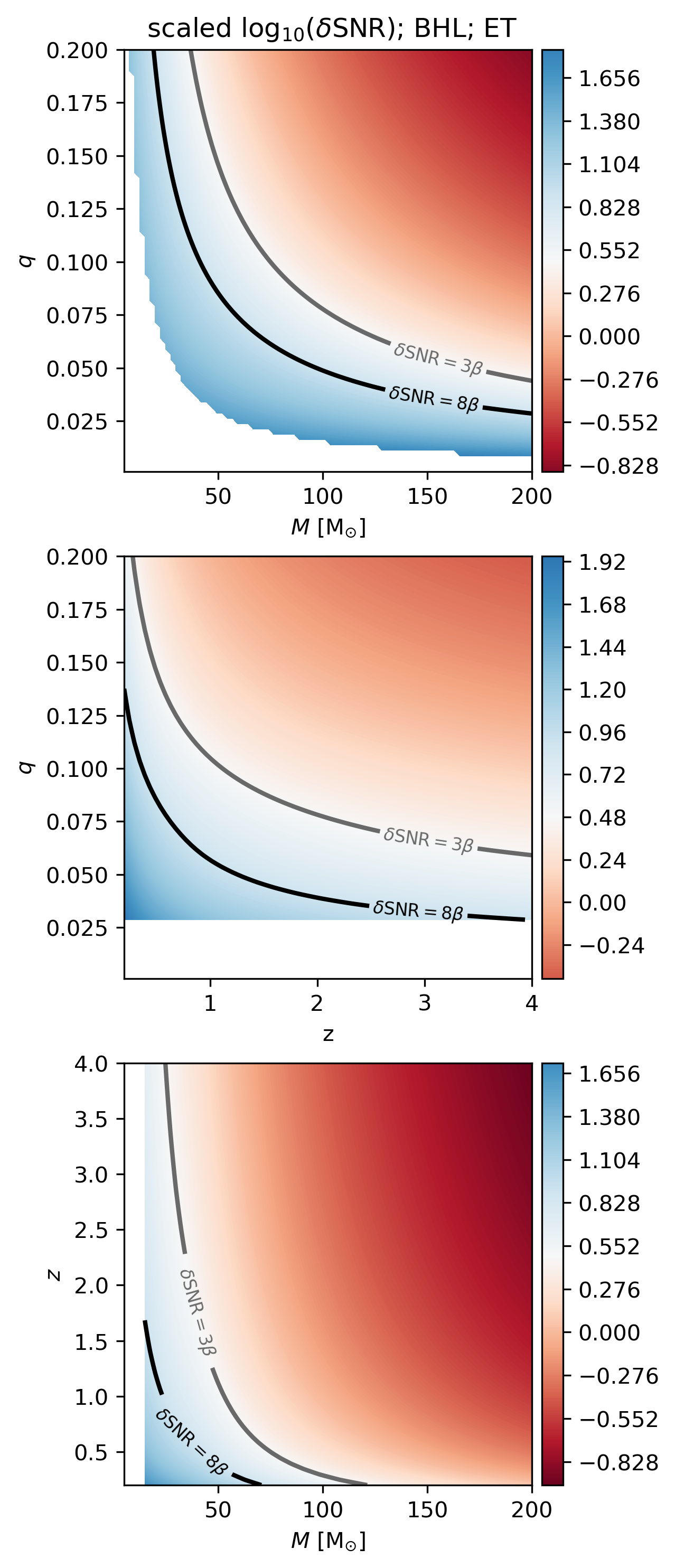}

    \caption{{A visualisation of the two dimensional slices of the $\delta$SNR cubes for various dephasing prescriptions with a fixed value of the parameter $\xi$ (for ET). The colorbar represents the logarithm of the $\delta$SNR results, though the absolute values do not matter here (as they can be rescaled with a choice of $\xi$.) Note how the contours vary purely as a function of binary parameters, highlighting how different EE preferentially affect binaries with certain parameters. Here, the blue regions of phase space are more likely to present detectable EE, all else being equal. Note that this particular choice of color scheme only serves as a visualisation and is not used for any results discussed in the main text. Similarly, the boundaries of the plots are simply chosen for visualisation purposes.}}
\end{figure*}

Fig. \ref{fig:app_BG} shows the two dimensional slices of $\delta$SNR parameter space referenced in section \ref{sec:GB_detectors}. These only serve as a visualisation of the preferred regions of parameter space in which EE are most likely to strongly affect an inspiralling binary.
\label{appendix_GB}

\section{A note on space based detectors}
\label{appendix_SB}
\begin{figure*}
\label{fig:app_lisa}
    \centering
\includegraphics[scale=0.51]{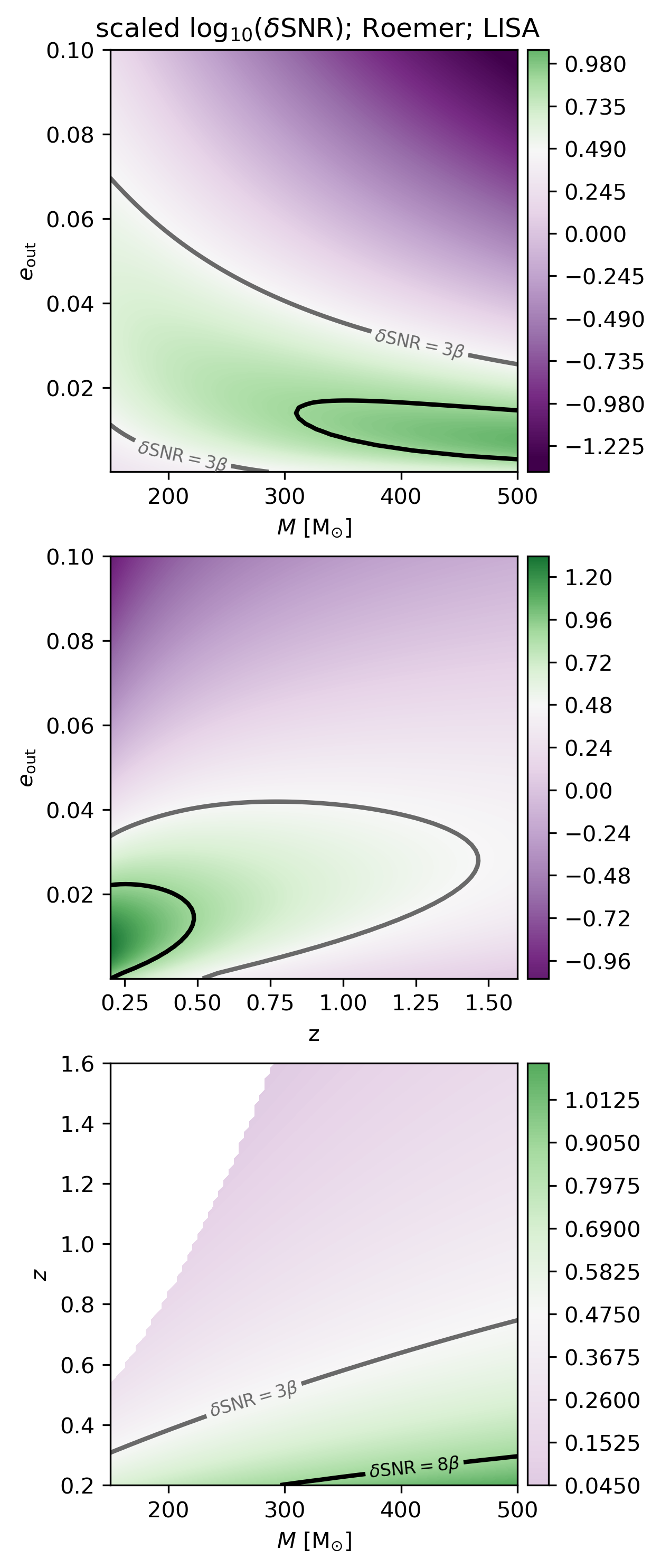}
\includegraphics[scale=0.51]{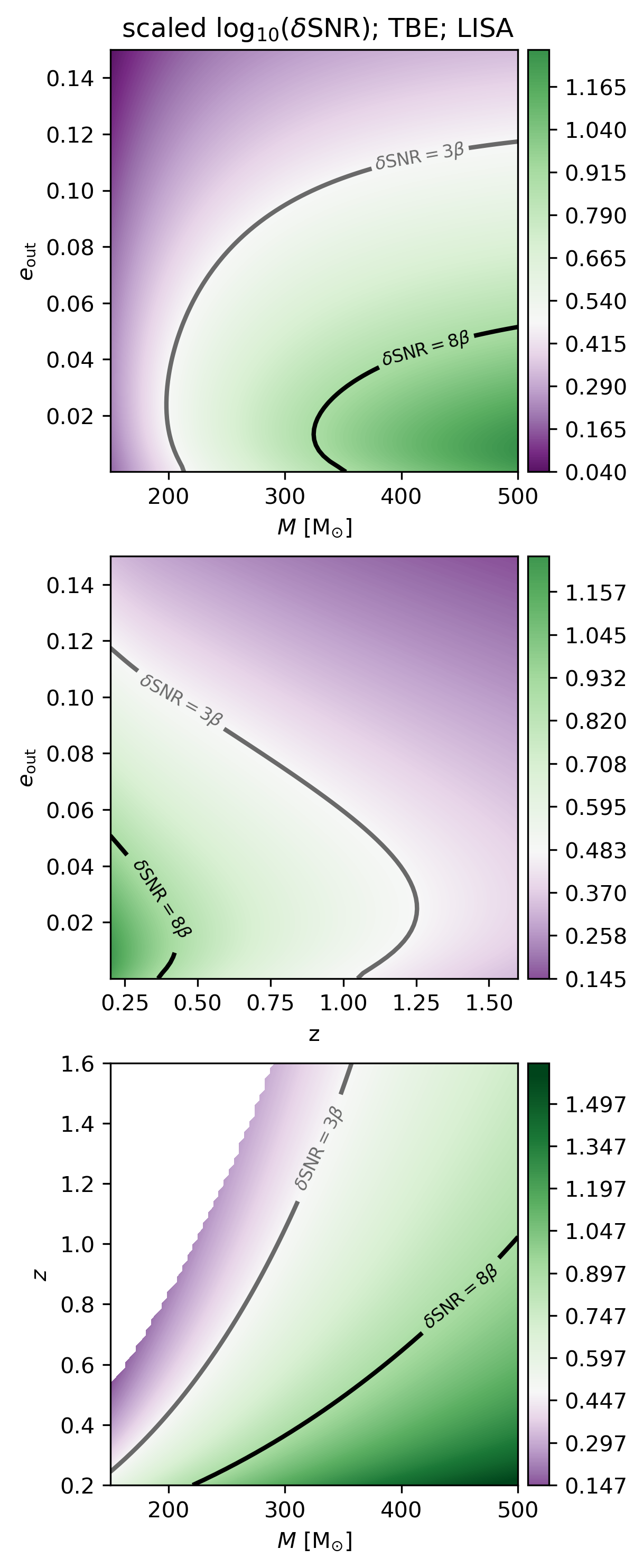}
\includegraphics[scale=0.51]{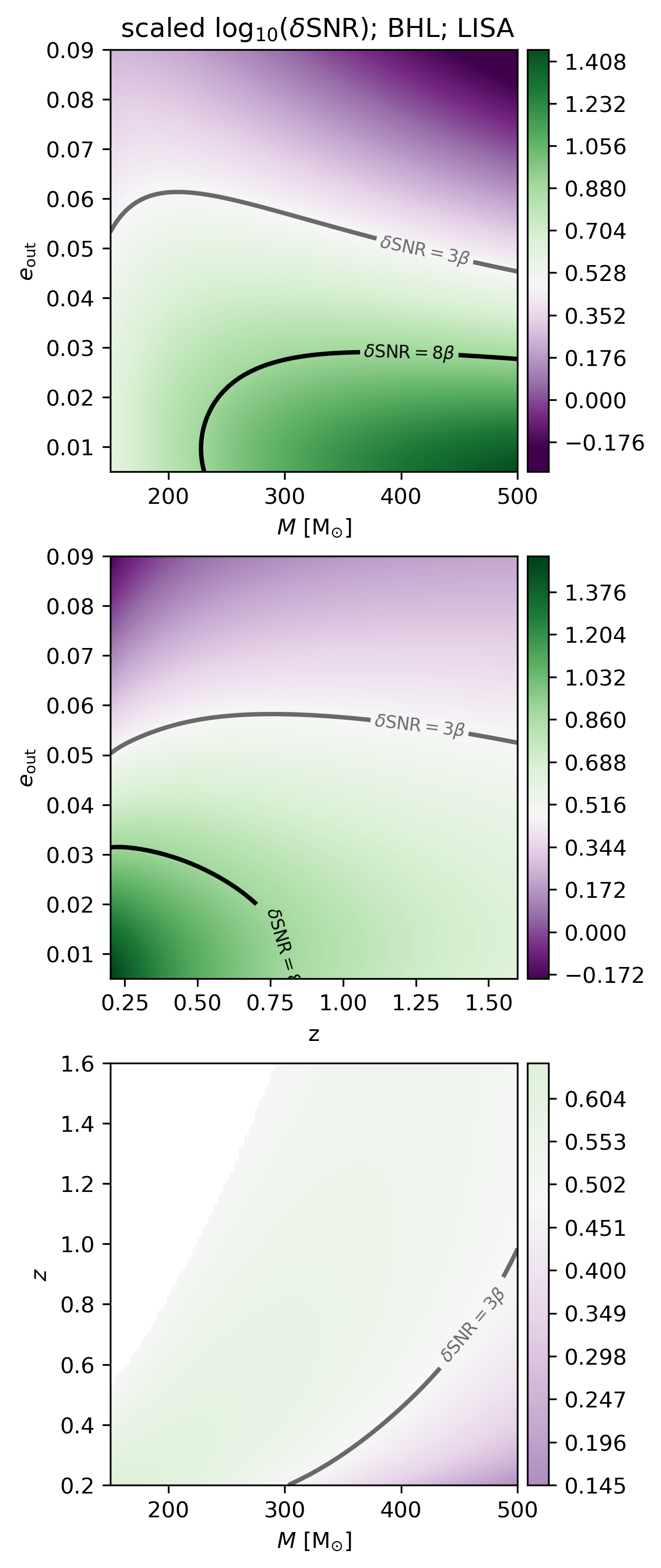}

    \caption{Two dimensional slices of the $\delta$SNR cubes for various dephasing prescriptions with a fixed value of the parameter $\xi$ (for LISA). Note how the contours vary purely as a function of binary parameters. Green regions of phase space are more likely to present detectable EE, all else being equal. Note that this particular choice of color scheme only serves as a visualisation and is not used for any results discussed in the main text. Similarly, the boundaries of the plots are simply chosen for visualisation purposes.}
\end{figure*}

It is interesting to repeat part of the procedures detailed in the main text for stellar mass BHs that will potentially show up in the milli-Hz band as pre-LVK sources. Here we take the LISA \citep{2024lisa} sensitivity curve as a baseline, and integrate the SNR and $\delta$SNR of dephased GW going backwards in frequency, from the moment they exit the band to four years prior. The presence of eccentricity for such sources is much more significant as in the ground based case, as an eccentric binary can chirp over a much wider range of frequencies over the limited observation time of a space-borne GW detector (with respect to the signal life-time). We do not discuss this in detail, as here it absolutely crucial to perform the estimates with fully eccentric waveforms. We note however the rich and interesting shapes of the slices of the $\delta$SNR results, in Fig. \ref{fig:app_lisa}. With the caveat of the rate and signal identification challenges of such sources, looking for EE in eccentric pre-LVK binaries in LISA is potentially extraordinarily promising \citep[see also][]{2022xuan}, and deserves more investigation. We note that the figure only serves as a visualisation and is not used for any results in the main text.


\bibliography{podistr}

\begin{thebibliography}{}
\expandafter\ifx\csname natexlab\endcsname\relax\def\natexlab#1{#1}\fi
\providecommand{\url}[1]{\href{#1}{#1}}
\providecommand{\dodoi}[1]{doi:~\href{http://doi.org/#1}{\nolinkurl{#1}}}
\providecommand{\doeprint}[1]{\href{http://ascl.net/#1}{\nolinkurl{http://ascl.net/#1}}}
\providecommand{\doarXiv}[1]{\href{https://arxiv.org/abs/#1}{\nolinkurl{https://arxiv.org/abs/#1}}}

\bibitem[{{Abbott} \& et~al.(2020)}]{lvcGW190814}
{Abbott}, R., \& et~al. 2020, \apjl, 896, L44, \dodoi{10.3847/2041-8213/ab960f}

\bibitem[{{Abbott} {et~al.}(2023){Abbott}, {Abbott}, {Acernese}, {Ackley}, {Adams}, {Adhikari}, {Adhikari}, {Adya}, {Affeldt}, {Agarwal}, {Agathos}, {Agatsuma}, {Aggarwal}, {Aguiar}, {Aiello}, {Ain}, {Ajith}, {Akutsu}, {de Alarc{\'o}n}, {Akcay}, {Albanesi}, {Allocca}, {Altin}, {Amato}, {Anand}, {Anand}, {Ananyeva}, {Anderson}, {Anderson}, {Ando}, {Andrade}, {Andres}, {Andri{\'c}}, {Angelova}, {Ansoldi}, {Antelis}, {Antier}, {Antonini}, {Appert}, {Arai}, {Arai}, {Arai}, {Araki}, {Araya}, {Araya}, {Areeda}, {Ar{\`e}ne}, {Aritomi}, {Arnaud}, {Arogeti}, {Aronson}, {Arun}, {Asada}, {Asali}, {Ashton}, {Aso}, {Assiduo}, {Aston}, {Astone}, {Aubin}, {Austin}, {Babak}, {Badaracco}, {Bader}, {Badger}, {Bae}, {Bae}, {Baer}, {Bagnasco}, {Bai}, {Baiotti}, {Baird}, {Bajpai}, {Ball}, {Ballardin}, {Ballmer}, {Balsamo}, {Baltus}, {Banagiri}, {Bankar}, {Barayoga}, {Barbieri}, {Barish}, {Barker}, {Barneo}, {Barone}, {Barr}, {Barsotti}, {Barsuglia}, {Barta}, {Bartlett}, {Barton}, {Bartos}, {Bassiri}, {Basti}, {Bawaj}, {Bayley},
  {Baylor}, {Bazzan}, {B{\'e}csy}, {Bedakihale}, {Bejger}, {Belahcene}, {Benedetto}, {Beniwal}, {Bennett}, {Bentley}, {Benyaala}, {Bergamin}, {Berger}, {Bernuzzi}, {Berry}, {Bersanetti}, {Bertolini}, {Betzwieser}, {Beveridge}, {Bhandare}, {Bhardwaj}, {Bhattacharjee}, {Bhaumik}, {Bilenko}, {Billingsley}, {Bini}, {Birney}, {Birnholtz}, {Biscans}, {Bischi}, {Biscoveanu}, {Bisht}, {Biswas}, {Bitossi}, {Bizouard}, {Blackburn}, {Blair}, {Blair}, {Blair}, {Bobba}, {Bode}, {Boer}, {Bogaert}, {Boldrini}, {Bonavena}, {Bondu}, {Bonilla}, {Bonnand}, {Booker}, {Boom}, {Bork}, {Boschi}, {Bose}, {Bose}, {Bossilkov}, {Boudart}, {Bouffanais}, {Bozzi}, {Bradaschia}, {Brady}, {Bramley}, {Branch}, {Branchesi}, {Brandt}, {Brau}, {Breschi}, {Briant}, {Briggs}, {Brillet}, {Brinkmann}, {Brockill}, {Brooks}, {Brooks}, {Brown}, {Brunett}, {Bruno}, {Bruntz}, {Bryant}, {Bulik}, {Bulten}, {Buonanno}, {Buscicchio}, {Buskulic}, {Buy}, {Byer}, {Cadonati}, {Cagnoli}, {Cahillane}, {Bustillo}, {Callaghan}, {Callister}, {Calloni}, {Cameron},
  {Camp}, {Canepa}, {Canevarolo}, {Cannavacciuolo}, {Cannon}, {Cao}, {Cao}, {Capocasa}, {Capote}, \& {Carapella}}]{Abbot2023}
{Abbott}, R., {Abbott}, T.~D., {Acernese}, F., {et~al.} 2023, Physical Review X, 13, 011048, \dodoi{10.1103/PhysRevX.13.011048}

\bibitem[{{Antognini} \& {Thompson}(2016)}]{2016MNRAS.456.4219A}
{Antognini}, J.~M.~O., \& {Thompson}, T.~A. 2016, \mnras, 456, 4219, \dodoi{10.1093/mnras/stv2938}

\bibitem[{{Antoni} {et~al.}(2019){Antoni}, {MacLeod}, \& {Ramirez-Ruiz}}]{Antoni:2019}
{Antoni}, A., {MacLeod}, M., \& {Ramirez-Ruiz}, E. 2019, \apj, 884, 22, \dodoi{10.3847/1538-4357/ab3466}

\bibitem[{{Antonini} {et~al.}(2016{\natexlab{a}}){Antonini}, {Chatterjee}, {Rodriguez}, {Morscher}, {Pattabiraman}, {Kalogera}, \& {Rasio}}]{2016ApJ...816...65A}
{Antonini}, F., {Chatterjee}, S., {Rodriguez}, C.~L., {et~al.} 2016{\natexlab{a}}, \apj, 816, 65, \dodoi{10.3847/0004-637X/816/2/65}

\bibitem[{{Antonini} {et~al.}(2016{\natexlab{b}}){Antonini}, {Chatterjee}, {Rodriguez}, {Morscher}, {Pattabiraman}, {Kalogera}, \& {Rasio}}]{antonini2016b}
---. 2016{\natexlab{b}}, \apj, 816, 65, \dodoi{10.3847/0004-637X/816/2/65}

\bibitem[{{Antonini} \& {Rasio}(2016)}]{2016antonini}
{Antonini}, F., \& {Rasio}, F.~A. 2016, \apj, 831, 187, \dodoi{10.3847/0004-637X/831/2/187}

\bibitem[{{Antonini} {et~al.}(2017){Antonini}, {Toonen}, \& {Hamers}}]{antonini2017}
{Antonini}, F., {Toonen}, S., \& {Hamers}, A.~S. 2017, \apj, 841, 77, \dodoi{10.3847/1538-4357/aa6f5e}

\bibitem[{{Arca Sedda} {et~al.}(2021){Arca Sedda}, {Li}, \& {Kocsis}}]{arcasedda2021}
{Arca Sedda}, M., {Li}, G., \& {Kocsis}, B. 2021, \aap, 650, A189, \dodoi{10.1051/0004-6361/202038795}

\bibitem[{{Armitage} \& {Natarajan}(2002)}]{ArmitageNatarajan:2005}
{Armitage}, P.~J., \& {Natarajan}, P. 2002, \apjl, 567, L9, \dodoi{10.1086/339770}

\bibitem[{{Askar} {et~al.}(2017){Askar}, {Szkudlarek}, {Gondek-Rosi{\'n}ska}, {Giersz}, \& {Bulik}}]{2017MNRAS.464L..36A}
{Askar}, A., {Szkudlarek}, M., {Gondek-Rosi{\'n}ska}, D., {Giersz}, M., \& {Bulik}, T. 2017, \mnras, 464, L36, \dodoi{10.1093/mnrasl/slw177}

\bibitem[{{Bae} {et~al.}(2014){Bae}, {Kim}, \& {Lee}}]{2014MNRAS.440.2714B}
{Bae}, Y.-B., {Kim}, C., \& {Lee}, H.~M. 2014, \mnras, 440, 2714, \dodoi{10.1093/mnras/stu381}

\bibitem[{{Banerjee} {et~al.}(2010){Banerjee}, {Baumgardt}, \& {Kroupa}}]{2010MNRAS.402..371B}
{Banerjee}, S., {Baumgardt}, H., \& {Kroupa}, P. 2010, \mnras, 402, 371, \dodoi{10.1111/j.1365-2966.2009.15880.x}

\bibitem[{{Barausse} {et~al.}(2014){Barausse}, {Cardoso}, \& {Pani}}]{2014barausse}
{Barausse}, E., {Cardoso}, V., \& {Pani}, P. 2014, \prd, 89, 104059, \dodoi{10.1103/PhysRevD.89.104059}

\bibitem[{{Barausse} \& {Rezzolla}(2008)}]{2008barausse}
{Barausse}, E., \& {Rezzolla}, L. 2008, \prd, 77, 104027, \dodoi{10.1103/PhysRevD.77.104027}

\bibitem[{{Bartos} {et~al.}(2017){Bartos}, {Kocsis}, {Haiman}, \& {M{\'a}rka}}]{2017ApJ...835..165B}
{Bartos}, I., {Kocsis}, B., {Haiman}, Z., \& {M{\'a}rka}, S. 2017, \apj, 835, 165, \dodoi{10.3847/1538-4357/835/2/165}

\bibitem[{Baruteau {et~al.}(2010)Baruteau, Cuadra, \& Lin}]{Baruteau:2010bk}
Baruteau, C., Cuadra, J., \& Lin, D. N.~C. 2010, \apj, 726, 28

\bibitem[{{Basu} {et~al.}(2024){Basu}, {Chatterjee}, \& {Mondal}}]{2024basu}
{Basu}, P., {Chatterjee}, S., \& {Mondal}, S. 2024, \mnras, 531, 1506, \dodoi{10.1093/mnras/stae1239}

\bibitem[{{Beckmann} {et~al.}(2018){Beckmann}, {Slyz}, \& {Devriendt}}]{Beckmann2018}
{Beckmann}, R.~S., {Slyz}, A., \& {Devriendt}, J. 2018, \mnras, 478, 995, \dodoi{10.1093/mnras/sty931}

\bibitem[{{Belczynski} {et~al.}(2002{\natexlab{a}}){Belczynski}, {Kalogera}, \& {Bulik}}]{2002belczynski}
{Belczynski}, K., {Kalogera}, V., \& {Bulik}, T. 2002{\natexlab{a}}, \apj, 572, 407, \dodoi{10.1086/340304}

\bibitem[{{Belczynski} {et~al.}(2002{\natexlab{b}}){Belczynski}, {Kalogera}, \& {Bulik}}]{belczynski2002}
---. 2002{\natexlab{b}}, \apj, 572, 407, \dodoi{10.1086/340304}

\bibitem[{{Bellovary} {et~al.}(2016){Bellovary}, {Mac Low}, {McKernan}, \& {Ford}}]{2016bellovary}
{Bellovary}, J.~M., {Mac Low}, M.-M., {McKernan}, B., \& {Ford}, K.~E.~S. 2016, \apjl, 819, L17, \dodoi{10.3847/2041-8205/819/2/L17}

\bibitem[{{Blanchet}(2014)}]{blanchet2014}
{Blanchet}, L. 2014, Living Reviews in Relativity, 17, 2, \dodoi{10.12942/lrr-2014-2}

\bibitem[{{Bondi}(1952)}]{bondi1952}
{Bondi}, H. 1952, \mnras, 112, 195, \dodoi{10.1093/mnras/112.2.195}

\bibitem[{{Bonetti} {et~al.}(2017){Bonetti}, {Barausse}, {Faye}, {Haardt}, \& {Sesana}}]{2017Bonetti}
{Bonetti}, M., {Barausse}, E., {Faye}, G., {Haardt}, F., \& {Sesana}, A. 2017, Classical and Quantum Gravity, 34, 215004, \dodoi{10.1088/1361-6382/aa8da5}

\bibitem[{{Cahillane} \& {Mansell}(2022)}]{2022ligopl}
{Cahillane}, C., \& {Mansell}, G. 2022, Galaxies, 10, 36, \dodoi{10.3390/galaxies10010036}

\bibitem[{{Camilloni} {et~al.}(2023){Camilloni}, {Grignani}, {Harmark}, {Orselli}, \& {Pica}}]{pica2023}
{Camilloni}, F., {Grignani}, G., {Harmark}, T., {Orselli}, M., \& {Pica}, D. 2023, arXiv e-prints, arXiv:2310.06894, \dodoi{10.48550/arXiv.2310.06894}

\bibitem[{{Caneva Santoro} {et~al.}(2024){Caneva Santoro}, {Roy}, {Vicente}, {Haney}, {Piccinni}, {Del Pozzo}, \& {Martinez}}]{2024santoro}
{Caneva Santoro}, G., {Roy}, S., {Vicente}, R., {et~al.} 2024, \prl, 132, 251401, \dodoi{10.1103/PhysRevLett.132.251401}

\bibitem[{{Capote} {et~al.}(2025){Capote}, {Jia}, {Aritomi}, {Nakano}, {Xu}, {Abbott}, {Abouelfettouh}, {Adhikari}, {Ananyeva}, {Appert}, {Apple}, {Arai}, {Aston}, {Ball}, {Ballmer}, {Barker}, {Barsotti}, {Berger}, {Betzwieser}, {Bhattacharjee}, {Billingsley}, {Biscans}, {Blair}, {Bode}, {Bonilla}, {Bossilkov}, {Branch}, {Brooks}, {Brown}, {Bryant}, {Cahillane}, {Cao}, {Clara}, {Collins}, {Compton}, {Cottingham}, {Coyne}, {Crouch}, {Csizmazia}, {Cumming}, {Dartez}, {Davis}, {Demos}, {Dohmen}, {Driggers}, {Dwyer}, {Effler}, {Ejlli}, {Etzel}, {Evans}, {Feicht}, {Frey}, {Frischhertz}, {Fritschel}, {Frolov}, {Fuentes-Garcia}, {Fulda}, {Fyffe}, {Ganapathy}, {Gateley}, {Gayer}, {Giaime}, {Giardina}, {Glanzer}, {Goetz}, {Goetz}, {Goodwin-Jones}, {Gras}, {Gray}, {Griffith}, {Grote}, {Guidry}, {Gurs}, {Hall}, {Hanks}, {Hanson}, {Heintze}, {Helmling-Cornell}, {Holland}, {Hoyland}, {Huang}, {Inoue}, {James}, {Jamies}, {Jennings}, {Jones}, {Kabagoz}, {Karat}, {Karki}, {Kasprzack}, {Kawabe}, {Kijbunchoo}, {King},
  {Kissel}, {Komori}, {Kontos}, {Kumar}, {Kuns}, {Landry}, {Lantz}, {Laxen}, {Lee}, {Lesovsky}, {Villarreal}, {Lormand}, {Loughlin}, {Macas}, {MacInnis}, {Makarem}, {Mannix}, {Mansell}, {Martin}, {Mason}, {Matichard}, {Mavalvala}, {Maxwell}, {McCarrol}, {McCarthy}, {McClelland}, {McCormick}, {McRae}, {Mera}, {Merilh}, {Meylahn}, {Mittleman}, {Moraru}, {Moreno}, {Mullavey}, {Nelson}, {Neunzert}, {Notte}, {Oberling}, {O'Hanlon}, {Osthelder}, {Ottaway}, {Overmier}, {Parker}, {Patane}, {Pele}, {Pham}, {Pirello}, {Pullin}, {Quetschke}, {Ramirez}, {Ransom}, {Reyes}, {Richardson}, {Robinson}, {Rollins}, {Romel}, {Romie}, {Ross}, {Ryan}, {Sadecki}, {Sanchez}, {Sanchez}, {Sanchez}, {Savage}, {Schaetzl}, {Schiworski}, {Schnabel}, {Schofield}, {Schwartz}, {Sellers}, {Shaffer}, {Short}, {Sigg}, {Slagmolen}, {Soike}, {Soni}, {Srivastava}, {Sun}, {Tanner}, {Thomas}, {Thomas}, {Thorne}, {Todd}, {Torrie}, {Traylor}, {Ubhi}, {Vajente}, {Vanosky}, {Vecchio}, {Veitch}, {Vibhute}, {von Reis}, {Warner}, {Weaver}, {Weiss},
  {Whittle}, {Willke}, {Wipf}, {Wright}, {Yamamoto}, {Zhang}, \& {Zucker}}]{2025ligopl}
{Capote}, E., {Jia}, W., {Aritomi}, N., {et~al.} 2025, \prd, 111, 062002, \dodoi{10.1103/PhysRevD.111.062002}

\bibitem[{{Caputo} {et~al.}(2020){Caputo}, {Sberna}, {Toubiana}, {Babak}, {Barausse}, {Marsat}, \& {Pani}}]{2020caputo}
{Caputo}, A., {Sberna}, L., {Toubiana}, A., {et~al.} 2020, \apj, 892, 90, \dodoi{10.3847/1538-4357/ab7b66}

\bibitem[{{Cardoso} {et~al.}(2022){Cardoso}, {Destounis}, {Duque}, {Macedo}, \& {Maselli}}]{2022cardoso}
{Cardoso}, V., {Destounis}, K., {Duque}, F., {Macedo}, R.~P., \& {Maselli}, A. 2022, \prl, 129, 241103, \dodoi{10.1103/PhysRevLett.129.241103}

\bibitem[{{Cardoso} \& {Macedo}(2020)}]{2020cardosoself}
{Cardoso}, V., \& {Macedo}, C. F.~B. 2020, \mnras, 498, 1963, \dodoi{10.1093/mnras/staa2396}

\bibitem[{{Cardoso} \& {Maselli}(2020)}]{2020cardoso}
{Cardoso}, V., \& {Maselli}, A. 2020, \aap, 644, A147, \dodoi{10.1051/0004-6361/202037654}

\bibitem[{{Carr} {et~al.}(2016){Carr}, {K{\"u}hnel}, \& {Sandstad}}]{2016PhRvD..94h3504C}
{Carr}, B., {K{\"u}hnel}, F., \& {Sandstad}, M. 2016, \prd, 94, 083504, \dodoi{10.1103/PhysRevD.94.083504}

\bibitem[{{Chakrabarti}(1993)}]{1993chakrabarti}
{Chakrabarti}, S.~K. 1993, \apj, 411, 610, \dodoi{10.1086/172863}

\bibitem[{{Chandramouli} \& {Yunes}(2022)}]{2022chandramouli}
{Chandramouli}, R.~S., \& {Yunes}, N. 2022, \prd, 105, 064009, \dodoi{10.1103/PhysRevD.105.064009}

\bibitem[{{Cole} {et~al.}(2022){Cole}, {Coogan}, {Kavanagh}, \& {Bertone}}]{2022cole}
{Cole}, P.~S., {Coogan}, A., {Kavanagh}, B.~J., \& {Bertone}, G. 2022, arXiv e-prints, arXiv:2207.07576.
\newblock \doarXiv{2207.07576}

\bibitem[{{Colpi} {et~al.}(2024){Colpi}, {Danzmann}, {Hewitson}, {Holley-Bockelmann}, {Jetzer}, {Nelemans}, {Petiteau}, {Shoemaker}, {Sopuerta}, {Stebbins}, {Tanvir}, {Ward}, {Weber}, {Thorpe}, {Daurskikh}, {Deep}, {Fern{\'a}ndez N{\'u}{\~n}ez}, {Garc{\'\i}a Marirrodriga}, {Gehler}, {Halain}, {Jennrich}, {Lammers}, {Larra{\~n}aga}, {Lieser}, {L{\"u}tzgendorf}, {Martens}, {Mondin}, {Piris Ni{\~n}o}, {Amaro-Seoane}, {Arca Sedda}, {Auclair}, {Babak}, {Baghi}, {Baibhav}, {Baker}, {Bayle}, {Berry}, {Berti}, {Boileau}, {Bonetti}, {Brito}, {Buscicchio}, {Calcagni}, {Capelo}, {Caprini}, {Caputo}, {Castelli}, {Chen}, {Chen}, {Chua}, {Davies}, {Derdzinski}, {Domcke}, {Doneva}, {Dvorkin}, {Mar{\'\i}a Ezquiaga}, {Gair}, {Haiman}, {Harry}, {Hartwig}, {Hees}, {Heffernan}, {Husa}, {Izquierdo}, {Karnesis}, {Klein}, {Korol}, {Korsakova}, {Kupfer}, {Laghi}, {Lamberts}, {Larson}, {Le Jeune}, {Lewicki}, {Littenberg}, {Madge}, {Mangiagli}, {Marsat}, {Vilchez}, {Maselli}, {Mathews}, {van de Meent}, {Muratore}, {Nardini}, {Pani},
  {Peloso}, {Pieroni}, {Pound}, {Quelquejay-Leclere}, {Ricciardone}, {Rossi}, {Sartirana}, {Savalle}, {Sberna}, {Sesana}, {Shoemaker}, {Slutsky}, {Sotiriou}, {Speri}, {Staab}, {Steer}, {Tamanini}, {Tasinato}, {Torrado}, {Torres-Orjuela}, {Toubiana}, {Vallisneri}, {Vecchio}, {Volonteri}, {Yagi}, \& {Zwick}}]{2024lisa}
{Colpi}, M., {Danzmann}, K., {Hewitson}, M., {et~al.} 2024, arXiv e-prints, arXiv:2402.07571, \dodoi{10.48550/arXiv.2402.07571}

\bibitem[{{Cutler} \& {Flanagan}(1994)}]{1994cutler}
{Cutler}, C., \& {Flanagan}, {\'E}.~E. 1994, \prd, 49, 2658, \dodoi{10.1103/PhysRevD.49.2658}

\bibitem[{{De Luca} {et~al.}(2021){De Luca}, {Franciolini}, {Pani}, \& {Riotto}}]{deluca2021}
{De Luca}, V., {Franciolini}, G., {Pani}, P., \& {Riotto}, A. 2021, arXiv e-prints, arXiv:2106.13769.
\newblock \doarXiv{2106.13769}

\bibitem[{{Dempsey} {et~al.}(2022){Dempsey}, {Li}, {Mishra}, \& {Li}}]{Dempsey3D:2022}
{Dempsey}, A.~M., {Li}, H., {Mishra}, B., \& {Li}, S. 2022, \apj, 940, 155, \dodoi{10.3847/1538-4357/ac9d92}

\bibitem[{{Derdzinski} {et~al.}(2021){Derdzinski}, {D'Orazio}, {Duffell}, {Haiman}, \& {MacFadyen}}]{Derdzinksi:2021}
{Derdzinski}, A., {D'Orazio}, D., {Duffell}, P., {Haiman}, Z., \& {MacFadyen}, A. 2021, \mnras, 501, 3540, \dodoi{10.1093/mnras/staa3976}

\bibitem[{{Destounis} {et~al.}(2022){Destounis}, {Kulathingal}, {Kokkotas}, \& {Papadopoulos}}]{2022destounis}
{Destounis}, K., {Kulathingal}, A., {Kokkotas}, K.~D., \& {Papadopoulos}, G.~O. 2022, arXiv e-prints, arXiv:2210.09357, \dodoi{10.48550/arXiv.2210.09357}

\bibitem[{{Dittmann} {et~al.}(2023{\natexlab{a}}){Dittmann}, {Dempsey}, \& {Li}}]{dittmann2024}
{Dittmann}, A.~J., {Dempsey}, A.~M., \& {Li}, H. 2023{\natexlab{a}}, arXiv e-prints, arXiv:2310.03832, \dodoi{10.48550/arXiv.2310.03832}

\bibitem[{{Dittmann} {et~al.}(2024){Dittmann}, {Dempsey}, \& {Li}}]{DittmannDempsey:2024}
---. 2024, \apj, 964, 61, \dodoi{10.3847/1538-4357/ad23ce}

\bibitem[{{Dittmann} {et~al.}(2023{\natexlab{b}}){Dittmann}, {Ryan}, \& {Miller}}]{Dittmann:Decoupling:2023}
{Dittmann}, A.~J., {Ryan}, G., \& {Miller}, M.~C. 2023{\natexlab{b}}, \apjl, 949, L30, \dodoi{10.3847/2041-8213/acd183}

\bibitem[{{D'Orazio} \& {Loeb}(2020)}]{DOrazioGWLens:2020}
{D'Orazio}, D.~J., \& {Loeb}, A. 2020, \prd, 101, 083031, \dodoi{10.1103/PhysRevD.101.083031}

\bibitem[{{Duque} {et~al.}(2024){Duque}, {Kejriwal}, {Sberna}, {Speri}, \& {Gair}}]{2024duque}
{Duque}, F., {Kejriwal}, S., {Sberna}, L., {Speri}, L., \& {Gair}, J. 2024, arXiv e-prints, arXiv:2411.03436, \dodoi{10.48550/arXiv.2411.03436}

\bibitem[{{Dyson} {et~al.}(2024){Dyson}, {Redondo-Yuste}, {van de Meent}, \& {Cardoso}}]{2024dyson}
{Dyson}, C., {Redondo-Yuste}, J., {van de Meent}, M., \& {Cardoso}, V. 2024, \prd, 109, 104038, \dodoi{10.1103/PhysRevD.109.104038}

\bibitem[{{Dyson} {et~al.}(2025){Dyson}, {Spieksma}, {Brito}, {van de Meent}, \& {Dolan}}]{2025dyson}
{Dyson}, C., {Spieksma}, T. F.~M., {Brito}, R., {van de Meent}, M., \& {Dolan}, S. 2025, arXiv e-prints, arXiv:2501.09806, \dodoi{10.48550/arXiv.2501.09806}

\bibitem[{{Ennoggi} {et~al.}(2025){Ennoggi}, {Campanelli}, {Zlochower}, {Noble}, {Krolik}, {Cattorini}, {Kalinani}, {Mewes}, {Chabanov}, {Ji}, \& {de Simone}}]{EnnoggiCampanelliNoble:2025}
{Ennoggi}, L., {Campanelli}, M., {Zlochower}, Y., {et~al.} 2025, arXiv e-prints, arXiv:2502.06389.
\newblock \doarXiv{2502.06389}

\bibitem[{{Evans} {et~al.}(2023){Evans}, {Corsi}, {Afle}, {Ananyeva}, {Arun}, {Ballmer}, {Bandopadhyay}, {Barsotti}, {Baryakhtar}, {Berger}, {Berti}, {Biscoveanu}, {Borhanian}, {Broekgaarden}, {Brown}, {Cahillane}, {Campbell}, {Chen}, {Daniel}, {Dhani}, {Driggers}, {Effler}, {Eisenstein}, {Fairhurst}, {Feicht}, {Fritschel}, {Fulda}, {Gupta}, {Hall}, {Hammond}, {Hannuksela}, {Hansen}, {Haster}, {Kacanja}, {Kamai}, {Kashyap}, {Shapiro Key}, {Khadkikar}, {Kontos}, {Kuns}, {Landry}, {Landry}, {Lantz}, {Li}, {Lovelace}, {Mandic}, {Mansell}, {Martynov}, {McCuller}, {Miller}, {Nitz}, {Owen}, {Palomba}, {Read}, {Phurailatpam}, {Reddy}, {Richardson}, {Rollins}, {Romano}, {Sathyaprakash}, {Schofield}, {Shoemaker}, {Sigg}, {Singh}, {Slagmolen}, {Sledge}, {Smith}, {Soares-Santos}, {Strunk}, {Sun}, {Tanner}, {van Son}, {Vitale}, {Willke}, {Yamamoto}, \& {Zucker}}]{2023arXiv230613745E}
{Evans}, M., {Corsi}, A., {Afle}, C., {et~al.} 2023, arXiv e-prints, arXiv:2306.13745, \dodoi{10.48550/arXiv.2306.13745}

\bibitem[{{Fabj} {et~al.}(2020){Fabj}, {Nasim}, {Caban}, {Ford}, {McKernan}, \& {Bellovary}}]{fabj2020}
{Fabj}, G., {Nasim}, S.~S., {Caban}, F., {et~al.} 2020, \mnras, 499, 2608, \dodoi{10.1093/mnras/staa3004}

\bibitem[{Fabj \& Samsing(2024)}]{Fabj24}
Fabj, G., \& Samsing, J. 2024, arXiv e-prints, arXiv:2402.16948.
\newblock \url{https://arxiv.org/pdf/2402.16948}

\bibitem[{Favata {et~al.}(2022)Favata, Kim, Arun, Kim, \& Lee}]{Favata:2021vhw}
Favata, M., Kim, C., Arun, K.~G., Kim, J., \& Lee, H.~W. 2022, Phys. Rev. D, 105, 023003, \dodoi{10.1103/PhysRevD.105.023003}

\bibitem[{{Gair} {et~al.}(2013){Gair}, {Vallisneri}, {Larson}, \& {Baker}}]{2013gair}
{Gair}, J.~R., {Vallisneri}, M., {Larson}, S.~L., \& {Baker}, J.~G. 2013, Living Rev. Relat., 16, 7, \dodoi{10.12942/lrr-2013-7}

\bibitem[{{Garg} {et~al.}(2024){Garg}, {Derdzinski}, {Tiwari}, {Gair}, \& {Mayer}}]{2024garg}
{Garg}, M., {Derdzinski}, A., {Tiwari}, S., {Gair}, J., \& {Mayer}, L. 2024, arXiv e-prints, arXiv:2402.14058, \dodoi{10.48550/arXiv.2402.14058}

\bibitem[{{Garg} {et~al.}(2022){Garg}, {Derdzinski}, {Zwick}, {Capelo}, \& {Mayer}}]{garg2022}
{Garg}, M., {Derdzinski}, A., {Zwick}, L., {Capelo}, P.~R., \& {Mayer}, L. 2022, \mnras, 517, 1339, \dodoi{10.1093/mnras/stac2711}

\bibitem[{{Goldreich} \& {Tremaine}(1980)}]{GoldreichTremain:1980}
{Goldreich}, P., \& {Tremaine}, S. 1980, \apj, 241, 425, \dodoi{10.1086/158356}

\bibitem[{{Goodman}(2003)}]{goo03}
{Goodman}, J. 2003, \mnras, 339, 937, \dodoi{10.1046/j.1365-8711.2003.06241.x}

\bibitem[{Greene \& Ho(2007)}]{Greene:2007dz}
Greene, J.~E., \& Ho, L.~C. 2007, \apj, 667, 131

\bibitem[{G{\"u}ltekin {et~al.}(2006)G{\"u}ltekin, Miller, \& Hamilton}]{2006ApJ...640..156G}
G{\"u}ltekin, K., Miller, M.~C., \& Hamilton, D.~P. 2006, \apj, 640, 156

\bibitem[{{Gupta} {et~al.}(2024){Gupta}, {Afle}, {Arun}, {Bandopadhyay}, {Baryakhtar}, {Biscoveanu}, {Borhanian}, {Broekgaarden}, {Corsi}, {Dhani}, {Evans}, {Hall}, {Hannuksela}, {Kacanja}, {Kashyap}, {Khadkikar}, {Kuns}, {Li}, {Miller}, {Harvey Nitz}, {Owen}, {Palomba}, {Pearce}, {Phurailatpam}, {Rajbhandari}, {Read}, {Romano}, {Sathyaprakash}, {Shoemaker}, {Singh}, {Vitale}, {Barsotti}, {Berti}, {Cahillane}, {Chen}, {Fritschel}, {Haster}, {Landry}, {Lovelace}, {McClelland}, {J J Slagmolen}, {R Smith}, {Soares-Santos}, {Sun}, {Tanner}, {Yamamoto}, \& {Zucker}}]{2024ligosharp}
{Gupta}, I., {Afle}, C., {Arun}, K.~G., {et~al.} 2024, Classical and Quantum Gravity, 41, 245001, \dodoi{10.1088/1361-6382/ad7b99}

\bibitem[{{Hamers}(2021)}]{hamers2021fit}
{Hamers}, A.~S. 2021, Research Notes of the American Astronomical Society, 5, 275, \dodoi{10.3847/2515-5172/ac3d98}

\bibitem[{{Hamers} \& {Thompson}(2019)}]{2019ApJ...883...23H}
{Hamers}, A.~S., \& {Thompson}, T.~A. 2019, \apj, 883, 23, \dodoi{10.3847/1538-4357/ab3b06}

\bibitem[{{Han} {et~al.}(2024){Han}, {Yang}, {Tagawa}, {Jiang}, {Shen}, {Yun}, {Zhang}, \& {Zhong}}]{2024Han}
{Han}, W.-B., {Yang}, S.-C., {Tagawa}, H., {et~al.} 2024, arXiv e-prints, arXiv:2401.01743, \dodoi{10.48550/arXiv.2401.01743}

\bibitem[{{Hendriks} {et~al.}(2024{\natexlab{a}}){Hendriks}, {Zwick}, \& {Samsing}}]{kai22024}
{Hendriks}, K., {Zwick}, L., \& {Samsing}, J. 2024{\natexlab{a}}, arXiv e-prints, arXiv:2408.04603, \dodoi{10.48550/arXiv.2408.04603}

\bibitem[{{Hendriks} {et~al.}(2024{\natexlab{b}}){Hendriks}, {Atallah}, {Martinez}, {Zevin}, {Zwick}, {Trani}, {Saini}, {Tak{\'a}tsy}, \& {Samsing}}]{kai2024}
{Hendriks}, K., {Atallah}, D., {Martinez}, M., {et~al.} 2024{\natexlab{b}}, arXiv e-prints, arXiv:2411.08572, \dodoi{10.48550/arXiv.2411.08572}

\bibitem[{{Inayoshi} {et~al.}(2017){Inayoshi}, {Hirai}, {Kinugawa}, \& {Hotokezaka}}]{inayoshi2017}
{Inayoshi}, K., {Hirai}, R., {Kinugawa}, T., \& {Hotokezaka}, K. 2017, \mnras, 468, 5020, \dodoi{10.1093/mnras/stx757}

\bibitem[{{Jo} {et~al.}(2021){Jo}, {Youn}, {Kim}, {Park}, {Hwang}, {Lee}, \& {Kim}}]{2021jo}
{Jo}, J.~U., {Youn}, S., {Kim}, S., {et~al.} 2021, \apss, 366, 18, \dodoi{10.1007/s10509-021-03925-7}

\bibitem[{{Katz} {et~al.}(2021){Katz}, {Chua}, {Speri}, {Warburton}, \& {Hughes}}]{2021katz}
{Katz}, M.~L., {Chua}, A. J.~K., {Speri}, L., {Warburton}, N., \& {Hughes}, S.~A. 2021, \prd, 104, 064047, \dodoi{10.1103/PhysRevD.104.064047}

\bibitem[{{Kavanagh} {et~al.}(2020){Kavanagh}, {Nichols}, {Bertone}, \& {Gaggero}}]{kavanagh2020}
{Kavanagh}, B.~J., {Nichols}, D.~A., {Bertone}, G., \& {Gaggero}, D. 2020, \prd, 102, 083006, \dodoi{10.1103/PhysRevD.102.083006}

\bibitem[{{Kimball} {et~al.}(2021){Kimball}, {Talbot}, {Berry}, {Zevin}, {Thrane}, {Kalogera}, {Buscicchio}, {Carney}, {Dent}, {Middleton}, {Payne}, {Veitch}, \& {Williams}}]{2021kimball}
{Kimball}, C., {Talbot}, C., {Berry}, C. P.~L., {et~al.} 2021, \apjl, 915, L35, \dodoi{10.3847/2041-8213/ac0aef}

\bibitem[{{Kocsis} {et~al.}(2011){Kocsis}, {Yunes}, \& {Loeb}}]{kocsis}
{Kocsis}, B., {Yunes}, N., \& {Loeb}, A. 2011, \prd, 84, 024032, \dodoi{10.1103/PhysRevD.84.024032}

\bibitem[{{Kozai}(1962)}]{koz62}
{Kozai}, Y. 1962, \aj, 67, 591, \dodoi{10.1086/108790}

\bibitem[{{Lai} \& {Mu{\~n}oz}(2022)}]{LaiMunoz:Review:2022}
{Lai}, D., \& {Mu{\~n}oz}, D.~J. 2022, arXiv e-prints, arXiv:2211.00028.
\newblock \doarXiv{2211.00028}

\bibitem[{Lee {et~al.}(2010)Lee, Ramirez-Ruiz, \& van~de Ven}]{Lee:2010in}
Lee, W.~H., Ramirez-Ruiz, E., \& van~de Ven, G. 2010, \apj, 720, 953

\bibitem[{{Levin}(2007)}]{2007levin}
{Levin}, Y. 2007, \mnras, 374, 515, \dodoi{10.1111/j.1365-2966.2006.11155.x}

\bibitem[{{Li} {et~al.}(2014){Li}, {Naoz}, {Kocsis}, \& {Loeb}}]{2014ApJ...785..116L}
{Li}, G., {Naoz}, S., {Kocsis}, B., \& {Loeb}, A. 2014, \apj, 785, 116, \dodoi{10.1088/0004-637X/785/2/116}

\bibitem[{{Li} \& {Lai}(2022)}]{LiLai:2022}
{Li}, R., \& {Lai}, D. 2022, \mnras, 517, 1602, \dodoi{10.1093/mnras/stac2577}

\bibitem[{{Li} {et~al.}(2021){Li}, {Dempsey}, {Li}, {Li}, \& {Li}}]{LiDempsey:2021}
{Li}, Y.-P., {Dempsey}, A.~M., {Li}, S., {Li}, H., \& {Li}, J. 2021, \apj, 911, 124, \dodoi{10.3847/1538-4357/abed48}

\bibitem[{{Lidov}(1962)}]{lid62}
{Lidov}, M.~L. 1962, \planss, 9, 719, \dodoi{10.1016/0032-0633(62)90129-0}

\bibitem[{{Lin} \& {Papaloizou}(1986)}]{LinPapaloizou:1986}
{Lin}, D.~N.~C., \& {Papaloizou}, J. 1986, \apj, 309, 846, \dodoi{10.1086/164653}

\bibitem[{{Liu} {et~al.}(2022){Liu}, {D'Orazio}, {Vigna-G{\'o}mez}, \& {Samsing}}]{2022liu}
{Liu}, B., {D'Orazio}, D.~J., {Vigna-G{\'o}mez}, A., \& {Samsing}, J. 2022, \prd, 106, 123010, \dodoi{10.1103/PhysRevD.106.123010}

\bibitem[{{Liu} \& {Lai}(2021)}]{2021MNRAS.502.2049L}
{Liu}, B., \& {Lai}, D. 2021, \mnras, 502, 2049, \dodoi{10.1093/mnras/stab178}

\bibitem[{{Liu} {et~al.}(2019){Liu}, {Lai}, \& {Wang}}]{2019ApJ...881...41L}
{Liu}, B., {Lai}, D., \& {Wang}, Y.-H. 2019, \apj, 881, 41, \dodoi{10.3847/1538-4357/ab2dfb}

\bibitem[{{Liu} {et~al.}(2015){Liu}, {Mu{\~n}oz}, \& {Lai}}]{liu2015}
{Liu}, B., {Mu{\~n}oz}, D.~J., \& {Lai}, D. 2015, \mnras, 447, 747, \dodoi{10.1093/mnras/stu2396}

\bibitem[{{Madau} \& {Dickinson}(2014)}]{2014madau}
{Madau}, P., \& {Dickinson}, M. 2014, \araa, 52, 415, \dodoi{10.1146/annurev-astro-081811-125615}

\bibitem[{{Maggiore}(2018)}]{2018maggiore}
{Maggiore}, M. 2018, {Gravitational Waves: Volume 2: Astrophysics and Cosmology}, \dodoi{10.1093/oso/9780198570899.001.0001}

\bibitem[{{Maggiore} {et~al.}(2020){Maggiore}, {Van Den Broeck}, {Bartolo}, {Belgacem}, {Bertacca}, {Bizouard}, {Branchesi}, {Clesse}, {Foffa}, {Garc{\'\i}a-Bellido}, {Grimm}, {Harms}, {Hinderer}, {Matarrese}, {Palomba}, {Peloso}, {Ricciardone}, \& {Sakellariadou}}]{2020maggioreet}
{Maggiore}, M., {Van Den Broeck}, C., {Bartolo}, N., {et~al.} 2020, \jcap, 2020, 050, \dodoi{10.1088/1475-7516/2020/03/050}

\bibitem[{{Mapelli}(2021)}]{mapelli2021review}
{Mapelli}, M. 2021, in Handbook of Gravitational Wave Astronomy, 4, \dodoi{10.1007/978-981-15-4702-7\_16-1}

\bibitem[{{Mar{\'\i}n Pina} {et~al.}(2025){Mar{\'\i}n Pina}, {Gieles}, {Andrade}, \& {Trani}}]{pin2025}
{Mar{\'\i}n Pina}, D., {Gieles}, M., {Andrade}, T., \& {Trani}, A.~A. 2025, arXiv e-prints, arXiv:2501.02907, \dodoi{10.48550/arXiv.2501.02907}

\bibitem[{{Martin Barandiaran} {et~al.}(2024){Martin Barandiaran}, {Kuroyanagi}, \& {Nesseris}}]{2024barandiaran}
{Martin Barandiaran}, M., {Kuroyanagi}, S., \& {Nesseris}, S. 2024, Classical and Quantum Gravity, 41, 095002, \dodoi{10.1088/1361-6382/ad36a7}

\bibitem[{{Martinez} {et~al.}(2020){Martinez}, {Fragione}, {Kremer}, {Chatterjee}, {Rodriguez}, {Samsing}, {Ye}, {Weatherford}, {Zevin}, {Naoz}, \& {Rasio}}]{martinez2020}
{Martinez}, M. A.~S., {Fragione}, G., {Kremer}, K., {et~al.} 2020, \apj, 903, 67, \dodoi{10.3847/1538-4357/abba25}

\bibitem[{{McKernan} {et~al.}(2017){McKernan}, {Ford}, {Bellovary}, {Leigh}, {Haiman}, {Kocsis}, {Lyra}, {MacLow}, {Metzger}, {O'Dowd}, {Endlich}, \& {Rosen}}]{2017arXiv170207818M}
{McKernan}, B., {Ford}, K.~E.~S., {Bellovary}, J., {et~al.} 2017, ArXiv e-prints.
\newblock \doarXiv{1702.07818}

\bibitem[{{Meiron} {et~al.}(2017){Meiron}, {Kocsis}, \& {Loeb}}]{2017meiron}
{Meiron}, Y., {Kocsis}, B., \& {Loeb}, A. 2017, \apj, 834, 200, \dodoi{10.3847/1538-4357/834/2/200}

\bibitem[{Moore {et~al.}(2016)Moore, Favata, Arun, \& Mishra}]{Moore:2016qxz}
Moore, B., Favata, M., Arun, K.~G., \& Mishra, C.~K. 2016, Phys. Rev. D, 93, 124061, \dodoi{10.1103/PhysRevD.93.124061}

\bibitem[{{Naoz} {et~al.}(2013){Naoz}, {Kocsis}, {Loeb}, \& {Yunes}}]{2013ApJ...773..187N}
{Naoz}, S., {Kocsis}, B., {Loeb}, A., \& {Yunes}, N. 2013, \apj, 773, 187, \dodoi{10.1088/0004-637X/773/2/187}

\bibitem[{{O'Leary} {et~al.}(2007){O'Leary}, {O'Shaughnessy}, \& {Rasio}}]{2007oleary}
{O'Leary}, R.~M., {O'Shaughnessy}, R., \& {Rasio}, F.~A. 2007, \prd, 76, 061504, \dodoi{10.1103/PhysRevD.76.061504}

\bibitem[{{O'Neill} {et~al.}(2025){O'Neill}, {Tiede}, {D'Orazio}, {Haiman}, \& {MacFadyen}}]{ONeillTiedeDOrazio:2025}
{O'Neill}, D., {Tiede}, C., {D'Orazio}, D.~J., {Haiman}, Z., \& {MacFadyen}, A. 2025, arXiv e-prints, arXiv:2501.11679, \dodoi{10.48550/arXiv.2501.11679}

\bibitem[{{Ostriker}(1999)}]{ostriker1999}
{Ostriker}, E.~C. 1999, \apj, 513, 252, \dodoi{10.1086/306858}

\bibitem[{{Park} {et~al.}(2017){Park}, {Kim}, {Lee}, {Bae}, \& {Belczynski}}]{2017MNRAS.469.4665P}
{Park}, D., {Kim}, C., {Lee}, H.~M., {Bae}, Y.-B., \& {Belczynski}, K. 2017, \mnras, 469, 4665, \dodoi{10.1093/mnras/stx1015}

\bibitem[{{Peters}(1964)}]{peters1964}
{Peters}, P.~C. 1964, Physical Review, 136, 1224, \dodoi{10.1103/PhysRev.136.B1224}

\bibitem[{Portegies~Zwart \& McMillan(2000)}]{2000ApJ...528L..17P}
Portegies~Zwart, S.~F., \& McMillan, S. L.~W. 2000, \apj, 528, L17

\bibitem[{{Ramirez-Ruiz} {et~al.}(2015){Ramirez-Ruiz}, {Trenti}, {MacLeod}, {Roberts}, {Lee}, \& {Saladino-Rosas}}]{2015ApJ...802L..22R}
{Ramirez-Ruiz}, E., {Trenti}, M., {MacLeod}, M., {et~al.} 2015, \apjl, 802, L22, \dodoi{10.1088/2041-8205/802/2/L22}

\bibitem[{{Randall} \& {Xianyu}(2018)}]{2018ApJ...864..134R}
{Randall}, L., \& {Xianyu}, Z.-Z. 2018, \apj, 864, 134, \dodoi{10.3847/1538-4357/aad7fe}

\bibitem[{{Randall} \& {Xianyu}(2019)}]{2019randall}
---. 2019, arXiv e-prints, arXiv:1902.08604, \dodoi{10.48550/arXiv.1902.08604}

\bibitem[{{Regimbau} {et~al.}(2012){Regimbau}, {Dent}, {Del Pozzo}, {Giampanis}, {Li}, {Robinson}, {Van Den Broeck}, {Meacher}, {Rodriguez}, {Sathyaprakash}, \& {W{\'o}jcik}}]{2012PhRvD..86l2001R}
{Regimbau}, T., {Dent}, T., {Del Pozzo}, W., {et~al.} 2012, \prd, 86, 122001, \dodoi{10.1103/PhysRevD.86.122001}

\bibitem[{{Riley} {et~al.}(2022){Riley}, {Agrawal}, {Barrett}, {Boyett}, {Broekgaarden}, {Chattopadhyay}, {Gaebel}, {Gittins}, {Hirai}, {Howitt}, {Justham}, {Khandelwal}, {Kummer}, {Lau}, {Mandel}, {de Mink}, {Neijssel}, {Riley}, {van Son}, {Stevenson}, {Vigna-G{\'o}mez}, {Vinciguerra}, {Wagg}, {Willcox}, \& {Team Compas}}]{compas2022}
{Riley}, J., {Agrawal}, P., {Barrett}, J.~W., {et~al.} 2022, \apjs, 258, 34, \dodoi{10.3847/1538-4365/ac416c}

\bibitem[{{Robson} {et~al.}(2019){Robson}, {Cornish}, \& {Liu}}]{2019robson}
{Robson}, T., {Cornish}, N.~J., \& {Liu}, C. 2019, Class. Quantum Gravity, 36, 105011, \dodoi{10.1088/1361-6382/ab1101}

\bibitem[{{Robson} {et~al.}(2018){Robson}, {Cornish}, {Tamanini}, \& {Toonen}}]{2018PhRvD..98f4012R}
{Robson}, T., {Cornish}, N.~J., {Tamanini}, N., \& {Toonen}, S. 2018, \prd, 98, 064012, \dodoi{10.1103/PhysRevD.98.064012}

\bibitem[{{Rodriguez} {et~al.}(2018){Rodriguez}, {Amaro-Seoane}, {Chatterjee}, {Kremer}, {Rasio}, {Samsing}, {Ye}, \& {Zevin}}]{2018PhRvD..98l3005R}
{Rodriguez}, C.~L., {Amaro-Seoane}, P., {Chatterjee}, S., {et~al.} 2018, \prd, 98, 123005, \dodoi{10.1103/PhysRevD.98.123005}

\bibitem[{{Rodriguez} \& {Antonini}(2018)}]{rodriguez2018}
{Rodriguez}, C.~L., \& {Antonini}, F. 2018, \apj, 863, 7, \dodoi{10.3847/1538-4357/aacea4}

\bibitem[{{Rodriguez} {et~al.}(2016{\natexlab{a}}){Rodriguez}, {Chatterjee}, \& {Rasio}}]{2016PhRvD..93h4029R}
{Rodriguez}, C.~L., {Chatterjee}, S., \& {Rasio}, F.~A. 2016{\natexlab{a}}, \prd, 93, 084029, \dodoi{10.1103/PhysRevD.93.084029}

\bibitem[{{Rodriguez} {et~al.}(2016{\natexlab{b}}){Rodriguez}, {Haster}, {Chatterjee}, {Kalogera}, \& {Rasio}}]{2016ApJ...824L...8R}
{Rodriguez}, C.~L., {Haster}, C.-J., {Chatterjee}, S., {Kalogera}, V., \& {Rasio}, F.~A. 2016{\natexlab{b}}, \apjl, 824, L8, \dodoi{10.3847/2041-8205/824/1/L8}

\bibitem[{{Rodriguez} {et~al.}(2015){Rodriguez}, {Morscher}, {Pattabiraman}, {Chatterjee}, {Haster}, \& {Rasio}}]{2015PhRvL.115e1101R}
{Rodriguez}, C.~L., {Morscher}, M., {Pattabiraman}, B., {et~al.} 2015, \prl, 115, 051101, \dodoi{10.1103/PhysRevLett.115.051101}

\bibitem[{{Romero-Shaw} {et~al.}(2021){Romero-Shaw}, {Lasky}, \& {Thrane}}]{2021ApJ...921L..31R}
{Romero-Shaw}, I., {Lasky}, P.~D., \& {Thrane}, E. 2021, \apjl, 921, L31, \dodoi{10.3847/2041-8213/ac3138}

\bibitem[{{Romero-Shaw} {et~al.}(2020){Romero-Shaw}, {Lasky}, {Thrane}, \& {Calder{\'o}n Bustillo}}]{2020shaw}
{Romero-Shaw}, I., {Lasky}, P.~D., {Thrane}, E., \& {Calder{\'o}n Bustillo}, J. 2020, \apjl, 903, L5, \dodoi{10.3847/2041-8213/abbe26}

\bibitem[{{Rowan} {et~al.}(2022){Rowan}, {Boekholt}, {Kocsis}, \& {Haiman}}]{rowan2023}
{Rowan}, C., {Boekholt}, T., {Kocsis}, B., \& {Haiman}, Z. 2022, arXiv e-prints, arXiv:2212.06133, \dodoi{10.48550/arXiv.2212.06133}

\bibitem[{{Rowan} {et~al.}(2024{\natexlab{a}}){Rowan}, {Whitehead}, {Boekholt}, {Kocsis}, \& {Haiman}}]{Rowan2024}
{Rowan}, C., {Whitehead}, H., {Boekholt}, T., {Kocsis}, B., \& {Haiman}, Z. 2024{\natexlab{a}}, \mnras, 527, 10448, \dodoi{10.1093/mnras/stad3641}

\bibitem[{{Rowan} {et~al.}(2025){Rowan}, {Whitehead}, {Fabj}, {Saini}, {Kocsis}, {Pessah}, \& {Samsing}}]{Rowan2025_trips}
{Rowan}, C., {Whitehead}, H., {Fabj}, G., {et~al.} 2025, arXiv e-prints, arXiv:2501.09017, \dodoi{10.48550/arXiv.2501.09017}

\bibitem[{{Rowan} {et~al.}(2024{\natexlab{b}}){Rowan}, {Whitehead}, \& {Kocsis}}]{Rowan2024_rates}
{Rowan}, C., {Whitehead}, H., \& {Kocsis}, B. 2024{\natexlab{b}}, arXiv e-prints, arXiv:2412.12086, \dodoi{10.48550/arXiv.2412.12086}

\bibitem[{{Roy} \& {Vicente}(2024)}]{2024soumen}
{Roy}, S., \& {Vicente}, R. 2024, arXiv e-prints, arXiv:2410.16388, \dodoi{10.48550/arXiv.2410.16388}

\bibitem[{{Rozner} \& {Perets}(2022)}]{rozner2022}
{Rozner}, M., \& {Perets}, H.~B. 2022, \apj, 931, 149, \dodoi{10.3847/1538-4357/ac6d55}

\bibitem[{{Ryan}(1995)}]{1995ryan}
{Ryan}, F.~D. 1995, \prd, 52, 5707, \dodoi{10.1103/PhysRevD.52.5707}

\bibitem[{{Sadowski} {et~al.}(2008){Sadowski}, {Belczynski}, {Bulik}, {Ivanova}, {Rasio}, \& {O'Shaughnessy}}]{2008Sadowski}
{Sadowski}, A., {Belczynski}, K., {Bulik}, T., {et~al.} 2008, \apj, 676, 1162, \dodoi{10.1086/528932}

\bibitem[{Saini(2024)}]{Saini:2023wdk}
Saini, P. 2024, Mon. Not. Roy. Astron. Soc., 528, 833, \dodoi{10.1093/mnras/stae037}

\bibitem[{Samsing(2018)}]{samsing2018a}
Samsing, J. 2018, \prd, 97, 103014, \dodoi{10.1103/PhysRevD.97.103014}

\bibitem[{{Samsing} {et~al.}(2018{\natexlab{a}}){Samsing}, {Askar}, \& {Giersz}}]{2018ApJ...855..124S}
{Samsing}, J., {Askar}, A., \& {Giersz}, M. 2018{\natexlab{a}}, \apj, 855, 124, \dodoi{10.3847/1538-4357/aaab52}

\bibitem[{{Samsing} \& {D'Orazio}(2018)}]{2018MNRAS.tmp.2223S}
{Samsing}, J., \& {D'Orazio}, D.~J. 2018, \mnras, \dodoi{10.1093/mnras/sty2334}

\bibitem[{{Samsing} {et~al.}(2020){Samsing}, {D'Orazio}, {Kremer}, {Rodriguez}, \& {Askar}}]{2020PhRvD.101l3010S}
{Samsing}, J., {D'Orazio}, D.~J., {Kremer}, K., {Rodriguez}, C.~L., \& {Askar}, A. 2020, \prd, 101, 123010, \dodoi{10.1103/PhysRevD.101.123010}

\bibitem[{{Samsing} {et~al.}(2019){Samsing}, {Hamers}, \& {Tyles}}]{2019PhRvD.100d3010S}
{Samsing}, J., {Hamers}, A.~S., \& {Tyles}, J.~G. 2019, \prd, 100, 043010, \dodoi{10.1103/PhysRevD.100.043010}

\bibitem[{{Samsing} {et~al.}(2024{\natexlab{a}}){Samsing}, {Hendriks}, {Zwick}, {D'Orazio}, \& {Liu}}]{2024samsing}
{Samsing}, J., {Hendriks}, K., {Zwick}, L., {D'Orazio}, D.~J., \& {Liu}, B. 2024{\natexlab{a}}, arXiv e-prints, arXiv:2403.05625, \dodoi{10.48550/arXiv.2403.05625}

\bibitem[{{Samsing} {et~al.}(2014{\natexlab{a}}){Samsing}, {MacLeod}, \& {Ramirez-Ruiz}}]{2014ApJ...784...71S}
{Samsing}, J., {MacLeod}, M., \& {Ramirez-Ruiz}, E. 2014{\natexlab{a}}, \apj, 784, 71, \dodoi{10.1088/0004-637X/784/1/71}

\bibitem[{{Samsing} {et~al.}(2014{\natexlab{b}}){Samsing}, {MacLeod}, \& {Ramirez-Ruiz}}]{samsing2014}
---. 2014{\natexlab{b}}, \apj, 784, 71, \dodoi{10.1088/0004-637X/784/1/71}

\bibitem[{{Samsing} {et~al.}(2018{\natexlab{b}}){Samsing}, {MacLeod}, \& {Ramirez-Ruiz}}]{Samsing2018}
---. 2018{\natexlab{b}}, \apj, 853, 140, \dodoi{10.3847/1538-4357/aaa715}

\bibitem[{{Samsing} \& {Ramirez-Ruiz}(2017)}]{2017ApJ...840L..14S}
{Samsing}, J., \& {Ramirez-Ruiz}, E. 2017, \apjl, 840, L14, \dodoi{10.3847/2041-8213/aa6f0b}

\bibitem[{{Samsing} {et~al.}(2025){Samsing}, {Zwick}, {Saini}, {Hendriks}, {Lo}, {Vujeva}, {Radev}, \& {Yu}}]{2025samsinglensing2}
{Samsing}, J., {Zwick}, L., {Saini}, P., {et~al.} 2025, arXiv e-prints, arXiv:2501.12494, \dodoi{10.48550/arXiv.2501.12494}

\bibitem[{{Samsing} {et~al.}(2022{\natexlab{a}}){Samsing}, {Bartos}, {D'Orazio}, {Haiman}, {Kocsis}, {Leigh}, {Liu}, {Pessah}, \& {Tagawa}}]{samsing2022}
{Samsing}, J., {Bartos}, I., {D'Orazio}, D.~J., {et~al.} 2022{\natexlab{a}}, \nat, 603, 237, \dodoi{10.1038/s41586-021-04333-1}

\bibitem[{{Samsing} {et~al.}(2022{\natexlab{b}}){Samsing}, {Bartos}, {D'Orazio}, {Haiman}, {Kocsis}, {Leigh}, {Liu}, {Pessah}, \& {Tagawa}}]{2022Natur.603..237S}
---. 2022{\natexlab{b}}, \nat, 603, 237, \dodoi{10.1038/s41586-021-04333-1}

\bibitem[{{Samsing} {et~al.}(2024{\natexlab{b}}){Samsing}, {Zwick}, {Saini}, {D'Orazio}, {Hendriks}, {Mar{\'\i}a Ezquiaga}, {Lo}, {Vujeva}, {Radev}, \& {Yu}}]{2024samsinglensing1}
{Samsing}, J., {Zwick}, L., {Saini}, P., {et~al.} 2024{\natexlab{b}}, arXiv e-prints, arXiv:2412.14159, \dodoi{10.48550/arXiv.2412.14159}

\bibitem[{{Santini} {et~al.}(2023){Santini}, {Gerosa}, {Cotesta}, \& {Berti}}]{2023santini}
{Santini}, A., {Gerosa}, D., {Cotesta}, R., \& {Berti}, E. 2023, \prd, 108, 083033, \dodoi{10.1103/PhysRevD.108.083033}

\bibitem[{{Sberna} {et~al.}(2022){Sberna}, {Babak}, {Marsat}, {Caputo}, {Cusin}, {Toubiana}, {Barausse}, {Caprini}, {Dal Canton}, {Sesana}, \& {Tamanini}}]{2022sberna}
{Sberna}, L., {Babak}, S., {Marsat}, S., {et~al.} 2022, \prd, 106, 064056, \dodoi{10.1103/PhysRevD.106.064056}

\bibitem[{Shakura \& Sunyaev(1973)}]{Shakura:1973uy}
Shakura, N.~I., \& Sunyaev, R.~A. 1973, \aap, 24, 337

\bibitem[{{Silsbee} \& {Tremaine}(2017{\natexlab{a}})}]{2017ApJ...836...39S}
{Silsbee}, K., \& {Tremaine}, S. 2017{\natexlab{a}}, \apj, 836, 39, \dodoi{10.3847/1538-4357/aa5729}

\bibitem[{{Silsbee} \& {Tremaine}(2017{\natexlab{b}})}]{silsbee2017}
---. 2017{\natexlab{b}}, \apj, 836, 39, \dodoi{10.3847/1538-4357/aa5729}

\bibitem[{{Sirko} \& {Goodman}(2003)}]{sirko2003}
{Sirko}, E., \& {Goodman}, J. 2003, \mnras, 341, 501, \dodoi{10.1046/j.1365-8711.2003.06431.x}

\bibitem[{{Speri} {et~al.}(2022){Speri}, {Antonelli}, {Sberna}, {Babak}, {Barausse}, {Gair}, \& {Katz}}]{2022speri}
{Speri}, L., {Antonelli}, A., {Sberna}, L., {et~al.} 2022, arXiv e-prints, arXiv:2207.10086.
\newblock \doarXiv{2207.10086}

\bibitem[{{Stegmann} {et~al.}(2024){Stegmann}, {Vigna-G{\'o}mez}, {Rantala}, {Wagg}, {Zwick}, {Renzo}, {van Son}, {de Mink}, \& {White}}]{2024stegmann}
{Stegmann}, J., {Vigna-G{\'o}mez}, A., {Rantala}, A., {et~al.} 2024, \apjl, 972, L19, \dodoi{10.3847/2041-8213/ad70bb}

\bibitem[{{Stone} {et~al.}(2017){Stone}, {Metzger}, \& {Haiman}}]{2017MNRAS.464..946S}
{Stone}, N.~C., {Metzger}, B.~D., \& {Haiman}, Z. 2017, \mnras, 464, 946, \dodoi{10.1093/mnras/stw2260}

\bibitem[{{Tagawa} {et~al.}(2020{\natexlab{a}}){Tagawa}, {Haiman}, \& {Kocsis}}]{2020ApJ...898...25T}
{Tagawa}, H., {Haiman}, Z., \& {Kocsis}, B. 2020{\natexlab{a}}, \apj, 898, 25, \dodoi{10.3847/1538-4357/ab9b8c}

\bibitem[{{Tagawa} {et~al.}(2020{\natexlab{b}}){Tagawa}, {Haiman}, \& {Kocsis}}]{tagawa2020}
---. 2020{\natexlab{b}}, \apj, 898, 25, \dodoi{10.3847/1538-4357/ab9b8c}

\bibitem[{Tak\'atsy {et~al.}(2019)Tak\'atsy, B\'ecsy, \& Raffai}]{Takatsy:2018euo}
Tak\'atsy, J., B\'ecsy, B., \& Raffai, P. 2019, Mon. Not. Roy. Astron. Soc., 486, 570, \dodoi{10.1093/mnras/stz820}

\bibitem[{{Tak{\'a}tsy} {et~al.}(2025){Tak{\'a}tsy}, {Zwick}, {Hendriks}, {Saini}, {Fabj}, \& {Samsing}}]{pedpaper}
{Tak{\'a}tsy}, J., {Zwick}, L., {Hendriks}, K., {et~al.} 2025, arXiv e-prints, arXiv:2505.09513, \dodoi{10.48550/arXiv.2505.09513}

\bibitem[{{Tanikawa}(2013)}]{2013MNRAS.435.1358T}
{Tanikawa}, A. 2013, \mnras, 435, 1358, \dodoi{10.1093/mnras/stt1380}

\bibitem[{{Thorpe} {et~al.}(2019){Thorpe}, {Ziemer}, {Thorpe}, {Livas}, {Conklin}, {Caldwell}, {Berti}, {McWilliams}, {Stebbins}, {Shoemaker}, {Ferrara}, {Larson}, {Shoemaker}, {Key}, {Vallisneri}, {Eracleous}, {Schnittman}, {Kamai}, {Camp}, {Mueller}, {Bellovary}, {Rioux}, {Baker}, {Bender}, {Cutler}, {Cornish}, {Hogan}, {Manthripragada}, {Ware}, {Natarajan}, {Numata}, {Sankar}, {Kelly}, {McKenzie}, {Slutsky}, {Spero}, {Hewitson}, {Francis}, {DeRosa}, {Yu}, {Hornschemeier}, \& {Wass}}]{2019lisa}
{Thorpe}, J.~I., {Ziemer}, J., {Thorpe}, I., {et~al.} 2019, in Bulletin of the American Astronomical Society, Vol.~51, 77.
\newblock \doarXiv{1907.06482}

\bibitem[{{Tiede} {et~al.}(2024){Tiede}, {D'Orazio}, {Zwick}, \& {Duffell}}]{2023Tiede}
{Tiede}, C., {D'Orazio}, D.~J., {Zwick}, L., \& {Duffell}, P.~C. 2024, \apj, 964, 46, \dodoi{10.3847/1538-4357/ad2613}

\bibitem[{{Tiede} {et~al.}(2020){Tiede}, {Zrake}, {MacFadyen}, \& {Haiman}}]{tiede2020}
{Tiede}, C., {Zrake}, J., {MacFadyen}, A., \& {Haiman}, Z. 2020, \apj, 900, 43, \dodoi{10.3847/1538-4357/aba432}

\bibitem[{{Tiwari} {et~al.}(2024){Tiwari}, {Vijaykumar}, {Kapadia}, {Fragione}, \& {Chatterjee}}]{2024MNRAS.527.8586T}
{Tiwari}, A., {Vijaykumar}, A., {Kapadia}, S.~J., {Fragione}, G., \& {Chatterjee}, S. 2024, \mnras, 527, 8586, \dodoi{10.1093/mnras/stad3749}

\bibitem[{{Toonen} {et~al.}(2018){Toonen}, {Perets}, \& {Hamers}}]{toonen2018}
{Toonen}, S., {Perets}, H.~B., \& {Hamers}, A.~S. 2018, \aap, 610, A22, \dodoi{10.1051/0004-6361/201731874}

\bibitem[{{Torres-Orjuela} {et~al.}(2019){Torres-Orjuela}, {Chen}, {Cao}, {Amaro-Seoane}, \& {Peng}}]{2019alejandro}
{Torres-Orjuela}, A., {Chen}, X., {Cao}, Z., {Amaro-Seoane}, P., \& {Peng}, P. 2019, \prd, 100, 063012, \dodoi{10.1103/PhysRevD.100.063012}

\bibitem[{{Toubiana} {et~al.}(2021){Toubiana}, {Sberna}, {Caputo}, {Cusin}, {Marsat}, {Jani}, {Babak}, {Barausse}, {Caprini}, {Pani}, {Sesana}, \& {Tamanini}}]{2021toubiana}
{Toubiana}, A., {Sberna}, L., {Caputo}, A., {et~al.} 2021, \prl, 126, 101105, \dodoi{10.1103/PhysRevLett.126.101105}

\bibitem[{{Trani} {et~al.}(2024{\natexlab{a}}){Trani}, {Leigh}, {Boekholt}, \& {Portegies Zwart}}]{2024A&A...689A..24T}
{Trani}, A.~A., {Leigh}, N. W.~C., {Boekholt}, T. C.~N., \& {Portegies Zwart}, S. 2024{\natexlab{a}}, \aap, 689, A24, \dodoi{10.1051/0004-6361/202449862}

\bibitem[{{Trani} {et~al.}(2024{\natexlab{b}}){Trani}, {Leigh}, {Boekholt}, \& {Portegies Zwart}}]{tra2024b}
---. 2024{\natexlab{b}}, \aap, 689, A24, \dodoi{10.1051/0004-6361/202449862}

\bibitem[{{Trani} {et~al.}(2023){Trani}, {Quaini}, \& {Colpi}}]{2023arXiv231213281T}
{Trani}, A.~A., {Quaini}, S., \& {Colpi}, M. 2023, arXiv e-prints, arXiv:2312.13281, \dodoi{10.48550/arXiv.2312.13281}

\bibitem[{{Trani} {et~al.}(2024{\natexlab{c}}){Trani}, {Quaini}, \& {Colpi}}]{trani2024}
---. 2024{\natexlab{c}}, \aap, 683, A135, \dodoi{10.1051/0004-6361/202347920}

\bibitem[{{Trani} {et~al.}(2022{\natexlab{a}}){Trani}, {Rastello}, {Carlo}, {Santoliquido}, {Tanikawa}, \& {Mapelli}}]{trani2022}
{Trani}, A.~A., {Rastello}, S., {Carlo}, U. N.~D., {et~al.} 2022{\natexlab{a}}, \mnras, \dodoi{10.1093/mnras/stac122}

\bibitem[{{Trani} {et~al.}(2022{\natexlab{b}}){Trani}, {Rastello}, {Di Carlo}, {Santoliquido}, {Tanikawa}, \& {Mapelli}}]{2022MNRAS.511.1362T}
{Trani}, A.~A., {Rastello}, S., {Di Carlo}, U.~N., {et~al.} 2022{\natexlab{b}}, \mnras, 511, 1362, \dodoi{10.1093/mnras/stac122}

\bibitem[{{Trani} {et~al.}(2021){Trani}, {Tanikawa}, {Fujii}, {Leigh}, \& {Kumamoto}}]{2021MNRAS.504..910T}
{Trani}, A.~A., {Tanikawa}, A., {Fujii}, M.~S., {Leigh}, N.~W.~C., \& {Kumamoto}, J. 2021, \mnras, 504, 910, \dodoi{10.1093/mnras/stab967}

\bibitem[{{Vigna-G{\'o}mez} {et~al.}(2021){Vigna-G{\'o}mez}, {Toonen}, {Ramirez-Ruiz}, {Leigh}, {Riley}, \& {Haster}}]{vignagomez2021}
{Vigna-G{\'o}mez}, A., {Toonen}, S., {Ramirez-Ruiz}, E., {et~al.} 2021, \apjl, 907, L19, \dodoi{10.3847/2041-8213/abd5b7}

\bibitem[{{Vijaykumar} {et~al.}(2024){Vijaykumar}, {Hanselman}, \& {Zevin}}]{2024vijakumar}
{Vijaykumar}, A., {Hanselman}, A.~G., \& {Zevin}, M. 2024, \apj, 969, 132, \dodoi{10.3847/1538-4357/ad4455}

\bibitem[{{Vitale} {et~al.}(2017){Vitale}, {Lynch}, {Sturani}, \& {Graff}}]{2017vitale}
{Vitale}, S., {Lynch}, R., {Sturani}, R., \& {Graff}, P. 2017, Classical and Quantum Gravity, 34, 03LT01, \dodoi{10.1088/1361-6382/aa552e}

\bibitem[{{von Zeipel}(1910)}]{zeipel1910}
{von Zeipel}, H. 1910, Astronomische Nachrichten, 183, 345, \dodoi{10.1002/asna.19091832202}

\bibitem[{{Wang} {et~al.}(2021){Wang}, {Stephan}, {Naoz}, {Hoang}, \& {Breivik}}]{2021Wang}
{Wang}, H., {Stephan}, A.~P., {Naoz}, S., {Hoang}, B.-M., \& {Breivik}, K. 2021, \apj, 917, 76, \dodoi{10.3847/1538-4357/ac088d}

\bibitem[{{Whitehead} {et~al.}(2023){Whitehead}, {Rowan}, {Boekholt}, \& {Kocsis}}]{2023whitehead}
{Whitehead}, H., {Rowan}, C., {Boekholt}, T., \& {Kocsis}, B. 2023, arXiv e-prints, arXiv:2309.11561, \dodoi{10.48550/arXiv.2309.11561}

\bibitem[{{Will}(2014)}]{2014will}
{Will}, C.~M. 2014, Classical and Quantum Gravity, 31, 244001, \dodoi{10.1088/0264-9381/31/24/244001}

\bibitem[{{Xuan} {et~al.}(2022){Xuan}, {Naoz}, \& {Chen}}]{2022xuan}
{Xuan}, Z., {Naoz}, S., \& {Chen}, X. 2022, arXiv e-prints, arXiv:2210.03129, \dodoi{10.48550/arXiv.2210.03129}

\bibitem[{{Yunes} {et~al.}(2009){Yunes}, {Arun}, {Berti}, \& {Will}}]{2009yunes}
{Yunes}, N., {Arun}, K.~G., {Berti}, E., \& {Will}, C.~M. 2009, \prd, 80, 084001, \dodoi{10.1103/PhysRevD.80.084001}

\bibitem[{{Zevin} {et~al.}(2017){Zevin}, {Pankow}, {Rodriguez}, {Sampson}, {Chase}, {Kalogera}, \& {Rasio}}]{zevin2017}
{Zevin}, M., {Pankow}, C., {Rodriguez}, C.~L., {et~al.} 2017, \apj, 846, 82, \dodoi{10.3847/1538-4357/aa8408}

\bibitem[{{Zevin} {et~al.}(2019{\natexlab{a}}){Zevin}, {Samsing}, {Rodriguez}, {Haster}, \& {Ramirez-Ruiz}}]{2019ApJ...871...91Z}
{Zevin}, M., {Samsing}, J., {Rodriguez}, C., {Haster}, C.-J., \& {Ramirez-Ruiz}, E. 2019{\natexlab{a}}, \apj, 871, 91, \dodoi{10.3847/1538-4357/aaf6ec}

\bibitem[{{Zevin} {et~al.}(2019{\natexlab{b}}){Zevin}, {Samsing}, {Rodriguez}, {Haster}, \& {Ramirez-Ruiz}}]{zevin2019}
---. 2019{\natexlab{b}}, \apj, 871, 91, \dodoi{10.3847/1538-4357/aaf6ec}

\bibitem[{{Zevin} {et~al.}(2021{\natexlab{a}}){Zevin}, {Bavera}, {Berry}, {Kalogera}, {Fragos}, {Marchant}, {Rodriguez}, {Antonini}, {Holz}, \& {Pankow}}]{zevin2020}
{Zevin}, M., {Bavera}, S.~S., {Berry}, C. P.~L., {et~al.} 2021{\natexlab{a}}, \apj, 910, 152, \dodoi{10.3847/1538-4357/abe40e}

\bibitem[{{Zevin} {et~al.}(2021{\natexlab{b}}){Zevin}, {Bavera}, {Berry}, {Kalogera}, {Fragos}, {Marchant}, {Rodriguez}, {Antonini}, {Holz}, \& {Pankow}}]{2021zevin}
---. 2021{\natexlab{b}}, \apj, 910, 152, \dodoi{10.3847/1538-4357/abe40e}

\bibitem[{{Zwick} {et~al.}(2020){Zwick}, {Capelo}, {Bortolas}, {Mayer}, \& {Amaro-Seoane}}]{2020MNRAS.495.2321Z}
{Zwick}, L., {Capelo}, P.~R., {Bortolas}, E., {Mayer}, L., \& {Amaro-Seoane}, P. 2020, \mnras, 495, 2321, \dodoi{10.1093/mnras/staa1314}

\bibitem[{{Zwick} {et~al.}(2021){Zwick}, {Capelo}, {Bortolas}, {V{\'a}zquez-Aceves}, {Mayer}, \& {Amaro-Seoane}}]{2021zwick}
{Zwick}, L., {Capelo}, P.~R., {Bortolas}, E., {et~al.} 2021, \mnras, 506, 1007, \dodoi{10.1093/mnras/stab1818}

\bibitem[{{Zwick} {et~al.}(2023){Zwick}, {Capelo}, \& {Mayer}}]{2023zwick}
{Zwick}, L., {Capelo}, P.~R., \& {Mayer}, L. 2023, \mnras, 521, 4645, \dodoi{10.1093/mnras/stad707}

\bibitem[{{Zwick} {et~al.}(2022){Zwick}, {Derdzinski}, {Garg}, {Capelo}, \& {Mayer}}]{2022zwick}
{Zwick}, L., {Derdzinski}, A., {Garg}, M., {Capelo}, P.~R., \& {Mayer}, L. 2022, \mnras, 511, 6143, \dodoi{10.1093/mnras/stac299}

\bibitem[{{Zwick} \& {Samsing}(2025)}]{zwicklensing}
{Zwick}, L., \& {Samsing}, J. 2025, arXiv e-prints, arXiv:2502.03547, \dodoi{10.48550/arXiv.2502.03547}

\bibitem[{{Zwick} {et~al.}(2024{\natexlab{a}}){Zwick}, {Tiede}, {Trani}, {Derdzinski}, {Haiman}, {D'Orazio}, \& {Samsing}}]{2024zwicknovel}
{Zwick}, L., {Tiede}, C., {Trani}, A.~A., {et~al.} 2024{\natexlab{a}}, \prd, 110, 103005, \dodoi{10.1103/PhysRevD.110.103005}

\bibitem[{{Zwick} {et~al.}(2024{\natexlab{b}}){Zwick}, {Tiede}, {Trani}, {Derdzinski}, {Haiman}, {D'Orazio}, \& {Samsing}}]{2024zwick}
---. 2024{\natexlab{b}}, \prd, 110, 103005, \dodoi{10.1103/PhysRevD.110.103005}

\end{thebibliography}


\end{document}